\documentclass[12pt,a4paper]{article}
\usepackage{mathtools}
\usepackage{mathrsfs}
\usepackage{amsmath}
\usepackage{tikz-feynman}
\usepackage[utf8x]{inputenc}
\usepackage{a mssymb}
\usepackage{slashed}
\usepackage{bbm}
\usepackage{bm}
\usepackage{color}
\usepackage[a4paper,top=3cm,bottom=3cm,left=2cm,right=3cm,bindingoffset=5mm]{geometry}
\usepackage{booktabs} 
\usepackage{tikz}
\usepackage{cancel}
\usepackage[bottom]{footmisc}
\usepackage{cite}
\usepackage{float}
\usepackage[disable]{todonotes}
\usepackage{jheppub}

\usepackage{hyperref}

\hypersetup{
colorlinks=true,         
linkcolor=blue,          
citecolor=red,        
urlcolor=red            
}

\renewcommand{\vec}[1]{\mathbf{#1}}

\newcommand\docNote[1]{
 \todo[backgroundcolor=blue!20!white,fancyline,
 bordercolor=white]{DOC: #1}}

\renewcommand{\[}{\begin{equation}\begin{aligned}}
\renewcommand{\]}{\end{aligned}\end{equation}}

\newcommand{\hel}{\eta} 
\renewcommand{\d}{\mathrm{d}}
\newcommand{\dd}{\hat{\mathrm{d}}}
\newcommand{\del}{\hat{\delta}}
\newcommand{\ket}[1]{| #1 \rangle}
\newcommand{\bra}[1]{\langle #1 |}

\renewcommand{\Re}{\operatorname{Re}}

\DeclareMathOperator{\sign}{sign}

\definecolor{allOrderBlue}{rgb}{0.4,0.5,1}

\newcommand{\ampGR}{\mathcal{M}}
\newcommand{\Xyt}{X}
\newcommand{\gauge}{n}
\newcommand{\maxwell}{\phi}

\title{Classical Solutions and their Double Copy in Split Signature}
\author[1]{Ricardo Monteiro,}
\author[2]{Donal O'Connell,}
\author[1]{David Peinador Veiga,}
\author[2]{Matteo Sergola}

\affiliation[1]{Centre for Research in String Theory, School of Physics and Astronomy, Queen Mary University of London, E1 4NS, United Kingdom}
\affiliation[2]{Higgs Centre for Theoretical Physics, School of Physics and Astronomy, The University of Edinburgh, EH9 3FD, Scotland}

\abstract{
The three-point amplitude is the key building block in the on-shell approach to scattering amplitudes. We show that the classical objects computed by massive three-point amplitudes in gauge theory and gravity are Newman-Penrose scalars in a split-signature spacetime, where three-point amplitudes can be defined for real kinematics. In fact, the quantum state set up by the particle is a coherent state fully determined by the three-point amplitude due to an eikonal-type exponentiation. Having identified this simplest classical solution from the perspective of scattering amplitudes, we explore the double copy of the Newman-Penrose scalars induced by the traditional double copy of amplitudes, and find that it coincides with the Weyl version of the classical double copy. We also exploit the Kerr-Schild version of the classical double copy to determine the exact spacetime metric in the gravitational case. Finally, we discuss the direct implication of these results for Lorentzian signature via analytic continuation.
}

\begin{document}
\maketitle

\section{Introduction}
\label{sec:intro}

Three-point scattering amplitudes are the atoms in our modern approach to computing interactions between particles in quantum field theory. Using
BCFW~\cite{Britto:2005fq} and generalised unitarity~\cite{Bern:1994zx,Bern:1994cg}, it is possible to construct the complete $S$-matrix for
Yang-Mills theory and (up to ultraviolet divergences) for general relativity from their respective three-point amplitudes. These amplitudes are gauge invariant 
and beautifully simple objects, completely specified by the helicities of the massless gluons and gravitons~\cite{Benincasa:2007xk}. This 
basic simplicity  carries over to the  case of massive particles, for any spin~\cite{ah3}. But in spite of all these virtues, three-point amplitudes 
have one big defect: they do not exist in Minkowski space. As for any $n$-point amplitude, the external particles involved in a three-point amplitude
must all be on shell. But there is no solution to the on-shell conditions in Minkowski space for three particles with different momenta.

Recent years have seen a surprising new application of scattering amplitudes in classical physics, motivated especially by gravitational
wave physics~\cite{Neill:2013wsa,Bjerrum-Bohr:2013bxa,Monteiro:2014cda,Bjerrum-Bohr:2014zsa,Luna:2016due,Damour:2016gwp,Goldberger:2016iau,Cachazo:2017jef,Guevara:2017csg,Damour:2017zjx,Luna:2017dtq,Laddha:2018rle,Laddha:2018vbn,Bjerrum-Bohr:2018xdl,Cheung:2018wkq,Kosower:2018adc,Guevara:2018wpp,Bern:2019nnu,Cristofoli:2019neg,Maybee:2019jus,Guevara:2019fsj,Bern:2019crd,Kalin:2019rwq,Kalin:2019inp,Aoude:2020onz,Cheung:2020gyp,Bern:2020buy,Cheung:2020sdj,Kalin:2020fhe,Haddad:2020que,Kalin:2020lmz,DiVecchia:2020ymx,Bern:2020uwk,Huber:2020xny}.
This classical application has motivated a renewed interest in the wider applications of amplitudes~\cite{Bjerrum-Bohr:2017dxw,Laddha:2018myi,Sahoo:2018lxl,Bautista:2019tdr,Brandhuber:2019qpg,Laddha:2019yaj,Arkani-Hamed:2019ymq,Damgaard:2019lfh,Bjerrum-Bohr:2019kec,Huang:2019cja,Huber:2019ugz,Saha:2019tub,Bern:2020gjj,Moynihan:2020gxj,Cristofoli:2020uzm,Parra-Martinez:2020dzs,Haddad:2020tvs,AccettulliHuber:2020oou,Moynihan:2020ejh,A:2020lub,Sahoo:2020ryf,delaCruz:2020bbn,Bonocore:2020xuj,Mogull:2020sak,Emond:2020lwi,Cheung:2020gbf,Mougiakakos:2020laz,Carrasco:2020ywq,Kim:2020cvf,Bjerrum-Bohr:2020syg,Gonzo:2020xza,delaCruz:2020cpc}. 
It is now evident that the tools of quantum field theory --- for example, the double copy --- have interesting implications for 
classical physics. Certain amplitudes are closely connected with specific classical concepts: for example, the four-point amplitude between
massive particles in gravity is closely related to the classical potential~\cite{Donoghue:1993eb,Donoghue:1994dn}. But the three-point amplitude has so far received no classical interpretation,
because it is not present in Minkowski space.

Of course, the fact that the three-point amplitude vanishes in Minkowski space is no obstacle for the programme of determining more complicated
amplitudes. BCFW taught us a simple trick: we analytically continue the momenta so that the on-shell conditions \emph{do} have a solution. We can
take the momenta to be complex-valued, or else continue to a spacetime with metric signature $(+,+,-,-)$.\footnote{In our conventions, the $(+)$-directions are timelike and the $(-)$-directions are spacelike.} This second option has some
conceptual virtues: we can choose real momenta and polarisation vectors; the spinor variables we frequently use exist and are real; the
chirality properties in a four-dimensional manifold with this split signature mean that the two types of spinors are independent. For related discussions of field theory in split signature, see \cite{Srednyak:2013ylj,Mason:2005qu,Barrett:1993yn}.

Another virtue of a real spacetime with signature $(+,+,-,-)$ is that real classical equations exist in this spacetime and their solutions can be
studied. In this paper, we find a classical interpretation for the three-point amplitude in a split-signature spacetime: it computes the 
Newman-Penrose scalars~\cite{Newman:1961qr} (a spinorial version of the curvature of the field) for the classical solution that is generated by the massive particle in the amplitude.
For example, the three-point amplitude between a massive scalar and a gauge boson computes the electromagnetic field strength 
of a static point charge in split signature. In gravity, the three-point amplitude between a massive scalar and a graviton 
computes the Weyl spinor of the split-signature analogue of the Schwarzschild solution. 
Solutions in split signature which are determined by three-point amplitudes are, from the perspective of scattering amplitudes, the simplest
non-trivial classical solutions.

The Newman-Penrose (NP) formalism can be illuminated by taking a spinorial approach to field theory. The Lorentz group in split signature is locally isomorphic to $\mathrm{SL}(2, \mathbb{R}) \otimes \mathrm{SL}(2, \mathbb{R})$, and the 
spinorial representations of $\mathrm{SO}(2, 2)$ are the (real) two-dimensional fundamental representations of each 
$\mathrm{SL}(2, \mathbb{R})$ factor.
In electrodynamics, for example, we can pass from the tensorial field strength $F_{\mu\nu}(x)$ to a spinorial equivalent known as
the Maxwell spinor $\maxwell_{\alpha\beta}(x)$. This is obtained by contracting the Lorentz indices of $F_{\mu\nu}(x)$ with matrices 
$\sigma^{\mu\nu}{}_{\alpha\beta}$ which are proportional to the Lorentz generators in the spinor representation 
(that is, the $\sigma^{\mu\nu}{}_{\alpha\beta}$ generate one of the $\mathrm{SL}(2, \mathbb{R})$ subgroups of the Lorentz group). We have
\[
\maxwell_{\alpha\beta}(x) = \sigma^{\mu\nu}{}_{\alpha\beta} F_{\mu\nu}(x)\,.
\label{eq:defOfMaxwell}
\]
In split signature, the Maxwell spinor is a real 
quantity, symmetric in its
spinor indices. There is a second Maxwell spinor associated with the spinor representation of the other chirality:
\[
\tilde \maxwell_{\dot\alpha\dot\beta}(x) = \tilde{\sigma}^{\mu\nu}{}_{\dot\alpha\dot\beta} F_{\mu\nu}(x)\,,
\]
where $\tilde\sigma^{\mu\nu}{}_{\dot \alpha \dot \beta}$ are again proportional to the Lorentz generators, but now of the other ``antichiral'' 
$\mathrm{SL}(2, \mathbb{R})$ subgroup.

To obtain Newman-Penrose scalars, we expand the Maxwell spinor (and its antichiral friend) on a basis of spinors. Let us consider the Maxwell
spinor due to some localised source, such as a point-like charge. Solving the field equations with a retarded boundary condition, we can introduce spinors at any spacetime point by taking the light-cone direction $k$ from the charge to the point. Using the
notation of spinor-helicity, the vector $k$ is also the bispinor $\ket{k} [k|$. To complete the basis of (chiral) spinors, we choose another 
spinor $\ket{\gauge}$. Now we may write out the Maxwell spinor in this basis:\footnote{Details of our notation can be found in appendix~\ref{sec:conventions}. For later convenience, our (anti)symmetrisation symbols do not include the $1/n!$ factor.}
\[
\label{eq:maxwellkn}
\maxwell_{\alpha\beta}(x) = \phi_0(x) \ket{\gauge}_\alpha \ket{\gauge}_\beta -\,\phi_1(x) \ket{k}_{(\alpha} \ket\gauge_{\beta)} + \phi_2(x) \ket{k}_\alpha \ket{k}_\beta \,.
\]
The three scalar fields $\phi_i(x)$ are Newman-Penrose scalars. There are three more NP scalars in the antichiral field strength: these are
the six different components of the field strength. In split signature, all the quantities in \eqref{eq:maxwellkn} are real, and the chiral quantities are independent from the antichiral ones.

In gravity, the story is very similar. We pass from the Weyl curvature $W_{\mu\nu\rho\sigma}(x)$
(via a frame) to a Weyl spinor $\Psi_{\alpha\beta\gamma\delta}(x)$, which is real and completely symmetric in its four spinor indices. Expanding the Weyl spinor
on our basis of spinors, we encounter five real NP scalars, namely
\[
\label{eq:weylkn}
\Psi_{\alpha\beta\gamma\delta}(x) = \Psi_0(x) &\ket{\gauge}_{\alpha} \ket{\gauge}_\beta \ket{\gauge}_\gamma \ket{\gauge}_{\delta}
-\frac1{6}\, \Psi_1(x) \ket{k}_{(\alpha} \ket{\gauge}_\beta \ket{\gauge}_\gamma \ket{\gauge}_{\delta)} \\
&+ \frac1{4}\, \Psi_2(x) \ket{k}_{(\alpha} \ket{k}_\beta \ket{\gauge}_\gamma \ket{\gauge}_{\delta)}
-\frac1{6}\, \Psi_3(x) \ket{k}_{(\alpha} \ket{k}_\beta \ket{k}_\gamma \ket{\gauge}_{\delta)} \\
&\hspace{150pt}+ \Psi_4(x) \ket{k}_{\alpha} \ket{k}_\beta \ket{k}_\gamma \ket{k}_{\delta} \,.
\]
Together with their compatriots in the antichiral Weyl spinor $\tilde \Psi$, these are the ten real components of the Weyl tensor.

In Minkowski space, the NP scalars have an important property known as peeling \cite{Sachs:1961zz,Newman:1961qr}. This is a hierarchy in their fall-off with large distance $r$
between the observer and the localised source. In electrodynamics, we have
\[
\phi_0(x) &= \phi_0^1(\bar x)\, \frac{1}{r^3} + \mathcal{O}(1/r^4) \,,\\
\phi_1(x) &= \phi_1^1(\bar x)\, \frac{1}{r^2} + \mathcal{O}(1/r^3) \,,\\
\phi_2(x) &= \phi_2^1(\bar x)\, \frac{1}{r^1} + \mathcal{O}(1/r^2) \,,
\]
where $\bar x$ denotes non-radial dependence.
Thus, the scalar $\phi_2(x)$ is the dominant component of the field at large distances: it describes the asymptotic radiation field. Meanwhile,
$\phi_1(x)$ is Coulombic. In gravity, the situation is very similar:
\[
\Psi_0(x) &= \Psi_0^1(\bar x)\, \frac{1}{r^5} + \mathcal{O}(1/r^6) \,,\\
\Psi_1(x) &= \Psi_1^1(\bar x)\, \frac{1}{r^4} + \mathcal{O}(1/r^5) \,,\\
\Psi_2(x) &= \Psi_2^1(\bar x)\, \frac{1}{r^3} + \mathcal{O}(1/r^4) \,,\\
\Psi_3(x) &= \Psi_3^1(\bar x)\, \frac{1}{r^2} + \mathcal{O}(1/r^3) \,,\\
\Psi_4(x) &= \Psi_4^1(\bar x)\, \frac{1}{r^1} + \mathcal{O}(1/r^2) \,.
\]
Asymptotic gravitational radiation is described by $\Psi_4(x)$, while $\Psi_2(x)$ describes a potential-type contribution, as in Schwarzschild.
We will see aspects of this structure in our split-signature examples.

The double copy relation between scattering amplitudes in gravity and in Yang-Mills theory \cite{klt,Bern:2008qj,Bern:2010ue,zvirev} is quite a surprise from the classical
geometric perspective on general relativity: geometrically, there seems to be little hint that gravity is some kind of ``square'' of
Yang-Mills theory. But it has always been clear that some aspects of gravity are analogues of aspects of gauge theory (or, in simple settings, of
electrodynamics), and the application of the double copy to classical solutions in recent years has provided a unified understanding to several such analogies~\cite{Monteiro:2011pc,Saotome:2012vy,Anastasiou:2014qba,Monteiro:2014cda,Luna:2015paa,Luna:2016due,Cardoso:2016ngt,Goldberger:2016iau,Luna:2016hge,Cardoso:2016amd,Luna:2017dtq,Adamo:2017nia,Ilderton:2018lsf,Anastasiou:2018rdx,Lee:2018gxc,Plefka:2018dpa,Berman:2018hwd,Luna:2018dpt,Andrzejewski:2019hub,Sabharwal:2019ngs,Cho:2019ype,Plefka:2019hmz,Godazgar:2019ikr,Bautista:2019evw,Bah:2019sda,Alawadhi:2019urr,Kim:2019jwm,Borsten:2019prq,Banerjee:2019saj,Goldberger:2019xef,Luna:2020adi,Cristofoli:2020hnk,Keeler:2020rcv,Bahjat-Abbas:2020cyb,Elor:2020nqe,Alawadhi:2020jrv,Alfonsi:2020lub,Adamo:2020qru,Borsten:2020xbt,Borsten:2020zgj,Chacon:2020fmr,Godazgar:2020zbv,Ferrero:2020vww,White:2020sfn,Prabhu:2020avf,Berman:2020xvs,Easson:2020esh,Lescano:2020nve}. For instance, the structure of the gravitational Newman-Penrose scalars is evidently analogous to that of the electromagnetic
NP scalars. This is particularly clear for special classes of solutions, such as the Petrov type N class, which has only $\Psi_4 \neq 0$ for an appropriate choice of spinor basis.
Then the Weyl spinor is simply
\[
\Psi_{\alpha\beta\gamma\delta}(x) = 
\Psi_4(x) \ket{k}_{\alpha} \ket{k}_\beta \ket{k}_\gamma \ket{k}_{\delta} \,.
\]
In electrodynamics, we can consider a similar situation where the Maxwell spinor is simply
\[
\maxwell_{\alpha\beta}(x) = \phi_2(x) \ket{k}_\alpha \ket{k}_\beta \,.
\]
Roughly speaking, type N spacetimes look like two copies of purely radiative electromagnetic solutions. A more careful 
analysis led to a sharp proposal of an exact ``Weyl'' double copy for special classes of solutions~\cite{Luna:2018dpt}, where
the Maxwell and Weyl spinors are related by
\[
\Psi_{\alpha\beta\gamma\delta}(x) = \frac{1}{S(x)} \phi_{(\alpha\beta}(x) \phi_{\gamma\delta)}(x) \,.
\label{eq:weylDoubleCopy}
\]
Here, $S(x)$ is a scalar field satisfying the (flat space) wave equation. The proposal was first proven for vacuum solutions of type D, which
have only $\Psi_2 \neq 0$, but has also been studied for vacuum solutions of type N~\cite{Godazgar:2020zbv}; see \cite{White:2020sfn} for the relation to the twistor correspondence in the linearised case.
In split signature, we will show that the double copy relation between the three-point amplitudes in gauge theory and gravity directly relates the 
Newman-Penrose scalars of the Coulomb charge and the Schwarzschild solution at linearised level.\docNote{Isn't this a bit of an undersell? Linearised
results equal exact results because of Kerr-Schild}
This relation between the Newman-Penrose scalars in gauge theory and gravity is directly expressed in the on-shell momentum space formalism of \cite{Kosower:2018adc}, but the translation to position space for these particular solutions precisely reproduces the Weyl 
double copy \eqref{eq:weylDoubleCopy}.

At the quantum level, it is natural to expect that the Coulomb field or the Schwarzschild field  should be 
described by a coherent state. For instance, in the Schwarzschild case, the metric would be given by the expectation value of the all-order metric quantum operator on the coherent state (this operator would include all higher-order perturbative terms). We show that the coherent state is uniquely described by the relevant three-point amplitude. This is a
gauge-invariant characterisation of the classical field. Thus, both the field strengths and the coherent state are determined by the same
data. This is satisfying: classically, knowledge of the field strength is complete knowledge of the field, so it should be that one can
determine the coherent state from the field strength. Indeed, this is the case. The structure of the coherent state we encounter is strongly
reminiscent of the eikonal exponentiation which is receiving renewed attention in the context of the dialog between scattering amplitudes
and classical physics. 

Our results concerning the classical double copy have direct implications for Lorentzian signature. Indeed, as we will see, the Newman-Penrose scalars we construct have a close Lorentzian analogue. The Coulombic $\phi_1(x)$ and the Schwarzschild-like $\Psi_2(x)$ that we compute from our coherent states in split signature are essentially trivial analytic continuations of their Minkowski-space counterparts.

The double copy between Coulomb and Schwarzschild is
expected to be exact, but our methods based on amplitudes are perturbative.
To go beyond perturbation theory, we use the Kerr-Schild double copy \cite{Monteiro:2014cda} to find the exact classical metric
set up by our static particle, subject to the precise boundary conditions we impose in split signature. 
We believe that this example is the first time that the double copy has been used to find a novel exact solution in gravity. While we could in principle obtain the exact solution using purely gravitational methods, some care would be required to ensure that the correct boundary conditions are imposed at non-linear level. Using the Kerr-Schild double copy, the boundary conditions in gravity are trivially imported from those of the `single copy' gauge theory solution.

This paper focuses on amplitudes and the Newman-Penrose formalism in split signature, and on implications for Lorentzian signature. Two companion papers~\cite{cgkoc,kerrEFT} will appear shortly, discussing
similar ideas in purely Minkowski space and an application to the effective dynamics of the Kerr black hole.

Our paper is organised as follows. In section~\ref{sec:quantum}, we explain
how to compute electromagnetic and gravitational field strengths using the methods of quantum field theory, making direct contact with 
three-point amplitudes. We also discuss the corresponding coherent states.
We point out a double copy between the Maxwell and Weyl spinors
in momentum space, induced directly by the corresponding
amplitudes. Building on this observation in section~\ref{ref:WeylDCAmps}, we determine the nature of
the double copy in position space by performing integrals over on-shell
momentum space. We recover the Weyl double copy, thereby directly connecting
the Weyl form of the classical double copy to scattering amplitudes.
In section~\ref{sec:classicalGR}, we use a Kerr-Schild Ansatz to determine the 
exact spacetime metric in the
gravitational context. The implications of our split-signature results for Minkowski space are described in section~\ref{sec:lorentzian}. Finally, section~\ref{sec:discussion} contains a
summary of our results with an overview of some of their implications.
We describe our conventions
in appendix~\ref{sec:conventions}, and provide a detailed exposition of our choice of retarded Green's function for split signature in appendix~\ref{sec:retardedGF}.

\section{Classical Solutions from Three-Point Amplitudes}
\label{sec:quantum}

To connect three-point amplitudes to Newman-Penrose scalars, all that is needed is a direct computation using the methods of quantum
field theory.  The first order of business, then, is to define the quantum fields we use in split signature. 

Given that our spacetime has two time directions, which we will denote as $t^1$ and $t^2$, there are two notions of energy. Correspondingly, 
the choice of vacuum is not unique. Much
of the interesting physics we exploit actually arises from this non-uniqueness. For our force-carrying ``messenger'' particles (photons or gravitons),
we impose the condition that the fields are in a vacuum state for $t^1 \rightarrow -\infty$. The corresponding mode expansion of the field operator
in the electromagnetic case is then\footnote{The notation is that $(a_\hel(k))^\dagger\equiv a^\dagger_\hel(k)$. Notice that the helicity polarisation vectors are real in split signature.}
\[
A^\mu(x)=\sum_{\hel=\pm} \int \d\Phi(k)\,\hbar^{-\frac{1}{2}} \left( a_\hel(k)\varepsilon_\hel ^\mu(k) e^{-i\frac{k\cdot x}{\hbar}}+a^\dagger_\hel (k) \varepsilon_{\hel}^{\mu}(k) e^{i\frac{k\cdot x}{\hbar}}
\right),
\]
where the position and momentum are given by $x = (t^1, t^2, \vec{x})$
and $k=(E^1, E^2, \vec{k})$, while the measure is
\begin{equation}
\d\Phi(k)=\dd^4 k \, \del(k^2)\Theta(E^1) \,.
\end{equation}
Details of our notation can be found in appendix~\ref{sec:conventions}.
The sum is over the helicity $\hel$. Notice that we have retained factors of $\hbar$; it will be reassuring to check that these factors 
drop out for classical quantities. The theta function ensures that quanta created around the vacuum have momenta directed into the future
with respect to $t^1$; in other words, they have positive energy with respect to this choice of time direction.

We also introduce a scalar particle which will be our source. In order for our calculation to be in the regime of validity of the classical 
approximation, we place our particle in a wave packet of the type discussed in detail in reference~\cite{Kosower:2018adc}. We will discuss the
properties of these wave packets in more detail shortly. For now, note simply that the wave packet is such that the uncertainties in
the position and the momentum of our source are small.
We will treat this scalar particle as a probe.

To benefit from the unusual possibilities of a split-signature spacetime, we choose the expectation value of the probe's momentum to be
$\langle p^\mu \rangle = m\, u^\mu = m\, (0, 1, 0, 0)$. Thus, the particle's worldline can be chosen to be the $t^2$ axis.
As a probe particle, we will not need a field operator for this state. It is enough to define the state itself:
\begin{equation}\label{state}
|\psi\rangle=\int \d\Phi(p) \, \varphi (p) \, |p\rangle, \,\,\,  \d\Phi(p)=\dd^4 p\, \del(p^2-m^2)\Theta(E_2) \,,
\end{equation}
where the wave function $\varphi(p)$ is sharply-peaked around the momentum $p^\mu = m u^\mu$. Note that in this case the theta function\footnote{We write 
$\Theta(E_2)$ rather than $\Theta(E^2)$ to emphasise that the second component
of the momentum vector is constrained, avoiding confusion with a squared energy. Recall that $E_2 = E^2$ in our conventions.}
enforces positive energy along $t^2$. For brevity of notation, we left it implied that a 
measure $\d\Phi(p)$ involves a factor $\Theta(E_2)$ while a measure $\d\Phi(k)$ involves $\Theta(E^1) = \Theta(E_1)$.

\subsection{The electromagnetic case}
\label{sec:quantumEM}

Now, let us investigate the electromagnetic field set up by endowing our probe with a charge $Q$. For large negative $t^1$ we have chosen a trivial electromagnetic field. To characterise the field
for other times $t^1$ we must perform a computation. As we will see, the result is non-trivial.

We evolve the state along $t^1$ with 
\begin{equation}
|\psi_\text{out}\rangle=\lim_{t^1\to \infty}U(-t^1, t^1)|\psi\rangle=S|\psi\rangle,
\end{equation}
and we measure the expectation value of the quantum operator
\begin{equation}
F^{\mu\nu}(x)=-i\sum_{\hel=\pm} \int \d\Phi(k)\hbar^{-\frac{3}{2}} \left( a_\hel(k) k^{[\mu}\varepsilon_\hel ^{\nu]} e^{-i\frac{k\cdot x}{\hbar}}-a^\dagger _\hel (k)k^{[\mu}\varepsilon_\hel ^{\nu]} e^{i\frac{k\cdot x}{\hbar}}
\right)\,.
\end{equation}
While the scattering picture may suggest that we reproduce the electromagnetic field only for large positive $t^1$, in fact we reproduce the field for any time $t^1$ 
much larger than any time scale characteristic of the scattering. In our case, the 
largest spacetime length associated with the scattering is the size of the wave packet of the source particle (the Compton wavelength of the particle is very small compared to the size of the wave packet \cite{Kosower:2018adc}). We will discuss these scales in more detail momentarily.

Defining the $T$ matrix via $S=1+iT$, we find that  this expectation value is
\[
\langle F^{\mu\nu} \rangle\equiv\langle \psi|S^\dagger F^{\mu\nu} S |\psi\rangle=  2\,\text{Re}\,i\langle \psi| F^{\mu\nu} T|\psi\rangle +\langle \psi|T^\dagger F^{\mu\nu}T|\psi\rangle \,.
\label{eq:fexpect}
\]
Notice that we imposed
\[
\langle \psi|F^{\mu\nu}|\psi\rangle=0 \,,
\]
which holds because of our boundary conditions (there are no photons in the initial state).

In the Minkowski case, the expectation value of the field of a static massive charge is of course the Coulomb field, and can be computed 
exactly. In our split-signature case the expectation value, although less familiar, is evidently some sort of analytic continuation of Coulomb. 
We will determine the field to all orders of 
perturbation theory below. Before doing so, however, it is instructive to compute the 
leading order field strength, closely following the methods of KMOC~\cite{Kosower:2018adc}.

\begin{figure}[t]
\begin{center}
\tikzset{every picture/.style={line width=0.75pt}} 

\begin{tikzpicture}[x=0.75pt,y=0.75pt,yscale=-1,xscale=1]

\draw    (160,830) -- (90,830) ;
\draw [shift={(125,830)}, rotate = 180] [fill={rgb, 255:red, 0; green, 0; blue, 0 }  ][line width=0.08]  [draw opacity=0] (5.36,-2.57) -- (0,0) -- (5.36,2.57) -- cycle    ;
\draw    (90,830) -- (39.44,798.5) ;
\draw [shift={(64.72,814.25)}, rotate = 211.93] [fill={rgb, 255:red, 0; green, 0; blue, 0 }  ][line width=0.08]  [draw opacity=0] (5.36,-2.57) -- (0,0) -- (5.36,2.57) -- cycle    ;
\draw    (90,830) .. controls (89.46,832.29) and (88.04,833.17) .. (85.75,832.63) .. controls (83.46,832.08) and (82.04,832.96) .. (81.49,835.25) .. controls (80.95,837.54) and (79.53,838.42) .. (77.24,837.88) .. controls (74.95,837.33) and (73.53,838.21) .. (72.98,840.5) .. controls (72.44,842.79) and (71.02,843.67) .. (68.73,843.13) .. controls (66.44,842.59) and (65.02,843.47) .. (64.47,845.76) .. controls (63.93,848.05) and (62.51,848.93) .. (60.22,848.38) .. controls (57.93,847.84) and (56.51,848.72) .. (55.96,851.01) .. controls (55.42,853.3) and (54,854.18) .. (51.71,853.63) .. controls (49.42,853.09) and (48,853.97) .. (47.45,856.26) .. controls (46.91,858.55) and (45.49,859.43) .. (43.2,858.88) .. controls (40.91,858.34) and (39.49,859.22) .. (38.94,861.51) -- (38.78,861.61) -- (38.78,861.61) ;
\draw    (59.44,796.5) -- (74.05,806.85) ;
\draw [shift={(76.5,808.58)}, rotate = 215.32] [fill={rgb, 255:red, 0; green, 0; blue, 0 }  ][line width=0.08]  [draw opacity=0] (5.36,-2.57) -- (0,0) -- (5.36,2.57) -- cycle    ;
\draw    (108.78,819.83) -- (127.78,819.83) ;
\draw [shift={(130.78,819.83)}, rotate = 180] [fill={rgb, 255:red, 0; green, 0; blue, 0 }  ][line width=0.08]  [draw opacity=0] (5.36,-2.57) -- (0,0) -- (5.36,2.57) -- cycle    ;
\draw    (59.44,859.39) -- (74.36,848.49) ;
\draw [shift={(76.78,846.72)}, rotate = 503.84] [fill={rgb, 255:red, 0; green, 0; blue, 0 }  ][line width=0.08]  [draw opacity=0] (5.36,-2.57) -- (0,0) -- (5.36,2.57) -- cycle    ;

\draw (44.67,778.72) node [anchor=north west][inner sep=0.75pt]  [font=\footnotesize]  {$p-k$};
\draw (115,800.39) node [anchor=north west][inner sep=0.75pt]  [font=\footnotesize]  {$p$};
\draw (68.11,858.06) node [anchor=north west][inner sep=0.75pt]  [font=\footnotesize]  {$k,\ \hel $};

\end{tikzpicture}

\end{center}
\caption{The three-point electromagnetic amplitude. Notice that the photon with polarization $\hel$ is incoming.}
\label{fig:3point}
\end{figure}
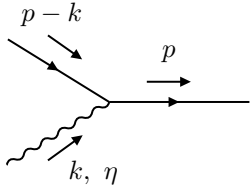

At leading order in perturbation theory, we can approximate
\begin{equation}
\langle F^{\mu\nu}(x) \rangle\simeq  2\,\text{Re}\,i\langle \psi| F^{\mu\nu}(x) T|\psi\rangle.
\label{eq:offensiveApproximation}
\end{equation}
Inserting the explicit initial state of equation~\eqref{state}, the expectation value becomes
\[
\label{ev}
\hspace{-11pt}
\langle F^{\mu\nu}&(x) \rangle=2 \hbar^{-\frac{3}{2}} \,\text{Re} \sum_{\hel} \int \d\Phi(k) \, \langle\psi|  a_\hel(k)T|\psi\rangle \,  k^{[\mu}\varepsilon_\hel ^{\nu]} \, e^{-i\frac{k\cdot x}{\hbar}}
\\=&
\,2\hbar^{-\frac{3}{2}}\,\text{Re} \sum_{\hel} \int \d\Phi (k)\d\Phi (p')\d\Phi (p) \, \varphi^*(p')\varphi(p)\, \langle p'|a_\hel(k)T|p\rangle \,  k^{[\mu}\varepsilon_\hel ^{\nu]} \, e^{-i\frac{k\cdot x}{\hbar}}.
\]
Expanding the matrix element $\langle p'|a_\hel(k)T|p\rangle$ appearing in equation~\eqref{ev} in terms of a three-point amplitude and
the momentum-conserving delta function, we can equivalently write
\[
\langle F^{\mu\nu}(x) \rangle=2 \hbar^{-\frac{3}{2}} \,\text{Re} \sum_{\hel} \int \d\Phi(k)& \d\Phi (p) \, \Theta(E^2 + k^2) \del(2 p \cdot k + k^2) 
\\
& \times
\varphi^*(p) \varphi(p+k)  \,
\mathcal{A}^{(3)}_{-\hel} (k) \, k^{[\mu}\varepsilon_{\hel} ^{\nu]}\,e^{-i{k}\cdot x/\hbar} \,.
\label{eq:fieldstrengthStillQuantum}
\]
where $\mathcal{A}^{(3)}_{\hel}(k)$ is the three-point scattering amplitude for the process shown in figure~\ref{fig:3point}.
The helicity labels of our amplitudes are for incoming messengers; since our photons are outgoing, we encounter the amplitude for the opposite 
helicity $-\hel$.

This expression for the field strength simplifies in the classical approximation. As argued by KMOC~\cite{Kosower:2018adc}, the classical approximation
is valid when the scales in our problem satisfy $x \gg \ell_w \gg \ell_c$, where $\ell_w$ is the length scale associated with the finite size of the
spatial wave packet, which controls the quantum uncertainty in the position of our source particle, while $\ell_c = \hbar / m$ is the (reduced)
Compton wavelength of the particle.\footnote{In KMOC, the role of the observer position $x$ was played by an impact parameter $b$.} Working
in Fourier space, we require that $k \ll 1/\ell_w \ll m$ (where $k$ is a messenger momentum). It is only when these inequalities are satisfied that 
our classical expressions are valid. We assume that the integrals appearing in equation~\eqref{eq:fieldstrengthStillQuantum} are 
defined (e.g.~with cutoffs) so that these inequalities are satisfied.

Taking advantage of the classical approximation, we can ignore the explicit theta function in equation~\eqref{eq:fieldstrengthStillQuantum},
since $k^2$ is a small momentum component compared to the large, positive classical energy $E^2$ of the massive particle, which is of order $m$. Similarly,
we can ignore the shift $k$ in the wave function $\varphi(p +k) \simeq \varphi(p)$, because this shift is small on the scale $1/\ell_w$ of the wave function.
It is also useful to introduce a wavenumber $\bar{k}$ associated with the momentum $k$ by $k=\hbar \bar{k}$, so that we have
\[
\langle F^{\mu\nu}(x) \rangle=2 \Re \sum_{\hel} \int \d\Phi(\bar{k})& \d\Phi (p) \, |\varphi(p)|^2 \, \del(2 p \cdot \bar k + \hbar \bar k^2) 
\sqrt{\hbar}\mathcal{A}^{(3)}_{-\hel} (\bar{k}) \, \bar{k}^{[\mu}\varepsilon_{\hel} ^{\nu]}\,e^{-i\bar{k}\cdot x} \,.
\]
Now, the wave function is sharply-peaked about an average (classical) momentum $mu^\mu$. The integral of this sharply-peaked function
over the amplitude, which is smooth near the peak momentum, sets the momenta appearing in the amplitude to $mu^\mu$, and at the
same time will broaden the explicit delta function. We can therefore drop the $\hbar \bar{k}^2$ shift in the delta function~\cite{Kosower:2018adc}, arriving at
\[
\langle F^{\mu\nu}(x) \rangle &= \frac{1}{m}\,\text{Re} \sum_{\hel} \int \d\Phi (\bar{k})\, \del_{\ell_w}(u\cdot \bar{k}) \, \sqrt{\hbar} \, \mathcal{A}^{(3)}_{-\hel}(\bar{k}) \, \bar k^{[\mu}\varepsilon_{\hel} ^{\nu]}\,e^{-i\bar{k}\cdot x} \,, \label{eq:fieldstrengthStep}
\]
where $\delta_{\ell_w}$ is a sharply-peaked function of width $\ell_w$. Below,
we assume that this length $\ell_w$ is negligible, so that we may think of our
source particle as being point-like. Since this is the largest spacetime scale
involved in the scattering calculation, our results for the behaviour of the field
are valid (in the classical point-source approximation) at all points away from the
source worldline. For example, neglecting $\ell_w$, the classical field strength is
\[
\langle F^{\mu\nu}(x) \rangle &= \frac{1}{m}\,\text{Re} \sum_{\hel} \int \d\Phi (\bar{k})\, \del(u\cdot \bar{k}) \, \sqrt{\hbar} \, \mathcal{A}^{(3)}_{-\hel}(\bar{k}) \, \bar k^{[\mu}\varepsilon_{\hel} ^{\nu]}\,e^{-i\bar{k}\cdot x} \,. \label{eq:fieldstrength}
\]
Notice that the field strength is given by the scattering amplitude, up to a universal (theory independent) integration and essential kinematic factors.

For our static charge in electromagnetism, the amplitude is the three-point scalar QED vertex,
\begin{equation}\label{3sqed}
\begin{split}
&\mathcal{A}^{(3)}_-(k)=-2 \frac{Q}{\sqrt{\hbar}}\,p\cdot \varepsilon_-({k})= \sqrt{2}m\,\frac{Q}{\sqrt{\hbar}} \, \frac{1}{X},\qquad \frac{1}{X}:=-\sqrt{2} u\cdot \varepsilon_-,\\&
\mathcal{A}^{(3)}_+(k)=-2 \frac{Q}{\sqrt{\hbar}}\,p\cdot \varepsilon_+({k})= -\sqrt{2}m\,\frac{Q}{\sqrt{\hbar}}X, \,\qquad X:=\sqrt{2}u\cdot \varepsilon_{+},
\end{split}
\end{equation}
where 
we have written $\mathcal{A}$ in terms of the kinematics-dependent $X$-factor introduced in \cite{ah3}.\footnote{We use the notation $X$ for this
factor rather than $x$ to avoid confusion with the position $x$.} Notice that the amplitude
depends on $k$ only through the polarisation vector $\varepsilon_\hel(k)$: it therefore does not depend on whether we treat $k$ as
a momentum or as a wave vector.

Taking the factor $1/\sqrt{\hbar}$ in the amplitude into account, we see that the $\hbar$ dependence of equation~\eqref{eq:fieldstrength} 
reassuringly drops out. This is obviously consistent with the computation of a classical quantity. Since all factors of 
$\hbar$ will similarly disappear for classical quantities in the remainder of the paper, we will henceforth set $\hbar = 1$, restoring it
only when necessary.

Our expressions simplify further if we pass from the field strength tensor to the associated spinorial quantity, the Maxwell spinor defined in 
equation~\eqref{eq:defOfMaxwell}, which we reproduce here for convenience:
\[
\maxwell_{\alpha\beta}(x) = \sigma^{\mu\nu}{}_{\alpha \beta} F_{\mu\nu}(x)\,.
\]
The $\sigma^{\mu\nu}$ matrices are symmetric on their spinor indices $\alpha$ and $\beta$. 
These matrices project two-forms onto their self-dual parts,\footnote{In our nomenclature, a two-form $F$ is self-dual if $* F = F$, and anti-self-dual if $* F = - F$.} and are proportional
to the generators of SL$(2, \mathbbm{R})$. (Details of our spinor conventions are given in appendix~\ref{sec:conventions}.) 
In view of the fact that the $\sigma^{\mu\nu}$ matrices matrices are real (as are their antichiral counterparts $\tilde \sigma^{\mu\nu}$), 
we can write the expectation value of the Maxwell spinor  as
\[
\langle \maxwell_{\alpha\beta}(x) \rangle &= \frac{1}{m}\,\text{Re} \sum_{\hel} \int \d\Phi ({k})\, \del(u\cdot {k}) \, \sigma_{\mu\nu\,{\alpha\beta}}  \,k^{[\mu}\varepsilon_\hel ^{\nu]}\,e^{-i{k}\cdot x} \, \mathcal{A}^{(3)}_{-\hel}(k) \,\\
&= -\frac{\sqrt{2}}{m}\,\text{Re} \int \d\Phi ({k})\, \del(u\cdot {k}) \, \ket{k}_\alpha \ket{k}_\beta \,e^{-i{k}\cdot x} \, \mathcal{A}^{(3)}_+(k) \,,
\label{eq:maxwellSpinor}
\]
where we have used the fact that a negative helicity plane wave has a self-dual field strength (equation~\eqref{eq:negativeHelFS}), while a 
positive helicity plane wave has an anti-self-dual field strength (equation~\eqref{eq:positiveHelFS}). For the other chirality, we similarly find
\[
\langle \tilde \maxwell_{\dot\alpha\dot\beta}(x) \rangle =+
\frac{\sqrt{2}}{m}\,\text{Re} \int \d\Phi ({k})\, \del(u\cdot {k}) \, [k|_{\dot\alpha} [k|_{\dot\beta} \,e^{-i{k}\cdot x} \, \mathcal{A}^{(3)}_-(k) \,.
\]
Thus, the two helicity amplitudes correspond directly to the two different chiralities of Maxwell spinor. In split signature, these spinorial field strengths are
real (as is evident in the particular case of our expressions) and independent. We will focus on 
the chiral case of $\maxwell_{\alpha\beta}$ below, though the story for $\tilde \maxwell_{\dot\alpha\dot\beta}$ is completely parallel.

More concretely, we can evaluate the field strength by inserting the standard expressions~\eqref{3sqed} for the amplitude. 
The Maxwell spinor becomes simply
\[
\langle \maxwell_{\alpha\beta}(x) \rangle &= 2\, Q \, \text{Re} \int \d\Phi ({k})\, \del(u\cdot {k}) \, \ket{k}_\alpha \ket{k}_\beta \,e^{-i{k}\cdot x} \, \Xyt \,.
\label{eq:maxwellSpinorX}
\]
In other words, the spinorial field strength is in essence an on-shell Fourier transform of the unique kinematic factor $\Xyt$.

\subsection{The coherent state}
\label{sec:coherent}

\begin{figure}[t]
\begin{center}

\tikzset{every picture/.style={line width=0.75pt}} 

\begin{tikzpicture}[x=0.75pt,y=0.75pt,yscale=-1,xscale=1]

\draw    (111,2888.33) -- (111,3034.33) ;
\draw [shift={(111,3033.33)}, rotate = 90] [fill={rgb, 255:red, 0; green, 0; blue, 0 }  ][line width=0.08]  [draw opacity=0] (6.25,-3) -- (0,0) -- (6.25,3) -- cycle    ;
\draw [shift={(111,2885.33)}, rotate = 90] [fill={rgb, 255:red, 0; green, 0; blue, 0 }  ][line width=0.08]  [draw opacity=0] (6.25,-3) -- (0,0) -- (6.25,3) -- cycle    ;
\draw    (111,3015.33) .. controls (112.67,3013.66) and (114.33,3013.66) .. (116,3015.33) .. controls (117.67,3017) and (119.33,3017) .. (121,3015.33) .. controls (122.67,3013.66) and (124.33,3013.66) .. (126,3015.33) .. controls (127.67,3017) and (129.33,3017) .. (131,3015.33) .. controls (132.67,3013.66) and (134.33,3013.66) .. (136,3015.33) .. controls (137.67,3017) and (139.33,3017) .. (141,3015.33) .. controls (142.67,3013.66) and (144.33,3013.66) .. (146,3015.33) .. controls (147.67,3017) and (149.33,3017) .. (151,3015.33) .. controls (152.67,3013.66) and (154.33,3013.66) .. (156,3015.33) .. controls (157.67,3017) and (159.33,3017) .. (161,3015.33) .. controls (162.67,3013.66) and (164.33,3013.66) .. (166,3015.33) .. controls (167.67,3017) and (169.33,3017) .. (171,3015.33) -- (171,3015.33) ;
\draw    (111,2995.33) .. controls (112.67,2993.66) and (114.33,2993.66) .. (116,2995.33) .. controls (117.67,2997) and (119.33,2997) .. (121,2995.33) .. controls (122.67,2993.66) and (124.33,2993.66) .. (126,2995.33) .. controls (127.67,2997) and (129.33,2997) .. (131,2995.33) .. controls (132.67,2993.66) and (134.33,2993.66) .. (136,2995.33) .. controls (137.67,2997) and (139.33,2997) .. (141,2995.33) .. controls (142.67,2993.66) and (144.33,2993.66) .. (146,2995.33) .. controls (147.67,2997) and (149.33,2997) .. (151,2995.33) .. controls (152.67,2993.66) and (154.33,2993.66) .. (156,2995.33) .. controls (157.67,2997) and (159.33,2997) .. (161,2995.33) .. controls (162.67,2993.66) and (164.33,2993.66) .. (166,2995.33) .. controls (167.67,2997) and (169.33,2997) .. (171,2995.33) -- (171,2995.33) ;
\draw    (111,2975.33) .. controls (112.67,2973.66) and (114.33,2973.66) .. (116,2975.33) .. controls (117.67,2977) and (119.33,2977) .. (121,2975.33) .. controls (122.67,2973.66) and (124.33,2973.66) .. (126,2975.33) .. controls (127.67,2977) and (129.33,2977) .. (131,2975.33) .. controls (132.67,2973.66) and (134.33,2973.66) .. (136,2975.33) .. controls (137.67,2977) and (139.33,2977) .. (141,2975.33) .. controls (142.67,2973.66) and (144.33,2973.66) .. (146,2975.33) .. controls (147.67,2977) and (149.33,2977) .. (151,2975.33) .. controls (152.67,2973.66) and (154.33,2973.66) .. (156,2975.33) .. controls (157.67,2977) and (159.33,2977) .. (161,2975.33) .. controls (162.67,2973.66) and (164.33,2973.66) .. (166,2975.33) .. controls (167.67,2977) and (169.33,2977) .. (171,2975.33) -- (171,2975.33) ;
\draw    (111,2905.33) .. controls (112.67,2903.66) and (114.33,2903.66) .. (116,2905.33) .. controls (117.67,2907) and (119.33,2907) .. (121,2905.33) .. controls (122.67,2903.66) and (124.33,2903.66) .. (126,2905.33) .. controls (127.67,2907) and (129.33,2907) .. (131,2905.33) .. controls (132.67,2903.66) and (134.33,2903.66) .. (136,2905.33) .. controls (137.67,2907) and (139.33,2907) .. (141,2905.33) .. controls (142.67,2903.66) and (144.33,2903.66) .. (146,2905.33) .. controls (147.67,2907) and (149.33,2907) .. (151,2905.33) .. controls (152.67,2903.66) and (154.33,2903.66) .. (156,2905.33) .. controls (157.67,2907) and (159.33,2907) .. (161,2905.33) .. controls (162.67,2903.66) and (164.33,2903.66) .. (166,2905.33) .. controls (167.67,2907) and (169.33,2907) .. (171,2905.33) -- (171,2905.33) ;
\draw    (111,2925.33) .. controls (112.67,2923.66) and (114.33,2923.66) .. (116,2925.33) .. controls (117.67,2927) and (119.33,2927) .. (121,2925.33) .. controls (122.67,2923.66) and (124.33,2923.66) .. (126,2925.33) .. controls (127.67,2927) and (129.33,2927) .. (131,2925.33) .. controls (132.67,2923.66) and (134.33,2923.66) .. (136,2925.33) .. controls (137.67,2927) and (139.33,2927) .. (141,2925.33) .. controls (142.67,2923.66) and (144.33,2923.66) .. (146,2925.33) .. controls (147.67,2927) and (149.33,2927) .. (151,2925.33) .. controls (152.67,2923.66) and (154.33,2923.66) .. (156,2925.33) .. controls (157.67,2927) and (159.33,2927) .. (161,2925.33) .. controls (162.67,2923.66) and (164.33,2923.66) .. (166,2925.33) .. controls (167.67,2927) and (169.33,2927) .. (171,2925.33) -- (171,2925.33) ;
\draw    (111,2906.61) -- (111,2875.33) ;
\draw    (111,2896.61) -- (111,3045.33) ;
\draw  [dash pattern={on 0.84pt off 2.51pt}]  (141,2930.33) -- (141.11,2971.94) ;
\draw    (294.33,2884.83) -- (294.33,3030.83) ;
\draw [shift={(294.33,3029.83)}, rotate = 90] [fill={rgb, 255:red, 0; green, 0; blue, 0 }  ][line width=0.08]  [draw opacity=0] (6.25,-3) -- (0,0) -- (6.25,3) -- cycle    ;
\draw [shift={(294.33,2881.83)}, rotate = 90] [fill={rgb, 255:red, 0; green, 0; blue, 0 }  ][line width=0.08]  [draw opacity=0] (6.25,-3) -- (0,0) -- (6.25,3) -- cycle    ;
\draw    (294.33,3011.83) .. controls (296,3010.16) and (297.66,3010.16) .. (299.33,3011.83) .. controls (301,3013.5) and (302.66,3013.5) .. (304.33,3011.83) .. controls (306,3010.16) and (307.66,3010.16) .. (309.33,3011.83) .. controls (311,3013.5) and (312.66,3013.5) .. (314.33,3011.83) .. controls (316,3010.16) and (317.66,3010.16) .. (319.33,3011.83) .. controls (321,3013.5) and (322.66,3013.5) .. (324.33,3011.83) .. controls (326,3010.16) and (327.66,3010.16) .. (329.33,3011.83) .. controls (331,3013.5) and (332.66,3013.5) .. (334.33,3011.83) .. controls (336,3010.16) and (337.66,3010.16) .. (339.33,3011.83) .. controls (341,3013.5) and (342.66,3013.5) .. (344.33,3011.83) .. controls (346,3010.16) and (347.66,3010.16) .. (349.33,3011.83) .. controls (351,3013.5) and (352.66,3013.5) .. (354.33,3011.83) -- (354.33,3011.83) ;
\draw    (294.33,2991.83) .. controls (296,2990.16) and (297.66,2990.16) .. (299.33,2991.83) .. controls (301,2993.5) and (302.66,2993.5) .. (304.33,2991.83) .. controls (306,2990.16) and (307.66,2990.16) .. (309.33,2991.83) .. controls (311,2993.5) and (312.66,2993.5) .. (314.33,2991.83) .. controls (316,2990.16) and (317.66,2990.16) .. (319.33,2991.83) .. controls (321,2993.5) and (322.66,2993.5) .. (324.33,2991.83) .. controls (326,2990.16) and (327.66,2990.16) .. (329.33,2991.83) .. controls (331,2993.5) and (332.66,2993.5) .. (334.33,2991.83) .. controls (336,2990.16) and (337.66,2990.16) .. (339.33,2991.83) .. controls (341,2993.5) and (342.66,2993.5) .. (344.33,2991.83) .. controls (346,2990.16) and (347.66,2990.16) .. (349.33,2991.83) .. controls (351,2993.5) and (352.66,2993.5) .. (354.33,2991.83) -- (354.33,2991.83) ;
\draw    (294.33,2971.83) .. controls (296,2970.16) and (297.66,2970.16) .. (299.33,2971.83) .. controls (301,2973.5) and (302.66,2973.5) .. (304.33,2971.83) .. controls (306,2970.16) and (307.66,2970.16) .. (309.33,2971.83) .. controls (311,2973.5) and (312.66,2973.5) .. (314.33,2971.83) .. controls (316,2970.16) and (317.66,2970.16) .. (319.33,2971.83) .. controls (321,2973.5) and (322.66,2973.5) .. (324.33,2971.83) .. controls (326,2970.16) and (327.66,2970.16) .. (329.33,2971.83) .. controls (331,2973.5) and (332.66,2973.5) .. (334.33,2971.83) .. controls (336,2970.16) and (337.66,2970.16) .. (339.33,2971.83) .. controls (341,2973.5) and (342.66,2973.5) .. (344.33,2971.83) .. controls (346,2970.16) and (347.66,2970.16) .. (349.33,2971.83) .. controls (351,2973.5) and (352.66,2973.5) .. (354.33,2971.83) -- (354.33,2971.83) ;
\draw    (294.33,2901.83) .. controls (296,2900.17) and (297.67,2900.18) .. (299.33,2901.85) .. controls (301,2903.52) and (302.66,2903.52) .. (304.33,2901.86) .. controls (306,2900.2) and (307.66,2900.2) .. (309.33,2901.87) .. controls (310.99,2903.54) and (312.66,2903.55) .. (314.33,2901.89) .. controls (316,2900.23) and (317.66,2900.23) .. (319.33,2901.9) .. controls (320.99,2903.57) and (322.66,2903.58) .. (324.33,2901.92) .. controls (326,2900.26) and (327.66,2900.26) .. (329.33,2901.93) .. controls (331,2903.6) and (332.66,2903.6) .. (334.33,2901.94) .. controls (336,2900.28) and (337.67,2900.29) .. (339.33,2901.96) .. controls (341,2903.63) and (342.66,2903.63) .. (344.33,2901.97) .. controls (346,2900.31) and (347.67,2900.32) .. (349.33,2901.99) .. controls (351,2903.66) and (352.66,2903.66) .. (354.33,2902) -- (354.5,2902) -- (354.5,2902) ;
\draw    (294.33,2921.83) .. controls (296,2920.16) and (297.66,2920.16) .. (299.33,2921.83) .. controls (301,2923.5) and (302.66,2923.5) .. (304.33,2921.83) .. controls (306,2920.16) and (307.66,2920.16) .. (309.33,2921.83) .. controls (311,2923.5) and (312.66,2923.5) .. (314.33,2921.83) .. controls (316,2920.16) and (317.66,2920.16) .. (319.33,2921.83) .. controls (321,2923.5) and (322.66,2923.5) .. (324.33,2921.83) .. controls (326,2920.16) and (327.66,2920.16) .. (329.33,2921.83) .. controls (331,2923.5) and (332.66,2923.5) .. (334.33,2921.83) .. controls (336,2920.16) and (337.66,2920.16) .. (339.33,2921.83) .. controls (341,2923.5) and (342.66,2923.5) .. (344.33,2921.83) .. controls (346,2920.16) and (347.66,2920.16) .. (349.33,2921.83) .. controls (351,2923.5) and (352.66,2923.5) .. (354.33,2921.83) -- (354.33,2921.83) ;
\draw    (294.33,2903.11) -- (294.33,2871.83) ;
\draw    (294.33,2893.11) -- (294.33,3041.83) ;
\draw  [dash pattern={on 0.84pt off 2.51pt}]  (324.33,2926.83) -- (324.44,2968.44) ;
\draw    (293.78,3002.44) .. controls (292.29,3004.39) and (290.55,3004.6) .. (288.58,3003.06) .. controls (287.2,3001.29) and (285.59,3001.05) .. (283.75,3002.34) .. controls (281.4,3003.14) and (279.99,3002.36) .. (279.52,3000.01) .. controls (279.68,2997.73) and (278.69,2996.39) .. (276.56,2995.98) .. controls (274.43,2994.67) and (274.11,2993.02) .. (275.61,2991.03) .. controls (277.45,2989.77) and (277.85,2988.16) .. (276.8,2986.21) .. controls (276.47,2983.73) and (277.51,2982.5) .. (279.92,2982.53) .. controls (282.11,2983.24) and (283.7,2982.6) .. (284.69,2980.63) .. controls (286.34,2978.85) and (287.96,2978.86) .. (289.54,2980.67) .. controls (290.85,2982.66) and (292.56,2983.13) .. (294.67,2982.08) -- (294.67,2982.08) ;

\draw (87.83,2862) node [anchor=north west][inner sep=0.75pt]    {$p'$};
\draw (269.17,2862) node [anchor=north west][inner sep=0.75pt]    {$p'$};
\draw (87.83,3034.5) node [anchor=north west][inner sep=0.75pt]    {$p$};
\draw (271.67,3029.5) node [anchor=north west][inner sep=0.75pt]    {$p$};
\draw (174,3002) node [anchor=north west][inner sep=0.75pt]    {$k_{\pi ( 1)}$};
\draw (174,2983) node [anchor=north west][inner sep=0.75pt]    {$k_{\pi ( 2)}$};
\draw (174,2961.2) node [anchor=north west][inner sep=0.75pt]    {$k_{\pi ( 3)}$};
\draw (174,2913.67) node [anchor=north west][inner sep=0.75pt]    {$k_{\pi ( n-1)}$};
\draw (174,2891.67) node [anchor=north west][inner sep=0.75pt]    {$k_{\pi ( n)}$};
\draw (358,2999.67) node [anchor=north west][inner sep=0.75pt]    {$k_{\pi ( 1)}$};
\draw (358,2979.17) node [anchor=north west][inner sep=0.75pt]    {$k_{\pi ( 2)}$};
\draw (358,2958.17) node [anchor=north west][inner sep=0.75pt]    {$k_{\pi ( 3)}$};
\draw (358,2910.67) node [anchor=north west][inner sep=0.75pt]    {$k_{\pi ( n-1)}$};
\draw (358,2889.67) node [anchor=north west][inner sep=0.75pt]    {$k_{\pi ( n)}$};
\end{tikzpicture}
\end{center}
\caption{Diagrams of the form shown on the left contribute to the radiation field to all orders in the coupling, but leading classical order. Loop effects, as shown in the diagram on the right, are quantum corrections.}
\label{fig:coherentState}
\end{figure}
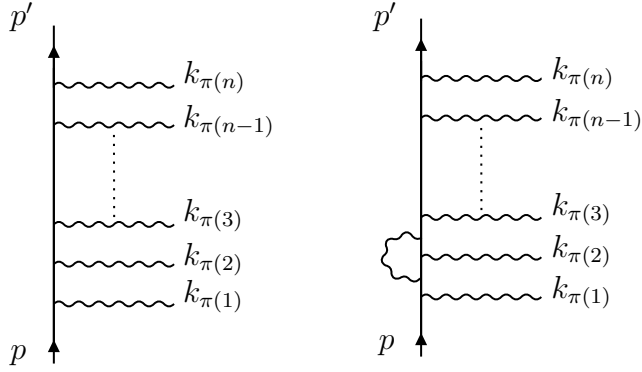

In the previous section, we found the classical electromagnetic field produced by a static source. Even in split signature, it is reasonable
to expect that this field should be very simple, so it is a little unsatisfying that we performed a perturbative approximation along the
way, at equation~\eqref{eq:offensiveApproximation}. 
Fortunately, it is not hard to determine the final quantum state to all orders in the perturbative coupling $Q$.
We only compute the classical approximation to the field, which (in this particular electromagnetic case) means that we should restrict to 
tree-level amplitudes.
The diagrams are shown in figure~\ref{fig:coherentState}. 

In fact, the final state of the electromagnetic field is coherent:
\[
\hspace{-4pt}
S \ket{\psi} = \frac{1}{\mathcal{N}} \int \d\Phi(p) \varphi(p) \exp \left[ \sum_\hel \int \d \Phi(k)\,  \del(2 p \cdot k)\, i \mathcal{A}^{(3)}_{-\hel}(k) \,a^\dagger_\hel(k) \right] \ket{p} \,,
\label{eq:advert}
\]
where $\mathcal{N}$ is a normalisation factor ensuring that $\bra{\psi} S^\dagger S \ket{\psi} = 1$.
The exponential structure of the state captures the intuition that the outgoing field contains a great many photons. It is also consistent with the 
intuition that coherent states are the
natural description of classical wave phenomena in quantum field theory.\footnote{More discussion of coherent states, amplitudes and classical
limits will appear in a forthcoming publication~\cite{cgkoc}.} The coherence of the state could also be demonstrated by taking advantage
of the linear coupling between the gauge field $A_\mu$ and a massive probe source worldline, so it comes as no surprise. However, it is
satisfying to see that the state is completely controlled by the on-shell three-point amplitude.

To see how the exponentiation in~\eqref{eq:advert} comes about in our approach, we expand the $S$ matrix acting on our initial state as
\[
S \ket{\psi}= \frac{1}{\mathcal{N}} (1 + i T_3 + i T_4 + \cdots) \ket{\psi} \,,
\]
where the $T_n$ are defined by
\[
T_{n+2} = \frac{1}{n!} \sum_{\hel_1, \ldots, \hel_n} \int & \d\Phi(p') \d\Phi(p) \prod_{i=1}^n \d\Phi(k_i) \, \mathcal{A}^{(n+2)}_{-\hel_1, \ldots, -\hel_n} (p \rightarrow p', k_1 \cdots k_n) \\
& \quad \times \del^4\left(p-p' - \sum k_i\right) \, a^\dagger_{\hel_1}(k_1) \cdots a^\dagger_{\hel_n} (k_n) \, a^\dagger(p') a(p) \,.
\label{eq:Tn+2def}
\]
That is, the $T_{n+2}$ are projections of the transition matrix $T$ onto final states with $n$ photons, in addition to the massive particle. We denote 
the creation and annihilation operators for the massive scalar state by $a^\dagger(p')$ and $a(p)$, respectively, as opposed to the photon creation 
operators $a^\dagger_{\hel_i}(k_i)$. 
Note that we include precisely one creation and one annihilation operator for our scalar, which is consistent with treating it as a probe source.
We omit all terms in $T_{n+2}$ containing photon annihilation operators since these would annihilate the initial state $\ket\psi$.
The factor $n!$ in equation~\eqref{eq:Tn+2def} is a symmetry factor associated with $n$ identical photons in the final state.

We begin by computing the action of $T_3$ and $T_4$ on $\ket{\psi}$ explicitly. It will then
be a small step to the general case and the exponential structure. First, the case of $T_3$ is straightforward:
\[
iT_3 \ket{\psi} &= \sum_{\hel} \int \d\Phi(p') \d\Phi(p) \d\Phi(k) \, \varphi(p) \,i \mathcal{A}^{(3)}_{-\hel}(k) \ket{p', k^{\hel}}\, \del^4\left(p-p' - k\right) \,\\
&= \sum_{\hel} \int \d\Phi(p) \d\Phi(k) \, \varphi(p+k) \, \Theta(E^2 + k^2) \del(2 p \cdot k) \,i \mathcal{A}^{(3)}_{-\hel}(k)  \ket{p, k^{\hel}} \,,
\]
where, in the second line, we integrated over $p$ with the help of a four-fold delta function, and we relabelled $p'$ to $p$. 
As we saw for the field strength, this expression simplifies when we compute in the domain of validity of the classical approximation.
Following the analysis of KMOC~\cite{Kosower:2018adc}, the shift $k$ in the wave function is negligible: $\varphi(p+k) \simeq \varphi(p)$ in the classical region. We again ignore the small photon momentum in the $\Theta$ function compared to the large, positive energy of the scalar. Thus, we find
\[
iT_3 \ket{\psi} = \sum_{\hel} \int \d\Phi(p) \d\Phi(k) \, \varphi(p) \,\del(2 p \cdot k) \, i\mathcal{A}^{(3)}_{-\hel}(k)\, a_{\hel}^\dagger(k)  \ket{p} \,.
\]
Comparing to the form for the coherent state we advertised in equation~\eqref{eq:advert}, we now see how the exponent can begin to emerge. 

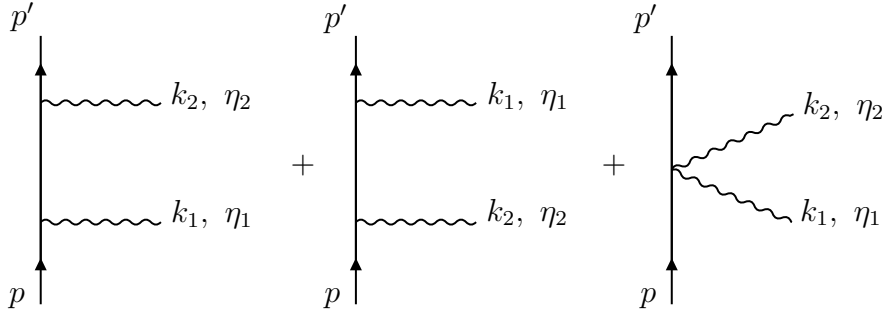
\begin{figure}[t]
\begin{center}
\tikzset{every picture/.style={line width=0.75pt}} 

\begin{tikzpicture}[x=0.75pt,y=0.75pt,yscale=-1,xscale=1]

\draw    (90,3109.18) -- (90,3205.18) ;
\draw [shift={(90,3204.18)}, rotate = 90] [fill={rgb, 255:red, 0; green, 0; blue, 0 }  ][line width=0.08]  [draw opacity=0] (6.25,-3) -- (0,0) -- (6.25,3) -- cycle    ;
\draw [shift={(90,3106.18)}, rotate = 90] [fill={rgb, 255:red, 0; green, 0; blue, 0 }  ][line width=0.08]  [draw opacity=0] (6.25,-3) -- (0,0) -- (6.25,3) -- cycle    ;
\draw    (90,3186.18) .. controls (91.67,3184.51) and (93.33,3184.51) .. (95,3186.18) .. controls (96.67,3187.85) and (98.33,3187.85) .. (100,3186.18) .. controls (101.67,3184.51) and (103.33,3184.51) .. (105,3186.18) .. controls (106.67,3187.85) and (108.33,3187.85) .. (110,3186.18) .. controls (111.67,3184.51) and (113.33,3184.51) .. (115,3186.18) .. controls (116.67,3187.85) and (118.33,3187.85) .. (120,3186.18) .. controls (121.67,3184.51) and (123.33,3184.51) .. (125,3186.18) .. controls (126.67,3187.85) and (128.33,3187.85) .. (130,3186.18) .. controls (131.67,3184.51) and (133.33,3184.51) .. (135,3186.18) .. controls (136.67,3187.85) and (138.33,3187.85) .. (140,3186.18) .. controls (141.67,3184.51) and (143.33,3184.51) .. (145,3186.18) .. controls (146.67,3187.85) and (148.33,3187.85) .. (150,3186.18) -- (150,3186.18) ;
\draw    (90,3126.18) .. controls (91.67,3124.51) and (93.33,3124.51) .. (95,3126.18) .. controls (96.67,3127.85) and (98.33,3127.85) .. (100,3126.18) .. controls (101.67,3124.51) and (103.33,3124.51) .. (105,3126.18) .. controls (106.67,3127.85) and (108.33,3127.85) .. (110,3126.18) .. controls (111.67,3124.51) and (113.33,3124.51) .. (115,3126.18) .. controls (116.67,3127.85) and (118.33,3127.85) .. (120,3126.18) .. controls (121.67,3124.51) and (123.33,3124.51) .. (125,3126.18) .. controls (126.67,3127.85) and (128.33,3127.85) .. (130,3126.18) .. controls (131.67,3124.51) and (133.33,3124.51) .. (135,3126.18) .. controls (136.67,3127.85) and (138.33,3127.85) .. (140,3126.18) .. controls (141.67,3124.51) and (143.33,3124.51) .. (145,3126.18) .. controls (146.67,3127.85) and (148.33,3127.85) .. (150,3126.18) -- (150,3126.18) ;
\draw    (90,3092.51) -- (90,3227.51) ;
\draw    (247.33,3109.18) -- (247.33,3205.18) ;
\draw [shift={(247.33,3204.18)}, rotate = 90] [fill={rgb, 255:red, 0; green, 0; blue, 0 }  ][line width=0.08]  [draw opacity=0] (6.25,-3) -- (0,0) -- (6.25,3) -- cycle    ;
\draw [shift={(247.33,3106.18)}, rotate = 90] [fill={rgb, 255:red, 0; green, 0; blue, 0 }  ][line width=0.08]  [draw opacity=0] (6.25,-3) -- (0,0) -- (6.25,3) -- cycle    ;
\draw    (247.33,3186.18) .. controls (249,3184.51) and (250.66,3184.51) .. (252.33,3186.18) .. controls (254,3187.85) and (255.66,3187.85) .. (257.33,3186.18) .. controls (259,3184.51) and (260.66,3184.51) .. (262.33,3186.18) .. controls (264,3187.85) and (265.66,3187.85) .. (267.33,3186.18) .. controls (269,3184.51) and (270.66,3184.51) .. (272.33,3186.18) .. controls (274,3187.85) and (275.66,3187.85) .. (277.33,3186.18) .. controls (279,3184.51) and (280.66,3184.51) .. (282.33,3186.18) .. controls (284,3187.85) and (285.66,3187.85) .. (287.33,3186.18) .. controls (289,3184.51) and (290.66,3184.51) .. (292.33,3186.18) .. controls (294,3187.85) and (295.66,3187.85) .. (297.33,3186.18) .. controls (299,3184.51) and (300.66,3184.51) .. (302.33,3186.18) .. controls (304,3187.85) and (305.66,3187.85) .. (307.33,3186.18) -- (307.33,3186.18) ;
\draw    (247.33,3126.18) .. controls (249,3124.51) and (250.66,3124.51) .. (252.33,3126.18) .. controls (254,3127.85) and (255.66,3127.85) .. (257.33,3126.18) .. controls (259,3124.51) and (260.66,3124.51) .. (262.33,3126.18) .. controls (264,3127.85) and (265.66,3127.85) .. (267.33,3126.18) .. controls (269,3124.51) and (270.66,3124.51) .. (272.33,3126.18) .. controls (274,3127.85) and (275.66,3127.85) .. (277.33,3126.18) .. controls (279,3124.51) and (280.66,3124.51) .. (282.33,3126.18) .. controls (284,3127.85) and (285.66,3127.85) .. (287.33,3126.18) .. controls (289,3124.51) and (290.66,3124.51) .. (292.33,3126.18) .. controls (294,3127.85) and (295.66,3127.85) .. (297.33,3126.18) .. controls (299,3124.51) and (300.66,3124.51) .. (302.33,3126.18) .. controls (304,3127.85) and (305.66,3127.85) .. (307.33,3126.18) -- (307.33,3126.18) ;
\draw    (247.33,3092.51) -- (247.33,3227.51) ;
\draw    (405,3109.18) -- (405,3205.18) ;
\draw [shift={(405,3204.18)}, rotate = 90] [fill={rgb, 255:red, 0; green, 0; blue, 0 }  ][line width=0.08]  [draw opacity=0] (6.25,-3) -- (0,0) -- (6.25,3) -- cycle    ;
\draw [shift={(405,3106.18)}, rotate = 90] [fill={rgb, 255:red, 0; green, 0; blue, 0 }  ][line width=0.08]  [draw opacity=0] (6.25,-3) -- (0,0) -- (6.25,3) -- cycle    ;
\draw    (405,3160.01) .. controls (407.19,3159.15) and (408.72,3159.82) .. (409.58,3162.01) .. controls (410.45,3164.2) and (411.98,3164.87) .. (414.17,3164.01) .. controls (416.36,3163.15) and (417.89,3163.82) .. (418.75,3166.01) .. controls (419.61,3168.2) and (421.14,3168.87) .. (423.33,3168.01) .. controls (425.52,3167.15) and (427.05,3167.82) .. (427.92,3170.01) .. controls (428.78,3172.2) and (430.31,3172.87) .. (432.5,3172.01) .. controls (434.69,3171.14) and (436.22,3171.81) .. (437.08,3174) .. controls (437.94,3176.19) and (439.47,3176.86) .. (441.66,3176) .. controls (443.85,3175.14) and (445.38,3175.81) .. (446.25,3178) .. controls (447.11,3180.19) and (448.64,3180.86) .. (450.83,3180) .. controls (453.02,3179.14) and (454.55,3179.81) .. (455.41,3182) .. controls (456.28,3184.19) and (457.81,3184.86) .. (460,3184) .. controls (462.19,3183.14) and (463.72,3183.81) .. (464.58,3186) -- (465,3186.18) -- (465,3186.18) ;
\draw    (405,3160.01) .. controls (405.81,3157.8) and (407.33,3157.1) .. (409.54,3157.91) .. controls (411.75,3158.72) and (413.27,3158.02) .. (414.08,3155.81) .. controls (414.89,3153.6) and (416.4,3152.9) .. (418.61,3153.72) .. controls (420.82,3154.53) and (422.34,3153.83) .. (423.15,3151.62) .. controls (423.96,3149.41) and (425.48,3148.71) .. (427.69,3149.52) .. controls (429.9,3150.33) and (431.42,3149.63) .. (432.23,3147.42) .. controls (433.04,3145.21) and (434.56,3144.51) .. (436.77,3145.32) .. controls (438.98,3146.13) and (440.49,3145.43) .. (441.3,3143.22) .. controls (442.11,3141.01) and (443.63,3140.31) .. (445.84,3141.12) .. controls (448.05,3141.93) and (449.57,3141.23) .. (450.38,3139.02) .. controls (451.19,3136.81) and (452.71,3136.11) .. (454.92,3136.92) .. controls (457.13,3137.74) and (458.65,3137.04) .. (459.46,3134.83) .. controls (460.27,3132.62) and (461.79,3131.92) .. (464,3132.73) -- (465.78,3131.9) -- (465.78,3131.9) ;
\draw    (405,3092.51) -- (405,3227.51) ;

\draw (73.83,3074) node [anchor=north west][inner sep=0.75pt]    {$p'$};
\draw (72.83,3218.01) node [anchor=north west][inner sep=0.75pt]    {$p$};
\draw (153.67,3173.51) node [anchor=north west][inner sep=0.75pt]    {$k_{1} ,\ \hel _{1}$};
\draw (153.33,3114.18) node [anchor=north west][inner sep=0.75pt]    {$k_{2} ,\ \hel _{2}$};
\draw (213.33,3150.57) node [anchor=north west][inner sep=0.75pt]    {$+$};
\draw (230.17,3075) node [anchor=north west][inner sep=0.75pt]    {$p'$};
\draw (230.17,3218.01) node [anchor=north west][inner sep=0.75pt]    {$p$};
\draw (311,3172.85) node [anchor=north west][inner sep=0.75pt]    {$k_{2} ,\ \hel _{2}$};
\draw (311.33,3114.18) node [anchor=north west][inner sep=0.75pt]    {$k_{1} ,\ \hel _{1}$};
\draw (368.67,3150.72) node [anchor=north west][inner sep=0.75pt]    {$+$};
\draw (384.83,3075) node [anchor=north west][inner sep=0.75pt]    {$p'$};
\draw (387.83,3218.01) node [anchor=north west][inner sep=0.75pt]    {$p$};
\draw (467.67,3172.85) node [anchor=north west][inner sep=0.75pt]    {$k_{1} ,\ \hel _{1}$};
\draw (469,3119.18) node [anchor=north west][inner sep=0.75pt]    {$k_{2} ,\ \hel _{2}$};

\end{tikzpicture}
\end{center}
\caption{The familiar Feynman diagrams for the four point scalar QED amplitude. In this figure, the photons are outgoing.}
\label{fig:fourPointDiags}
\end{figure}
The four-point case requires a little more work on the actual amplitude. Working at the textbook level of Feynman diagrams (using the notation in figure~\ref{fig:fourPointDiags}), we find
\[
i \mathcal{A}^{(4)} = -i Q^2 \frac{4 p \cdot \varepsilon_{-\hel_1} (k_1) \, p' \cdot \varepsilon_{-\hel_2}(k_2)}{2 k_1 \cdot p + i \epsilon} + &iQ^2 \frac{4 p \cdot \varepsilon_{-\hel_2} (k_2) \, p' \cdot \varepsilon_{-\hel_1}(k_1)}{2 k_1 \cdot p' - i \epsilon} \\
&\hspace{50pt}+ 2iQ^2 \varepsilon_{-\hel_1} (k_1) \cdot \varepsilon_{-\hel_2} (k_2) \,.
\]
Now, of these three terms the last is suppressed relative to the other two in the classical approximation. 
The suppression factor is of order $p \cdot k / m^2$, which is of order the energy of a single photon in units of the mass of the particle. 
(Equivalently, the suppression factor is $\hbar \bar k / m$, where $\bar k$ is a typical component of the wave vector of the photon. From this 
perspective, the contact term is explicitly down by a factor $\hbar$.) We therefore neglect the contact diagram. In terms of a more modern unitarity-based construction of the amplitude, this means that
we can simply ``sew'' three-point amplitudes to compute the dominant part of the four-point amplitude relevant for this computation.\footnote{It
may be worth emphasising that a one-loop computation of a classical observable such as the impulse also involves the four-point tree amplitude.
But in that case the contact term is absolutely necessary to recover the correct classical result, and in fact the terms we are concentrating
on cancel.}

We can make this sewing completely manifest in our four-point amplitude by writing
\[
k_1 \cdot p' &= k_1 \cdot p + \mathcal{O}(\hbar) \,,\qquad
p' \cdot \varepsilon(k) = p \cdot \varepsilon(k)+ \mathcal{O}(\hbar) \,,
\]
and neglecting the $\hbar$ corrections. (In dimensionless terms, these corrections are again suppressed by factors of the photon energy over the particle mass.) It
is then a matter of algebra to see that
\[
i \mathcal{A}^{(4)} &= \del(2 p \cdot k_1) \, (-2 i Q \, p \cdot \varepsilon_{-\hel_1}(k_1) )  (-2 i Q \, p \cdot \varepsilon_{-\hel_2}(k_2) ) \\
&= \del(2 p \cdot k_1) \, i\mathcal{A}^{(3)}_{-\hel_1}(k_1) \, i\mathcal{A}^{(3)}_{-\hel_2}(k_2)  \,.
\label{eq:fourPt}
\]
We picked up a delta function from the sum of two propagators. It is perhaps worth pausing to note that the two photon emissions are
completely uncorrelated from one another.

Now we can compute the action of $T_4$ on our initial state. Using the definition~\eqref{eq:Tn+2def} of $T_4$ and the fact that
\[
a(p) \ket{\psi} = \varphi(p) |0\rangle\,,
\]
we find
\[
iT_4 \ket{\psi} = \frac12 \sum_{\hel_1, \hel_2} \int \d\Phi(p') \d\Phi(p) \d\Phi(k_1)  \d\Phi&(k_2)  \, \varphi(p) \,  i\mathcal{A}^{(4)}_{-\hel_1, -\hel_2} (p \rightarrow p', k_1^{\hel_1} k_2^{\hel_2} )
\\ &\; \times \del^4(p - p' - k_1 - k_2) \ket{p' \; k_{1}^{\hel_1} k_{2}^{\hel_2}} \,.
\]
The integration over the momentum $p$ is trivial using the explicit four-fold delta function. The measure $\d\Phi(p)$ contains a theta 
function, requiring
that the $E^2$ component of $p' + k_1 + k_2$ is positive. Since the $d\Phi(p')$ measure already requires the relevant energy of $p'$ to be positive,
and the photon energies are small compared to the mass, we can ignore this theta function. 
We also encounter the wave function evaluated at $p' + k_1 + k_2$; since the photon energies are small compared to the width of the wave function,
we may approximate $\varphi(p'+k_1+k_2) \simeq \varphi(p')$.
Finally, $\d\Phi(p)$ contains a delta function requiring
\[
p^2 = (p' + k_1 + k_2)^2 = m^2 \,.
\]
Since $p'^2 = m^2$, this becomes a factor
\begin{equation*}
\del (2 p' \cdot (k_1 + k_2) + (k_1 + k_2)^2 )
\end{equation*}
in $T_4 \ket{\psi}$. 
Once again, we may neglect this shift of the delta function, as it is small compared
to the width of the broadened $\delta_{\ell_w}$ function resulting from integrating
against the wave function~\cite{Kosower:2018adc}. Neglecting this width $\ell_w$, we find
\[
iT_4 \ket{\psi} = \frac12 \sum_{\hel_1, \hel_2} \int \d\Phi(p) \d\Phi(k_1) \d\Phi(k_2) \, \varphi(p) \, i& \mathcal{A}^{(4)}_{-\hel_1, -\hel_2}(p+k_1+k_2 \rightarrow p, k_1 k_2 )
\\ & \times  \del(2 p \cdot (k_1 + k_2) )\ket{p \, k_{1}^{\hel_1} k_{2}^{\hel_2}} \,,
\]
where we relabelled the momentum $p'$ to $p$.
Now we may use our result~\eqref{eq:fourPt} for the four-point amplitude, arriving at
\[
iT_4 \ket{\psi} &= \frac12 \sum_{\hel_1, \hel_2} \int \d\Phi(p) \d\Phi(k_1) \d\Phi(k_2) \, \varphi(p) \, \del(2 p \cdot k_1 ) \del(2 p \cdot k_2 ) \\
&\hspace{180pt} \times i\mathcal{A}^{(3)}_{-\hel_1}(k_1) \, i\mathcal{A}^{(3)}_{-\hel_2}(k_2) \ket{p\, k_1^{\hel_1} k_2^{\hel_2}} \\
& = \, \frac12 \int \d\Phi(p) \, \varphi(p) \left( \sum_\hel \int \d \Phi(k) \del(2 p \cdot k) i \mathcal{A}^{(3)}_{-\hel}(k) a^\dagger_\hel(k) \right)^2 \ket{p} \,,
\]
consistent with the exponential structure of the coherent state in equation~\eqref{eq:advert}.

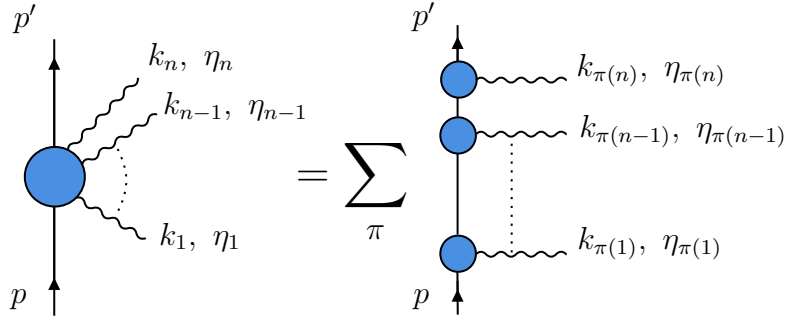
\begin{figure}[t]
\begin{center}

\tikzset{every picture/.style={line width=0.75pt}} 

\begin{tikzpicture}[x=0.75pt,y=0.75pt,yscale=-1,xscale=1]

\draw  [color={rgb, 255:red, 0; green, 0; blue, 0 }  ,draw opacity=1 ][fill={rgb, 255:red, 74; green, 144; blue, 226 }  ,fill opacity=1 ] (100.26,3351.3) .. controls (106.1,3345.43) and (106.07,3335.93) .. (100.2,3330.09) .. controls (94.32,3324.25) and (84.83,3324.27) .. (78.98,3330.15) .. controls (73.14,3336.02) and (73.17,3345.52) .. (79.04,3351.36) .. controls (84.92,3357.2) and (94.41,3357.18) .. (100.26,3351.3) -- cycle ;
\draw    (89.33,3283.67) -- (89.33,3325.67) ;
\draw [shift={(89.33,3280.67)}, rotate = 90] [fill={rgb, 255:red, 0; green, 0; blue, 0 }  ][line width=0.08]  [draw opacity=0] (6.25,-3) -- (0,0) -- (6.25,3) -- cycle    ;
\draw    (89.33,3355.67) -- (89.33,3392.43) ;
\draw [shift={(89.33,3391.43)}, rotate = 90] [fill={rgb, 255:red, 0; green, 0; blue, 0 }  ][line width=0.08]  [draw opacity=0] (6.25,-3) -- (0,0) -- (6.25,3) -- cycle    ;
\draw    (100.26,3351.3) .. controls (102.54,3350.72) and (103.97,3351.57) .. (104.55,3353.86) .. controls (105.13,3356.15) and (106.56,3357) .. (108.85,3356.41) .. controls (111.14,3355.83) and (112.57,3356.68) .. (113.15,3358.97) .. controls (113.73,3361.26) and (115.16,3362.11) .. (117.45,3361.53) .. controls (119.73,3360.95) and (121.16,3361.8) .. (121.74,3364.08) .. controls (122.32,3366.37) and (123.75,3367.22) .. (126.04,3366.64) .. controls (128.33,3366.05) and (129.76,3366.9) .. (130.34,3369.19) .. controls (130.92,3371.48) and (132.35,3372.33) .. (134.64,3371.75) -- (135,3371.97) -- (135,3371.97) ;
\draw    (103.26,3334.8) .. controls (103.69,3332.49) and (105.06,3331.54) .. (107.38,3331.97) .. controls (109.7,3332.4) and (111.07,3331.46) .. (111.5,3329.14) .. controls (111.93,3326.82) and (113.3,3325.88) .. (115.62,3326.31) .. controls (117.94,3326.74) and (119.31,3325.8) .. (119.74,3323.48) .. controls (120.17,3321.16) and (121.55,3320.22) .. (123.87,3320.65) .. controls (126.19,3321.08) and (127.56,3320.14) .. (127.99,3317.82) .. controls (128.42,3315.5) and (129.79,3314.56) .. (132.11,3314.99) .. controls (134.43,3315.42) and (135.8,3314.48) .. (136.23,3312.16) .. controls (136.66,3309.84) and (138.04,3308.9) .. (140.36,3309.33) -- (140.6,3309.17) -- (140.6,3309.17) ;
\draw    (96.2,3327.42) .. controls (96.17,3325.07) and (97.33,3323.87) .. (99.68,3323.84) .. controls (102.04,3323.81) and (103.2,3322.61) .. (103.17,3320.25) .. controls (103.13,3317.89) and (104.29,3316.69) .. (106.65,3316.66) .. controls (109,3316.63) and (110.16,3315.43) .. (110.13,3313.08) .. controls (110.1,3310.72) and (111.26,3309.52) .. (113.62,3309.49) .. controls (115.97,3309.46) and (117.13,3308.26) .. (117.1,3305.91) .. controls (117.07,3303.55) and (118.23,3302.35) .. (120.59,3302.32) .. controls (122.95,3302.29) and (124.11,3301.09) .. (124.07,3298.73) .. controls (124.04,3296.38) and (125.2,3295.18) .. (127.55,3295.15) .. controls (129.91,3295.12) and (131.07,3293.92) .. (131.04,3291.56) -- (131.5,3291.08) -- (131.5,3291.08) ;
\draw  [dash pattern={on 0.84pt off 2.51pt}]  (121,3326.77) .. controls (126.2,3337.57) and (127.4,3345.97) .. (120.6,3359.17) ;
\draw    (89.33,3325.67) -- (89.33,3271.57) ;
\draw    (89.33,3355.67) -- (89.33,3410.63) ;
\draw  [color={rgb, 255:red, 0; green, 0; blue, 0 }  ,draw opacity=1 ][fill={rgb, 255:red, 74; green, 144; blue, 226 }  ,fill opacity=1 ] (297.31,3385.79) .. controls (300.73,3382.18) and (300.64,3376.41) .. (297.12,3372.9) .. controls (293.59,3369.4) and (287.97,3369.49) .. (284.55,3373.1) .. controls (281.14,3376.71) and (281.22,3382.48) .. (284.75,3385.99) .. controls (288.27,3389.49) and (293.9,3389.4) .. (297.31,3385.79) -- cycle ;
\draw  [color={rgb, 255:red, 0; green, 0; blue, 0 }  ,draw opacity=1 ][fill={rgb, 255:red, 74; green, 144; blue, 226 }  ,fill opacity=1 ] (297.31,3326.29) .. controls (300.73,3322.68) and (300.64,3316.91) .. (297.12,3313.4) .. controls (293.59,3309.9) and (287.97,3309.99) .. (284.55,3313.6) .. controls (281.14,3317.21) and (281.22,3322.98) .. (284.75,3326.49) .. controls (288.27,3329.99) and (293.9,3329.9) .. (297.31,3326.29) -- cycle ;
\draw  [color={rgb, 255:red, 0; green, 0; blue, 0 }  ,draw opacity=1 ][fill={rgb, 255:red, 74; green, 144; blue, 226 }  ,fill opacity=1 ] (297.71,3298.19) .. controls (301.13,3294.58) and (301.04,3288.81) .. (297.52,3285.3) .. controls (293.99,3281.8) and (288.37,3281.89) .. (284.95,3285.5) .. controls (281.54,3289.11) and (281.62,3294.88) .. (285.15,3298.39) .. controls (288.67,3301.89) and (294.3,3301.8) .. (297.71,3298.19) -- cycle ;
\draw    (290.67,3329.08) -- (290.67,3369.75) ;
\draw    (290.67,3393.08) -- (290.67,3398.33) ;
\draw [shift={(290.67,3397.33)}, rotate = 90] [fill={rgb, 255:red, 0; green, 0; blue, 0 }  ][line width=0.08]  [draw opacity=0] (6.25,-3) -- (0,0) -- (6.25,3) -- cycle    ;
\draw    (290.67,3273.23) -- (290.67,3282.63) ;
\draw [shift={(290.67,3270.23)}, rotate = 90] [fill={rgb, 255:red, 0; green, 0; blue, 0 }  ][line width=0.08]  [draw opacity=0] (6.25,-3) -- (0,0) -- (6.25,3) -- cycle    ;
\draw    (290.67,3301.17) -- (290.67,3310.63) ;
\draw    (299.87,3379.73) .. controls (301.54,3378.06) and (303.2,3378.06) .. (304.87,3379.73) .. controls (306.54,3381.4) and (308.2,3381.4) .. (309.87,3379.73) .. controls (311.54,3378.06) and (313.2,3378.06) .. (314.87,3379.72) .. controls (316.54,3381.39) and (318.2,3381.39) .. (319.87,3379.72) .. controls (321.54,3378.05) and (323.2,3378.05) .. (324.87,3379.71) .. controls (326.54,3381.38) and (328.2,3381.38) .. (329.87,3379.71) .. controls (331.54,3378.04) and (333.2,3378.04) .. (334.87,3379.71) .. controls (336.54,3381.37) and (338.2,3381.37) .. (339.87,3379.7) .. controls (341.54,3378.03) and (343.2,3378.03) .. (344.87,3379.7) -- (345.13,3379.7) -- (345.13,3379.7) ;
\draw    (299.87,3319.33) .. controls (301.54,3317.66) and (303.2,3317.66) .. (304.87,3319.33) .. controls (306.54,3321) and (308.2,3321) .. (309.87,3319.33) .. controls (311.54,3317.66) and (313.2,3317.66) .. (314.87,3319.32) .. controls (316.54,3320.99) and (318.2,3320.99) .. (319.87,3319.32) .. controls (321.54,3317.65) and (323.2,3317.65) .. (324.87,3319.31) .. controls (326.54,3320.98) and (328.2,3320.98) .. (329.87,3319.31) .. controls (331.54,3317.64) and (333.2,3317.64) .. (334.87,3319.31) .. controls (336.54,3320.97) and (338.2,3320.97) .. (339.87,3319.3) .. controls (341.54,3317.63) and (343.2,3317.63) .. (344.87,3319.3) -- (345.13,3319.3) -- (345.13,3319.3) ;
\draw    (299.87,3292.13) .. controls (301.54,3290.46) and (303.2,3290.46) .. (304.87,3292.13) .. controls (306.54,3293.8) and (308.2,3293.8) .. (309.87,3292.13) .. controls (311.54,3290.46) and (313.2,3290.46) .. (314.87,3292.12) .. controls (316.54,3293.79) and (318.2,3293.79) .. (319.87,3292.12) .. controls (321.54,3290.45) and (323.2,3290.45) .. (324.87,3292.11) .. controls (326.54,3293.78) and (328.2,3293.78) .. (329.87,3292.11) .. controls (331.54,3290.44) and (333.2,3290.44) .. (334.87,3292.11) .. controls (336.54,3293.77) and (338.2,3293.77) .. (339.87,3292.1) .. controls (341.54,3290.43) and (343.2,3290.43) .. (344.87,3292.1) -- (345.13,3292.1) -- (345.13,3292.1) ;
\draw  [dash pattern={on 0.84pt off 2.51pt}]  (317.67,3324.1) -- (317.67,3380.1) ;
\draw    (290.58,3388.58) -- (290.75,3410.08) ;
\draw    (290.67,3264.47) -- (290.67,3282.63) ;

\draw (66.2,3395.77) node [anchor=north west][inner sep=0.75pt]    {$p$};
\draw (67.6,3252.37) node [anchor=north west][inner sep=0.75pt]    {$p'$};
\draw (140.23,3362.33) node [anchor=north west][inner sep=0.75pt]    {$k_{1} ,\ \hel _{1}$};
\draw (142.23,3296.53) node [anchor=north west][inner sep=0.75pt]    {$k_{n-1} ,\ \hel _{n-1}$};
\draw (134.53,3271.83) node [anchor=north west][inner sep=0.75pt]    {$k_{n} ,\ \hel _{n}$};
\draw (349.07,3364.5) node [anchor=north west][inner sep=0.75pt]    {$k_{\pi ( 1)} ,\ \hel _{\pi ( 1)}$};
\draw (350.07,3308) node [anchor=north west][inner sep=0.75pt]    {$k_{\pi ( n-1)} ,\ \hel _{\pi ( n-1)}$};
\draw (348.57,3277.5) node [anchor=north west][inner sep=0.75pt]    {$k_{\pi ( n)} ,\ \hel _{\pi ( n)}$};
\draw (208.67,3320) node [anchor=north west][inner sep=0.75pt]  [font=\Large]  {$\displaystyle=
\sum _{\pi }$};
\draw (267.2,3395.77) node [anchor=north west][inner sep=0.75pt]    {$p$};
\draw (266.6,3251.37) node [anchor=north west][inner sep=0.75pt]    {$p'$};

\end{tikzpicture}

\end{center}
\caption{The dominant term in the $n+2$ point amplitude can be obtained by sewing $n$ three-point amplitudes. The full amplitude is
obtained by summing over permutations $\pi$ of the $n$ outgoing photon lines.}
\label{fig:nPointSewing}
\end{figure}

Now we turn to the general term, evaluating $T_{n+2} \ket{\psi}$. We can make use of the knowledge gained from the four-point example, including
the fact that the leading term in the $(n+2)$-point amplitude can be obtained by sewing $n$ three-point amplitudes. We must nevertheless sum
over permutations of the external photon momenta as shown in figure~\ref{fig:nPointSewing}. The dominant term in the amplitude is
\[
i \mathcal{A}^{(n+2)} = \left(\prod_{i=1}^n i \mathcal{A}^{(3)}_{-\hel_i}(k_i) \right)\sum_{\pi} \frac{i}{2 p\cdot k_{\pi(1)} + i \epsilon} &\frac{i}{2 p\cdot (k_{\pi(1)} + k_{\pi(2)}) + i \epsilon} \cdots \\
&\hspace{-40pt}\times\frac{i}{2 p\cdot (k_{\pi(1)} + k_{\pi(2)} + \cdots k_{\pi(n-1)}) + i \epsilon} \,.
\label{eq:n+2step}
\]
The sum is over permutations $\pi$ of the $n$ final-state photons. 

At four points, the sum over sewings led to a delta function, and the same 
happens here. We can state the result most simply at the level of $T_{n+2} \ket{\psi}$, which can be written as
\[
iT_{n+2} \ket{\psi} = \frac{1}{n!}  \sum_{\hel_1, \ldots, \hel_n} \int & \d\Phi(p) \prod_{i=1}^n \d\Phi(k_i) \,\varphi(p) \, \del\left(2 p \cdot \sum_{j=1}^n k_j\right) i\mathcal{A}^{(n+2)} \,\ket{p \; k_1^{\hel_1} \cdots k_n^{\hel_n} } \,,
\]
using the properties of the wave function, and neglecting terms suppressed in the classical region.
We may now simplify the sum in equation~\eqref{eq:n+2step} using the result
\[
\del\left(\sum_{i=1}^n \omega _i\right) \sum_{\pi} \frac{i}{\omega_{\pi(1)} + i \epsilon} \frac{i}{\omega_{\pi(1)} + \omega_{\pi(2)} + i \epsilon} \cdots \frac{i}{\omega_{\pi(1)} + \omega_{\pi(2)} + \cdots \omega_{\pi(n-1)} + i \epsilon} \\
= \del(\omega_1) \del(\omega_2) \cdots \del(\omega_n) \,.
\]
This result, which is an on-shell analogue of the eikonal identity, is proven (for example) in appendix A of reference~\cite{Saotome:2012vy}. We find
that
\[
iT_{n+2} \ket{\psi} = \frac{1}{n!} \int \d\Phi(p) \,\varphi(p) \left( \sum_\hel \int \d \Phi(k)\, \del(2 p \cdot k) \,i\mathcal{A}^{(3)}_{-\hel}(k)\, a^\dagger_\hel(k) \right)^n \ket{p} \,.
\]
Performing the sum over $n$, we confirm the exponential structure of the state in equation~\eqref{eq:advert}. 

What about the normalisation factor of the coherent state~\eqref{eq:advert}?
As usual, to ensure a correct normalisation we need to include disconnected vacuum bubble diagrams. It is simpler to demand that the factor 
$\mathcal{N}$ appearing in equation~\eqref{eq:advert} is such that $S^\dagger S = 1$, and this is the procedure we adopt.

Now that we have seen that the final state is indeed given by equation~\eqref{eq:advert}, let us return to the evaluation of the expectation value of the field strength. The computation is simplified when we recall that (as usual for a coherent state) the annihilation operator acts as a derivative on the state:
\[
a_\hel(k) S \ket{\psi} &= \del(2 p \cdot k) \, i\mathcal{A}^{(3)}_{-\hel}(k) \, S\ket{\psi} \\
&= \frac{\delta}{\delta a^\dagger_\hel(k)} S\ket{\psi} \,.
\]
The field strength is therefore
\[
\bra{\psi} S^\dagger \, F^{\mu\nu}(x) \, S\ket{\psi} &= -2\Re i \sum_\hel \int \d\Phi(k) \, \bra{\psi} S^\dagger\, a_\hel(k) \, S \ket{\psi} \,  k^{[\mu}\varepsilon_\hel ^{\nu]} \, e^{-i k \cdot x} \\
&= \frac{1}m\Re \sum_\hel \int \d\Phi(k) \, \del(u \cdot k) \mathcal{A}^{(3)}_{-\hel}(k) \, k^{[\mu}\varepsilon_\hel ^{\nu]} \, e^{-i k \cdot x} \,.
\label{eq:fieldStrengthAllOrder}
\]
Notice that this agrees with our previous expression, equation~\eqref{eq:fieldstrength}, which we now see is correct to all orders in the classical
limit. Similarly, the Maxwell spinor is
\[
\bra{\psi} S^\dagger \, \maxwell_{\alpha\beta}(x) \, S \ket{\psi} &=-
\frac{\sqrt2}{m} \Re \int \d\Phi(k) \, \del(u \cdot k) \, \ket{k}_\alpha\ket{k}_\beta \, e^{-i k \cdot x} \mathcal{A}^{(3)}_+(k) \\
&= 2 \,Q \, \text{Re} \int \d\Phi ({k})\, \del(u\cdot {k}) \, \ket{k}_\alpha \ket{k}_\beta \,e^{-i{k}\cdot x} \, \Xyt \,,
\label{eq:ampMaxwellSpinor}
\]
in agreement with our earlier equations~\eqref{eq:maxwellSpinor} and~\eqref{eq:maxwellSpinorX}.

It is worth pausing to comment on this agreement. Classically, the field strength fully characterises the (electromagnetic) radiation field. We are
using a quantum-mechanical formalism, but we now see that it is still true that knowledge of the field strength is also knowledge of the full state
of the electromagnetic field, once we add the extra piece of information that this state is coherent. Mathematically, the field strength operator 
essentially differentiates the exponential form of the coherent state once, pulling down the parameter of the state. 
The structure of this computation is strongly reminiscent of eikonal methods which have also been of interest as a method of  
connecting classical field theory to scattering amplitudes~\cite{Amati:1987wq,tHooft:1987vrq,Muzinich:1987in,Amati:1987uf,Amati:1990xe,Amati:1992zb,Kabat:1992tb,Laenen:2008gt,DAppollonio:2010krb,Melville:2013qca,Akhoury:2013yua,Luna:2016idw,Bjerrum-Bohr:2017dxw,Collado:2018isu,KoemansCollado:2019ggb,Bjerrum-Bohr:2019kec,DiVecchia:2019myk,DiVecchia:2019kta,Cristofoli:2020uzm,Bern:2020buy,Parra-Martinez:2020dzs,DiVecchia:2020ymx,Parnachev:2020zbr}.

\subsection{The gravitational case and the momentum-space Weyl double copy}
\label{sec:gravity}

We have seen that a coherent state, equation~\eqref{eq:advert}, beautifully captures the radiation field in the electromagnetic case. What about gravity? 

Graviton self-interactions could spoil the exponentiation present in electromagnetism. Clearly there are additional diagrams in gravity, for example
at four points we could encounter the diagram
\begin{equation*}
\tikzset{every picture/.style={line width=0.75pt}} 
\begin{tikzpicture}[x=0.75pt,y=0.75pt,yscale=-1,xscale=1]
\draw  [color={rgb, 255:red, 0; green, 0; blue, 0 }  ,draw opacity=1 ][fill={rgb, 255:red, 74; green, 144; blue, 226 }  ,fill opacity=1 ] (98.13,3512.59) .. controls (103.58,3507.11) and (103.56,3498.25) .. (98.07,3492.79) .. controls (92.59,3487.34) and (83.73,3487.37) .. (78.27,3492.85) .. controls (72.82,3498.33) and (72.85,3507.2) .. (78.33,3512.65) .. controls (83.81,3518.1) and (92.68,3518.08) .. (98.13,3512.59) -- cycle ;
\draw    (88.45,3458.17) -- (88.5,3488.33) ;
\draw [shift={(88.44,3455.17)}, rotate = 89.9] [fill={rgb, 255:red, 0; green, 0; blue, 0 }  ][line width=0.08]  [draw opacity=0] (6.25,-3) -- (0,0) -- (6.25,3) -- cycle    ;
\draw    (88,3516.17) -- (88,3548.25) ;
\draw [shift={(88,3547.25)}, rotate = 90] [fill={rgb, 255:red, 0; green, 0; blue, 0 }  ][line width=0.08]  [draw opacity=0] (6.25,-3) -- (0,0) -- (6.25,3) -- cycle    ;
\draw    (88.5,3488.33) -- (88.44,3438.72) ;
\draw    (88,3516.17) -- (88,3562.75) ;
\draw    (101.97,3502.09) .. controls (103.64,3500.42) and (105.31,3500.43) .. (106.97,3502.1) .. controls (108.63,3503.77) and (110.3,3503.78) .. (111.97,3502.12) .. controls (113.64,3500.46) and (115.31,3500.47) .. (116.97,3502.14) .. controls (118.64,3503.81) and (120.3,3503.81) .. (121.97,3502.15) .. controls (123.64,3500.49) and (125.31,3500.5) .. (126.97,3502.17) .. controls (128.64,3503.84) and (130.3,3503.84) .. (131.97,3502.18) .. controls (133.64,3500.52) and (135.31,3500.53) .. (136.97,3502.2) -- (137.24,3502.2) -- (137.24,3502.2)(101.96,3505.09) .. controls (103.63,3503.42) and (105.3,3503.43) .. (106.96,3505.1) .. controls (108.62,3506.77) and (110.29,3506.78) .. (111.96,3505.12) .. controls (113.63,3503.46) and (115.3,3503.47) .. (116.96,3505.14) .. controls (118.63,3506.81) and (120.29,3506.81) .. (121.96,3505.15) .. controls (123.63,3503.49) and (125.3,3503.5) .. (126.96,3505.17) .. controls (128.63,3506.84) and (130.29,3506.84) .. (131.96,3505.18) .. controls (133.63,3503.52) and (135.3,3503.53) .. (136.96,3505.2) -- (137.23,3505.2) -- (137.23,3505.2) ;
\draw  [color={rgb, 255:red, 0; green, 0; blue, 0 }  ,draw opacity=1 ][fill={rgb, 255:red, 74; green, 144; blue, 226 }  ,fill opacity=1 ] (161.33,3514.19) .. controls (166.78,3508.71) and (166.76,3499.85) .. (161.27,3494.39) .. controls (155.79,3488.94) and (146.93,3488.97) .. (141.47,3494.45) .. controls (136.02,3499.93) and (136.05,3508.8) .. (141.53,3514.25) .. controls (147.01,3519.7) and (155.88,3519.68) .. (161.33,3514.19) -- cycle ;
\draw    (162.28,3513.04) .. controls (164.63,3512.81) and (165.91,3513.87) .. (166.14,3516.22) .. controls (166.36,3518.57) and (167.64,3519.63) .. (169.99,3519.41) .. controls (172.34,3519.19) and (173.62,3520.25) .. (173.84,3522.6) .. controls (174.07,3524.95) and (175.35,3526.01) .. (177.7,3525.78) .. controls (180.05,3525.56) and (181.33,3526.62) .. (181.55,3528.97) .. controls (181.77,3531.32) and (183.05,3532.38) .. (185.4,3532.16) .. controls (187.75,3531.93) and (189.03,3532.99) .. (189.26,3535.34) .. controls (189.48,3537.69) and (190.76,3538.75) .. (193.11,3538.53) .. controls (195.46,3538.31) and (196.74,3539.37) .. (196.96,3541.72) -- (197.62,3542.26) -- (197.62,3542.26)(160.37,3515.35) .. controls (162.72,3515.13) and (164,3516.19) .. (164.23,3518.54) .. controls (164.45,3520.89) and (165.73,3521.95) .. (168.08,3521.72) .. controls (170.43,3521.5) and (171.71,3522.56) .. (171.93,3524.91) .. controls (172.15,3527.26) and (173.43,3528.32) .. (175.78,3528.1) .. controls (178.13,3527.87) and (179.41,3528.93) .. (179.64,3531.28) .. controls (179.86,3533.63) and (181.14,3534.69) .. (183.49,3534.47) .. controls (185.84,3534.24) and (187.12,3535.3) .. (187.34,3537.65) .. controls (187.57,3540) and (188.85,3541.06) .. (191.2,3540.84) .. controls (193.55,3540.62) and (194.83,3541.68) .. (195.05,3544.03) -- (195.71,3544.57) -- (195.71,3544.57) ;
\draw    (161.24,3494.31) .. controls (161.29,3491.96) and (162.49,3490.8) .. (164.85,3490.85) .. controls (167.2,3490.9) and (168.41,3489.75) .. (168.46,3487.4) .. controls (168.51,3485.05) and (169.72,3483.89) .. (172.07,3483.94) .. controls (174.42,3483.99) and (175.63,3482.83) .. (175.68,3480.48) .. controls (175.73,3478.13) and (176.94,3476.97) .. (179.29,3477.02) .. controls (181.64,3477.07) and (182.85,3475.91) .. (182.9,3473.56) .. controls (182.95,3471.21) and (184.16,3470.05) .. (186.51,3470.1) .. controls (188.86,3470.15) and (190.07,3469) .. (190.13,3466.65) .. controls (190.18,3464.3) and (191.39,3463.14) .. (193.74,3463.19) -- (194.63,3462.33) -- (194.63,3462.33)(163.31,3496.48) .. controls (163.36,3494.12) and (164.57,3492.97) .. (166.92,3493.02) .. controls (169.27,3493.07) and (170.48,3491.91) .. (170.53,3489.56) .. controls (170.58,3487.21) and (171.79,3486.05) .. (174.14,3486.1) .. controls (176.49,3486.15) and (177.7,3485) .. (177.76,3482.65) .. controls (177.81,3480.3) and (179.02,3479.14) .. (181.37,3479.19) .. controls (183.72,3479.24) and (184.93,3478.08) .. (184.98,3475.73) .. controls (185.03,3473.38) and (186.24,3472.22) .. (188.59,3472.27) .. controls (190.94,3472.32) and (192.15,3471.16) .. (192.2,3468.81) .. controls (192.25,3466.46) and (193.46,3465.3) .. (195.81,3465.35) -- (196.7,3464.5) -- (196.7,3464.5) ;
\end{tikzpicture}
\end{equation*}
which involves a graviton three-point interaction. However, self-interactions of gravitons are suppressed compared to the dominant diagram
\begin{equation*}
\tikzset{every picture/.style={line width=0.75pt}} 
\begin{tikzpicture}[x=0.75pt,y=0.75pt,yscale=-1,xscale=1]
\draw  [color={rgb, 255:red, 0; green, 0; blue, 0 }  ,draw opacity=1 ][fill={rgb, 255:red, 74; green, 144; blue, 226 }  ,fill opacity=1 ] (297.94,3545.02) .. controls (303.39,3539.53) and (303.37,3530.67) .. (297.89,3525.22) .. controls (292.4,3519.77) and (283.54,3519.79) .. (278.09,3525.27) .. controls (272.63,3530.76) and (272.66,3539.62) .. (278.14,3545.07) .. controls (283.62,3550.52) and (292.49,3550.5) .. (297.94,3545.02) -- cycle ;
\draw    (301.64,3534.26) .. controls (303.31,3532.59) and (304.98,3532.6) .. (306.64,3534.27) .. controls (308.3,3535.94) and (309.97,3535.95) .. (311.64,3534.29) .. controls (313.31,3532.63) and (314.97,3532.63) .. (316.64,3534.3) .. controls (318.3,3535.97) and (319.97,3535.98) .. (321.64,3534.32) .. controls (323.31,3532.66) and (324.97,3532.66) .. (326.64,3534.33) .. controls (328.3,3536) and (329.97,3536.01) .. (331.64,3534.35) .. controls (333.31,3532.69) and (334.98,3532.7) .. (336.64,3534.37) -- (336.9,3534.37) -- (336.9,3534.37)(301.63,3537.26) .. controls (303.3,3535.59) and (304.97,3535.6) .. (306.63,3537.27) .. controls (308.29,3538.94) and (309.96,3538.95) .. (311.63,3537.29) .. controls (313.3,3535.63) and (314.96,3535.63) .. (316.63,3537.3) .. controls (318.29,3538.97) and (319.96,3538.98) .. (321.63,3537.32) .. controls (323.3,3535.66) and (324.96,3535.66) .. (326.63,3537.33) .. controls (328.29,3539) and (329.96,3539.01) .. (331.63,3537.35) .. controls (333.3,3535.69) and (334.97,3535.7) .. (336.63,3537.37) -- (336.9,3537.37) -- (336.9,3537.37) ;
\draw  [color={rgb, 255:red, 0; green, 0; blue, 0 }  ,draw opacity=1 ][fill={rgb, 255:red, 74; green, 144; blue, 226 }  ,fill opacity=1 ] (298.29,3489.76) .. controls (303.75,3484.28) and (303.72,3475.41) .. (298.24,3469.96) .. controls (292.76,3464.51) and (283.89,3464.53) .. (278.44,3470.02) .. controls (272.99,3475.5) and (273.01,3484.36) .. (278.5,3489.82) .. controls (283.98,3495.27) and (292.84,3495.24) .. (298.29,3489.76) -- cycle ;
\draw    (288.33,3494.08) -- (288.33,3521.58) ;
\draw    (302.64,3479.26) .. controls (304.31,3477.59) and (305.98,3477.6) .. (307.64,3479.27) .. controls (309.3,3480.94) and (310.97,3480.95) .. (312.64,3479.29) .. controls (314.31,3477.63) and (315.97,3477.63) .. (317.64,3479.3) .. controls (319.3,3480.97) and (320.97,3480.98) .. (322.64,3479.32) .. controls (324.31,3477.66) and (325.97,3477.66) .. (327.64,3479.33) .. controls (329.3,3481) and (330.97,3481.01) .. (332.64,3479.35) .. controls (334.31,3477.69) and (335.98,3477.7) .. (337.64,3479.37) -- (337.9,3479.37) -- (337.9,3479.37)(302.63,3482.26) .. controls (304.3,3480.59) and (305.97,3480.6) .. (307.63,3482.27) .. controls (309.29,3483.94) and (310.96,3483.95) .. (312.63,3482.29) .. controls (314.3,3480.63) and (315.96,3480.63) .. (317.63,3482.3) .. controls (319.29,3483.97) and (320.96,3483.98) .. (322.63,3482.32) .. controls (324.3,3480.66) and (325.96,3480.66) .. (327.63,3482.33) .. controls (329.29,3484) and (330.96,3484.01) .. (332.63,3482.35) .. controls (334.3,3480.69) and (335.97,3480.7) .. (337.63,3482.37) -- (337.9,3482.37) -- (337.9,3482.37) ;
\draw    (288,3548.5) -- (287.85,3561.25) ;
\draw [shift={(287.86,3560.25)}, rotate = 90.69] [fill={rgb, 255:red, 0; green, 0; blue, 0 }  ][line width=0.08]  [draw opacity=0] (6.25,-3) -- (0,0) -- (6.25,3) -- cycle    ;
\draw    (287.83,3450.25) -- (287.83,3460.25) ;
\draw [shift={(287.83,3447.25)}, rotate = 90] [fill={rgb, 255:red, 0; green, 0; blue, 0 }  ][line width=0.08]  [draw opacity=0] (6.25,-3) -- (0,0) -- (6.25,3) -- cycle    ;
\draw    (288,3548.5) -- (287.83,3574.25) ;
\draw    (287.83,3439.25) -- (287.83,3465.58) ;
\end{tikzpicture}
\end{equation*}
where the gravitons connect directly to the massive line. The reason is simply that the graviton self-interaction involves powers of the momenta of
the gravitons, while the coupling to the massive line involves the particle mass. Since the particle mass is large compared to the graviton momenta,
we may neglect graviton self-interactions. We may also neglect contact vertices (as in electromagnetism) for the same reason.

This does not mean that all self-interactions of the gravitational field are eliminated. The metric quantum operator has a perturbative expansion which includes these self-interactions. The expectation value of this all-order operator on our coherent state reproduces the classical metric. Notice that the coherent state is gauge invariant, while the quantum operator may not be (in quantum gravity, only asymptotic observables may be associated with gauge-invariant operators). This procedure would allow us to perturbatively construct the Schwarzschild metric, along the lines of \cite{Duff:1973zz,Jakobsen:2020ksu,Mougiakakos:2020laz} but in a manifestly on-shell formalism; see also \cite{Neill:2013wsa} for an alternative approach based on an intermediate matching with an effective theory of sources coupled to gravitons. We leave this programme for future work.

The computation of the final state $S \ket{\psi}$ proceeds in the gravitational case precisely as in the electromagnetic case. Writing
the gravitational three-point amplitude as $\ampGR^{(3)}$, we find that
\[
\hspace{-4pt}
S \ket{\psi} = \frac{1}{\mathcal{N}}\int \d\Phi(p) \varphi(p) \exp \left[ \sum_\hel \int \d \Phi(k)\, \del(2 p \cdot k)\,i \ampGR^{(3)}_{-\hel}(k) \, a^\dagger_\hel(k) \right] \ket{p} \,,
\label{eq:grState}
\]
where once more $\mathcal{N}$ is a normalisation factor.

We can now compute the gravitational field strength in the classical limit. We place an observer far from the source; then
the gravitational field is weak, and we can work in the formalism of linearised quantum gravity.
The graviton field operator is
\[
h^{\mu\nu}(x)= 2 \Re \sum_\hel \int  \d\Phi(k) \, a_\hel(k) \varepsilon_\hel^{\mu}(k)\varepsilon_\hel^\nu(k) \, e^{-i {k\cdot x}} \,,
\]
where we have written the polarisation tensor of a graviton as the outer product of polarisation vectors $\varepsilon_\hel^\mu(k)$.
The Weyl tensor $W_{\mu\nu\rho\sigma}(x)$ in empty space equals the curvature tensor $R_{\mu\nu\rho\sigma}(x)$, which in linearized 
gravity is 
\[
R^{\mu\nu\rho\sigma}(x)=\frac{\kappa}{2}\left(\partial^\sigma \partial^{[\mu}h^{\nu]\rho}+\partial^\rho \partial^{[\nu}h^{\mu]\sigma} \right)\,,
\label{eq:Riemann}
\]
where $\kappa = \sqrt{32\pi G}$.
Thus, the Weyl tensor operator is
\[
W^{\mu\nu\rho\sigma}(x)
&=\kappa \Re \sum_\hel \int \d\Phi(k) \, a_\hel(k) \varepsilon_\eta^{[\mu}(k) k^{\nu]} \varepsilon_\eta^{[\rho}(k) k^{\sigma]} \, e^{-i{k\cdot x}}\,.
\]
It is now very easy to compute the expectation value of the Weyl tensor, taking advantage once again of the fact that the action of $a_\hel(k)$ on
the coherent state is the same as a functional derivative with respect to the creation operator:
\[
\langle W^{\mu\nu\rho\sigma}(x) \rangle & \equiv \bra{\psi} S^\dagger \, W^{\mu\nu\rho\sigma}(x) \, S \ket{\psi} \\
&=\kappa \Re \sum_\hel \int \d\Phi(k) \, \del(2 p \cdot k) \, i\ampGR^{(3)}_{-\hel}(k) \, \varepsilon_\eta^{[\mu}(k) k^{\nu]} \varepsilon_\eta^{[\rho}(k) k^{\sigma]} \, e^{-i{k\cdot x}} \,.
\label{eq:grWeylTensor}
\]
The expectation value of the Weyl spinor is obtained by contracting with $\sigma^{\mu\nu}$ matrices, leading to 
\[
\bra{\psi} S^\dagger \, \Psi_{\alpha\beta\gamma\delta}(x) \, S \ket{\psi} &= 2\kappa \Re \int \d\Phi(k) \, \del(2 p \cdot k) \, i\ampGR^{(3)}_{+}(k) \, \ket{k}_\alpha \ket{k}_\beta\ket{k}_\gamma \ket{k}_\delta \,e^{-i{k}\cdot x}  \,.
\label{eq:qftWeylSpinor}
\]

This expression for the Weyl tensor is very interesting from the perspective of the double copy. Comparing to the Maxwell
spinor~\eqref{eq:ampMaxwellSpinor}, note that the amplitudes $\mathcal{A}^{(3)}_+(k)$ and $\ampGR^{(3)}_{+}(k)$ are related by
the double copy. Let us define momentum-space versions of the Maxwell and Weyl spinors by
\[
\label{eq:FTspinors}
\langle \maxwell_{\alpha\beta}(x) \rangle &= -
\Re \int \d\Phi(k) \, \del(2 p \cdot k) \, \maxwell_{\alpha\beta}(k)\,e^{-i{k}\cdot x} \,\,, \\
\langle \Psi_{\alpha\beta\gamma\delta}(x) \rangle &= \kappa \Re \int \d\Phi(k) \, \del(2 p \cdot k) \,\Psi_{\alpha\beta\gamma\delta}(k)\,e^{-i{k}\cdot x} \,,
\]
so that
\[
\phi_{\alpha\beta}(k) &= 2\sqrt2\,\ket{k}_\alpha \ket{k}_\beta \, \mathcal{A}^{(3)}_+(k)   \, , \\
\Psi_{\alpha\beta\gamma\delta}(k)&= 2 \ket{k}_\alpha \ket{k}_\beta \ket{k}_\gamma \ket{k}_\delta \, i\ampGR^{(3)}_{+}(k)  \,.
\]
Let us also consider the scalar field theory analogue,
\[
\langle S(x) \rangle =
\Re \int \d\Phi(k) \, \del(2 p \cdot k) \, S(k)\, e^{-i{k}\cdot x} \,\,,
\]
where the three-point amplitude $iS(k)$ is a real constant. Notice that $\langle S(x) \rangle$ manifestly satisfies the wave equation; more precisely, it is the Green's function, as we shall see later. With this scalar counterpart in hand, we obtain an on-shell momentum-space analogue of the position-space  Weyl double copy~\eqref{eq:weylDoubleCopy},
\[
\label{eq:weylDoubleCopymom}
\Psi_{\alpha\beta\gamma\delta}(k) = \frac{1}{S(k)} \phi_{(\alpha\beta}(k) \phi_{\gamma\delta)}(k) \,,
\]
which follows from the double copy relating the three-point amplitudes in gauge theory and in gravity.

We will verify in the next section the position-space version of the expression above, i.e.~after performing the on-shell momentum integrals in \eqref{eq:FTspinors}.
These integrals affect the algebraic structure of the spinors. Notice that the momentum-space Weyl and Maxwell spinors are of type N, if we use the analogue of the position-space characterisation of Weyl spinors. This is
consistent with the intuition that the on-shell three-point amplitudes describe radiation of messengers. However, there is something of a 
puzzle: the field strength of a point charge should have a Coulomb term, and a point mass should have a type D Schwarzschild term. 
In fact, these terms are present in the Weyl~\eqref{eq:qftWeylSpinor} and Maxwell~\eqref{eq:ampMaxwellSpinor} spinors. They emerge when
we perform the Fourier integrals to determine the position-space form of the field strength spinors, as we now show.

\section{The Position-Space Fields and Weyl Double Copy}
\label{ref:WeylDCAmps}

In the previous section, we saw that quantum field theory relates the Maxwell and Weyl spinors for a static charge and mass, respectively, by the
double copy, at least in Fourier space. We would like to express these quantities in position space. In fact, it is not
hard to perform the Fourier integrals to arrive at explicit expressions in position space, where the double copy will still be manifest.

\subsection{The Maxwell spinor in position space}

We will discuss the case of electrodynamics explicitly. Starting from the field strength expectation, equation~\eqref{eq:fieldStrengthAllOrder},
we insert the explicit scalar QED three-point amplitudes to find
\[
\langle F^{\mu\nu}(x) \rangle 
&\equiv \bra{\psi} S^\dagger \, F^{\mu\nu}(x) \, S\ket{\psi}  \\
&= -2 Q\Re \sum_\hel \int \d\Phi(k) \, \del(k \cdot u) \, e^{-i k \cdot x} \,
k^{[\mu} \varepsilon_\hel^{\nu]} \, \varepsilon_{-\hel} \cdot u \, .
\]
This expression simplifies if we resolve the proper velocity onto a Newman-Penrose-like basis of vectors given by $k^\mu$, $\varepsilon_\pm^\mu$ and a gauge 
choice $\gauge^\mu$, such that  $k \cdot \gauge \neq 0$ while $\gauge \cdot \varepsilon_\pm = 0$. Since $k \cdot u = 0$ on the support 
of the integration, the velocity is
\[
u^\mu = \frac{u \cdot \gauge}{k \cdot \gauge} \, k^\mu - \varepsilon_- \cdot u \, \varepsilon_+^\mu - \varepsilon_+ \cdot u \, \varepsilon_-^\mu \,.
\label{eq:velocityExpansion}
\]
Consequently, the field strength is given by the simple formula
\[
\langle F^{\mu\nu}(x) \rangle = 2 Q \Re \int \d\Phi(k) \, \del(k \cdot u)  \, e^{-i k \cdot x}\, k^{[\mu} u^{\nu]} \,.
\]

Before we perform any integrations, let us pause to interpret this formula. Note that we may write
\[
\langle F^{\mu\nu}(x) \rangle = 2 Q \partial^{[\mu} u^{\nu]} \Re i \int \d\Phi(k) \, \del(k \cdot u) \, e^{-i k \cdot x}   \,.
\]
We recognise the definition of the field strength as the (antisymmetrised) derivative of the gauge potential,
\[
\langle A^\mu(x) \rangle = \Re 2 iQ \, u^{\mu} \int \d\Phi(k) \, \del(k \cdot u) \, e^{-i k \cdot x}  \,.
\label{eq:Aexpectation}
\]
To interpret this formula, it's worth briefly digressing to discuss our situation from a classical perspective.

Consider solving the Maxwell equation with a static point charge
\[
\partial_\mu F^{\mu\nu}(x) = \int d\tau \,Q u^\nu \,\delta^4(x-u\tau) \,,
\]
where $u^\mu = (0,1,0,0)$, with the boundary condition that the electromagnetic field vanishes for $t^1 < 0$. Choosing Lorenz gauge, we can write the 
solution as a familiar Fourier integral:
\[
A^\mu(x) = - \int \dd^4 k \, \del(k \cdot u) \,e^{-ik \cdot x} \frac{1}{k^2} Q u^\mu \,.
\label{eq:Astep}
\]
As usual, we need to define the $k$ integral taking our boundary conditions into account. These boundary conditions are also familiar: they
are just traditional retarded boundary conditions. The only novelty lies the signature of the metric. But even the unfamiliar pattern of signs in
split signature disappears for the problem at hand, because of the factor
\begin{equation*}
\del(k \cdot u) = \del(k_2) 
\end{equation*}
in the measure. Consequently, the second component of the wave vector $k^\mu$ is guaranteed to be zero. We end up with an integral
of Minkowskian type, but in $1+2$ dimensions. This is a consequence of translation invariance in the $t^2$ direction.

Treating the $k$ integration as a contour integral, the only 
poles in the integration of equation~\eqref{eq:Astep} occur when
\[
(k^1)^2 = \vec{k}^2 \,,
\]
where $\vec{k} = (k^3, k^4)$ are the spatial components of the wave vector. Taking the sign of the exponent in equation~\eqref{eq:Astep} into
account, retarded boundary conditions are obtained by displacing the poles below the real axis:
\[
\frac{1}{k^2} \rightarrow \frac{1}{k^2_\textrm{ret}} = \frac{1}{(k^1 + i \epsilon)^2 + (k^2)^2 - (k^3)^2 - (k^4)^2}  \,,
\]
while advanced boundary conditions correspond to
\[
\frac{1}{k^2} \rightarrow \frac{1}{k^2_\textrm{adv}} = \frac{1}{(k^1 - i \epsilon)^2 + (k^2)^2 - (k^3)^2 - (k^4)^2}  \,.
\]
Notice that
\[
\frac{1}{k^2_\textrm{ret}} -  \frac{1}{k^2_\textrm{adv}} &= \frac{1}{k^2 + i (k^1) \epsilon} - \frac{1}{k^2 - i (k^1) \epsilon} 
&=- i \sign(k^1) \del(k^2) \,,
\]
where, in the first equality, we have written $(k^1)$ for the first component of the 4-vector $k$ and have
freely rescaled $\epsilon$ by positive quantities (as is conventional, we take $\epsilon \rightarrow 0$ from above at the end of our calculation). 

Returning to the gauge field of equation~\eqref{eq:Astep}, we have
\[
A^\mu(x) &= - \int \dd^4 k \, \del(k \cdot u) \,e^{-ik \cdot x} \left(-i \sign(k^1) \del(k^2) + \frac{1}{k^2_\textrm{adv}} \right) Q u^\mu \\
&= i \int \dd^4 k \, \del(k \cdot u) \,e^{-ik \cdot x}  \, \sign(k^1) \del(k^2)  \, Q u^\mu \\
&= i \int \d\Phi(k) \, \del(k \cdot u) \, Qu^\mu \left( e^{-i k \cdot x} - e^{i k \cdot x} \right) \,.
\label{eq:gaugeClassicalStep}
\]
We dropped the advanced term because, with our boundary conditions, the position $x$ has  positive $t^1$. But equation~\eqref{eq:gaugeClassicalStep} is just the 
result we found from the quantum
expectation~\eqref{eq:Aexpectation}. Thus, our quantum mechanical methods are computing the complete gauge field, as expected.

Given that we have made contact with a classical situation, we can use classical intuition to perform the Fourier integrals.
The integrals to be performed in equation~\eqref{eq:Astep} are the same as the integrals in the computation of the retarded Green's function
in $1+2$ dimensions. We discuss this Green's function in appendix~\ref{sec:retardedGF}. We find
\[
A^\mu(x) = \frac{Q u^\mu}{2\pi} \Theta(t^1) \frac{\Theta(x^2 - (x \cdot u)^2)}{\sqrt{x^2 - (x \cdot u)^2}} \,.
\]
In many respects, this result is familiar: it is just the usual 1/`distance' fall-off. There is no other possibility: the dimensional
analysis requires this behaviour with distance. The key new feature in split signature is the theta function $\Theta(x^2 - (x \cdot u)^2)$.
To see why, let's differentiate to compute the field strength, which is\footnote{We assume that the point $x$ is not on the 
worldline of the source particle, so we drop a term in the field strength involving $\delta(t^1) \Theta(x^2 - (x \cdot u)^2)$, which is only non-vanishing
on this worldline.}
\[
F^{\mu\nu}(x) = -\frac{Q\, \Theta(t^1) \, x^{[\mu} u^{\nu]}}{2\pi(x^2 -(x\cdot u)^2)^{1/2}} \left(
\frac{\Theta(x^2 - (x \cdot u)^2)}{x^2 - (x\cdot u)^2}
- 2\,  \delta\big(x^2 - (x \cdot u)^2\big)  
 \right) \,.
\label{eq:fieldstrengthPosition}
\]
The term involving the $\Theta(x^2 - (x \cdot u)^2)$ is the familiar Coulomb field. However, there is an additional $\delta$ function describing
the impulsive radiation field when the charge ``appears'' from the point of view of the observer.\docNote{This remark doesn't have context: diagram?}
Although the radiation field looks very singular classically, this should not really trouble us: the delta function distribution is only present
in the approximation that the source wave function is treated as of zero size. In reality, this wave function must have some spatial size $\ell_w$,
and the delta function will be broadened into a smooth function when this width is taken into account.

It will also be interesting to investigate the Maxwell spinor generated by our set up, especially when comparing to the Weyl spinor in the
gravitational case. First, let's break up our field strength into two terms,
\[
\label{eq:FF1F2}
F_{\mu\nu}(x) = F^{(1)}_{\mu\nu}(x) + F^{(2)}_{\mu\nu}(x) \,,
\]
where
\[
F^{(1)}_{\mu\nu}(x) &= - \frac{Q\, \Theta(t^1) \, x_{[\mu} u_{\nu]}}{2\pi(x^2 -(x\cdot u)^2)^{3/2}} \Theta(x^2 - (x \cdot u)^2) \,,\\
F^{(2)}_{\mu\nu}(x) &= \frac{Q\, \Theta(t^1) \, x_{[\mu} u_{\nu]}}{\pi(x^2 -(x\cdot u)^2)^{1/2}} \delta(x^2 - (x \cdot u)^2)  \,.
\]
It's natural to define the ``radial distance" (i.e.~its analogue under analytic continuation)
\[
\rho^2 = x^2 - (x\cdot u)^2 
\]
and the associated vector
\[
K_\mu = x_\mu - (x\cdot u) u_\mu  \,.
\]
The Maxwell spinor $\maxwell^{(1)}_{\alpha\beta}(x)$ associated with the Coulombic field strength $F^{(1)}$ is\footnote{Here, $\maxwell^{(1)}_{\alpha\beta}$ corresponds to the middle term on the right-hand side of \eqref{eq:maxwellkn}. Likewise, $\maxwell^{(2)}_{\alpha\beta}$ further below corresponds to the last term in \eqref{eq:maxwellkn}.}
\[
\label{eq:phi1}
\maxwell^{(1)}_{\alpha\beta}(x) = -\frac{Q \Theta(t^1)}{2\pi \rho^3} \sigma^{\mu\nu}{}_{\alpha\beta} K_{[\mu} u_{\nu]} \Theta (\rho^2) \,.
\]
Meanwhile, on the support of the delta function factor in $F^{(2)}$, the vector $K_\mu$ becomes null. Furthermore, a simple computation shows
that, in general,
\[
K \cdot u = 0 \,.
\]
Therefore, we may erect a Newman-Penrose basis using the vector $K$, an arbitrary gauge choice, and two ``polarisation'' vectors $\varepsilon_\pm(K)$ which can be taken to be the standard spinor-helicity vectors associated with the ``on-shell momentum'' $K$; these are defined explicitly in equation~\eqref{eq:polVectorDef}. In this basis we may once again decompose the proper velocity using the obvious
analogue of equation~\eqref{eq:velocityExpansion}. It follows that the Maxwell spinor is
\[
\label{eq:psi2deg}
\phi^{(2)}_{\alpha\beta}(x) = \frac{Q}{\pi\,\rho}\, \Theta(t^1)\, X \, \ket{K}_\alpha \ket{K}_\beta \, \delta(\rho^2) \,.
\]
Evidently, $\phi^{(2)}_{\alpha\beta}(x)$ has the structure expected for the radiative part of the field strength. We will encounter an analogous situation in gravity.

\subsection{The Weyl spinor and the double copy in position space}

We can perform the Fourier integrals for gravity in exact analogy with the electromagnetic case. Beginning from the Weyl
tensor~\eqref{eq:grWeylTensor}, we insert the explicit amplitudes
\[
\ampGR^{(3)}_{\hel}(k) = -\kappa\, m^2 (u \cdot \varepsilon_\hel(k))^2
\]
to find that 
\[
\langle W^{\mu\nu\rho\sigma}(x) \rangle = -\Re i \kappa^2 m^2 \int \d\Phi(k) \, \del(2 k \cdot p)\, e^{-i k \cdot x} &\left[(\varepsilon_+\cdot u)^2  k^{[\mu} \varepsilon_-^{\nu]} k^{[\rho} \varepsilon_-^{\sigma]}
\right. \\ &\left. \quad
+ (\varepsilon_-\cdot u)^2 k^{[\mu} \varepsilon_+^{\nu]} k^{[\rho} \varepsilon_+^{\sigma]} \right] \,.
\label{eq:explicitWeyl}
\]
Again, this expression is easily interpreted in the classical theory.
We define the metric perturbation by
\[
g_{\mu\nu} = \eta_{\mu\nu} + \kappa h_{\mu\nu} \,.
\]
By solving the linearised Einstein equation in De Donder gauge, we find that the metric perturbation is
\[
h_{\mu\nu}(x) = - 2\Re i \kappa m^2 \int \d\Phi(k) \, \del(2 k \cdot p) \, e^{-i k \cdot x} \left(u_\mu u_\nu - \frac12 \eta_{\mu\nu} \right) \,.
\label{eq:metricPert}
\]
The Riemann tensor~\eqref{eq:Riemann} is explicitly
\[
R^{\mu\nu\rho\sigma}(x)= \Re i \kappa^2 m^2 \int \d\Phi(k) \, \del(2 k \cdot p)\, e^{-i k \cdot x} \left( k^{[\mu} u^{\nu]} k^{[\sigma} u^{\rho]} - \frac12 
k^{[\mu} \eta^{\nu][\rho} k^{\sigma]} \right) \,.
\]
It is not obvious in this form that the traces of the Riemann tensor vanish. Of course, they must do so since our observer at $x$ is in empty
space. In fact, it is possible to simplify the tensor structure of this Riemann tensor by resolving the vector $u$ onto the Newman-Penrose-like 
basis of $k$, $\epsilon_\pm(k)$ and $\gauge$ as in equation~\eqref{eq:velocityExpansion}. The flat metric tensor in this basis is
\[
\eta^{\mu\nu} = \frac{1}{k \cdot \gauge} k^{(\mu} \gauge^{\nu)} - \varepsilon_+^{(\mu} \varepsilon_-^{\nu)} \,.
\]
It follows that 
\[
k^{[\mu} \eta^{\nu][\rho} k^{\sigma]} = -k^{[\mu} \varepsilon_+^{\nu]} \varepsilon_-^{[\rho} k^{\sigma]} -k^{[\mu} \varepsilon_-^{\nu]} \varepsilon_+^{[\rho} k^{\sigma]} \,,\label{eq:antisymketak}
\]
while the other tensor structure in the Riemann tensor simplifies to
\[
k^{[\mu} u^{\nu]} k^{[\sigma} u^{\rho]} = (\varepsilon_+\cdot u)^2  k^{[\mu} \varepsilon_-^{\nu]} k^{[\sigma} \varepsilon_-^{\rho]} &+ (\varepsilon_-\cdot u)^2 k^{[\mu} \varepsilon_+^{\nu]} k^{[\sigma} \varepsilon_+^{\rho]}  \\
&-\frac12 k^{[\mu} \varepsilon_-^{\nu]}k^{[\sigma} \varepsilon_+^{\rho]}- \frac12 k^{[\mu} \varepsilon_+^{\nu]}k^{[\sigma} \varepsilon_-^{\rho]} \,. \label{eq:antisymkuku}
\]
Combining, the Riemann tensor manifestly has no traces and we recover the Weyl tensor
of equation~\eqref{eq:explicitWeyl}.

Now it is easy to perform the Fourier integrals, for example at the level of the metric perturbation, which yields
\[
h_{\mu\nu}(x) = - \frac{\kappa m}{4\pi} \Theta(t^1) \frac{\Theta(x^2 - (x \cdot u)^2)}{\sqrt{x^2 - (x \cdot u)^2}} \left( u_\mu u_\nu - \frac12 \eta_{\mu\nu} \right).\label{eq:gravitonInPosSpace}
\]
The expectation value of the (linearised) Weyl tensor can be computed by differentiation. There are various terms, depending on whether derivatives
act on the delta functions or the $1/\rho$ fall-off factors. 
Analogously to \eqref{eq:FF1F2}, we can write the Weyl tensor as 
\[
\label{eq:weylposspace}
W_{\mu\nu\rho\sigma}=W^{(2)}_{\mu\nu\rho\sigma}+W^{(3)}_{\mu\nu\rho\sigma}+W^{(4)}_{\mu\nu\rho\sigma}~,
\]
with
\begin{align}
W^{(2)}_{\mu\nu\rho\sigma}&=
\frac{3\kappa^2\,m\,\Theta(t^1)\,\Theta(\rho^2)}{32\pi\,\rho^5}\; w_{\mu\nu\rho\sigma}\,,
\\[1.5em]
W^{(3)}_{\mu\nu\rho\sigma}&=
-\frac{\kappa^2\,m\,\Theta(t^1)\,\delta(\rho^2)}{8\pi\,\rho^3}\; w_{\mu\nu\rho\sigma}\,,\\[1.5em]
W^{(4)}_{\mu\nu\rho\sigma}&=
\frac{\kappa^2\,m\,\Theta(t^1)\, \delta'(\rho^2)}{8\pi \rho}\; w_{\mu\nu\rho\sigma}\,,
\end{align}
where
\begin{equation}
w\,{}^{\mu\nu}{}_{\rho\sigma} = 4\,K^{[\mu}u^{\nu]}K_{[\rho}u_{\sigma]}+2\,K^{[\mu}\delta^{\nu]}_{[\rho}K_{\sigma]}+2\,\rho^2\,u^{[\mu}\delta^{\nu]}_{[\rho}u_{\sigma]}+\frac{2\rho^2}{3}\delta^{[\mu}_{[\rho}\,\delta^{\nu]}_{\sigma]} \,.
\end{equation}
The corresponding Weyl spinor is
\[
\Psi_{\alpha\beta\gamma\delta}=\Psi^{(2)}_{\alpha\beta\gamma\delta}+\Psi^{(3)}_{\alpha\beta\gamma\delta}+\Psi^{(4)}_{\alpha\beta\gamma\delta}~,
\quad \text{with} \quad
\Psi^{(i)}_{\alpha\beta\gamma\delta}=W^{(i)}_{\mu\nu\rho\sigma}\sigma^{\mu\nu}_{\alpha\beta}\sigma^{\mu\nu}_{\gamma\delta}~.
\]
If we now compare these expressions with the ones for the Maxwell spinor obtained in the previous subsection, we find the position-space double copy relations
\begin{align}
\frac{\Theta(\rho^2)\,\Theta(t^1)}{2\pi\,\rho}\,\Psi^{(2)}_{\alpha\beta\gamma\delta}&= \frac{3\kappa^2\,m }{4\cdot4!\,Q^2\,}\phi^{(1)}_{(\alpha\beta}\phi^{(1)}_{\gamma\delta)}~,
\\
\frac{\Theta(\rho^2)\,\Theta(t^1)}{2\pi\,\rho}\,\Psi^{(3)}_{\alpha\beta\gamma\delta}&= \frac{\kappa^2\,m }{2\cdot4!\,Q^2}\phi^{(1)}_{(\alpha\beta}\phi^{(2)}_{\gamma\delta)}~,
\\
\frac{\delta^2(\rho^2)\,\Theta(t^1)}{2\pi\,\rho}\,\Psi^{(4)}_{\alpha\beta\gamma\delta} &= \frac{\kappa^2\,m\, \delta'(\rho^2)}{4\cdot4!\,Q^2}
\phi^{(2)}_{(\alpha\beta}\phi^{(2)}_{\gamma\delta)}~.
\end{align}
Notice that, in the first two lines, the relation \eqref{eq:weylDoubleCopy} is satisfied with
$$
S(x)=\frac{\Theta(t_1)\Theta(\rho^2)}{2\pi\,\rho}~,
$$
up to numerical factors.
The clearest example is that of $\Psi^{(2)}_{\alpha\beta\gamma\delta}$, which is the only one that has support in the interior of the future light-cone, as opposed to just the future light-cone itself. Hence, it satisfies on its own the Bianchi identity in that region. Indeed, its analytic continuation is the linearised Weyl tensor of the Lorentzian Schwarzschild solution, in the same way that the term $F^{(1)}_{\mu\nu}$ in \eqref{eq:FF1F2} is associated to the Coulomb solution. Therefore, the terms corresponding to the interior of the light-cone satisfy the position space Weyl double copy for type D solutions, equation \eqref{eq:weylDoubleCopy}, as discussed for the Lorentzian solutions in \cite{Luna:2018dpt}.

The spinors $\Psi^{(3)}_{\alpha\beta\gamma\delta}$ and $\Psi^{(4)}_{\alpha\beta\gamma\delta}$ are distributional, and supported only on the future light-cone, where $K_\mu$ is null. Analogously to \eqref{eq:psi2deg}, they are both proportional to $|K\rangle_\alpha|K\rangle_\beta|K\rangle_\gamma|K\rangle_\delta$.\footnote{Notice that $\phi^{(1)}_{\alpha\beta}$ degenerates on the light-cone (its principal rank-1 spinors coincide), and it becomes proportional to $\phi^{(2)}_{\alpha\beta}$.} They look very singular, and they do not satisfy the Bianchi identity on their own on the light-cone, since this identity receives contributions from the three terms. Nevertheless, a type of double copy is still evident in position space, satisfying the expectation of the type N position-space Weyl double copy \cite{Godazgar:2020zbv}.  In fact, this follows from \eqref{eq:weylDoubleCopymom}, when the on-shell momentum integrands of \eqref{eq:FTspinors} are evaluated only  at $k_\mu\propto K_\mu$.

\section{The Kerr-Schild Double Copy and the Exact Metric}
\label{sec:classicalGR}

In the previous sections, we computed the linearised metric and curvature generated by a massive particle.
It is actually straightforward for us to compute the exact metric. To do so, we exploit the Kerr-Schild double copy. 
As a reminder of the Kerr-Schild double copy, recall that, in the case of Lorentzian $(1,3)$ signature, we start with a Green's function
\begin{equation}
\Phi^{(L)}=  \frac1{4\pi\sqrt{x^2+y^2+z^2}} \,,
\end{equation}
satisfying
\begin{equation}
-(\partial_x^2+\partial_y^2+\partial_z^2) \, \Phi^{(L)} = 
 \delta(x)\delta(y)\delta(z)\,.
\end{equation} 
The Coulomb solution is
\begin{equation}
A^{(L)} = Q\, \Phi^{(L)}\, \d t\,,
\end{equation}
or in ``Kerr-Schild'' gauge,
\begin{align}
A^{(L,KS)} &= Q\, \Phi^{(L)}\, \d t - \frac{Q}{4\pi}\, \d \log \frac{\sqrt{x^2+y^2+z^2}}{r_0}
= Q\, \Phi^{(L)}\, L^{(L)} \,, \nonumber \\
L^{(L)} & =\d t - \frac{x\d x+y\d y+z\d z}{\sqrt{x^2+y^2+z^2}}
\,,
\end{align}
where $K^{(L)}$ is null and $r_0$ is a constant needed for dimensional purposes.

The (vacuum) double copy of this solution is the Schwarzschild solution, which can be written in Kerr-Schild coordinates as
\begin{equation}
\d s^2_{(L)} = \d t^2-\d x^2-\d y^2-\d z^2 - \frac{\kappa^2 m}{4} \, \Phi^{(L)}\, L^{(L)} L^{(L)}\,.
\end{equation}
It can also be written in static coordinates as
\begin{equation}
\d s^2_{(L)} = \left(1- \frac{\kappa^2 m}{4} \,\Phi^{(L)}\right)\d t'^2-\d x^2-\d y^2-\d z^2
- \frac{\frac{\kappa^2 m}{4} \,\Phi^{(L)}}{1-\frac{\kappa^2 m}{4} \,\Phi^{(L)}}
\frac{(x \d x+y \d y+z\d z)^2}{x^2+y^2+z^2}\,,
\end{equation}
with
\begin{equation}
\d t'=\d t + \frac{\frac{\kappa^2 m}{4} \,\Phi^{(L)}}{1-\frac{\kappa^2 m}{4} \,\Phi^{(L)}} \,
\frac{x \d x+y \d y+z\d z}{\sqrt{x^2+y^2+z^2}}\,.
\end{equation}
The commonly-seen Schwarzschild coordinates are obtained by changing from rectangular to spherical coordinates,
\begin{equation}
\d s^2_{(L)} = \left(1- \frac{\kappa^2 m}{4} \,\Phi^{(L)}\right)\d t'^2-\frac{\d r^2}{1- \frac{\kappa^2 m}{4} \,\Phi^{(L)}}-r^2(\d\theta^2+\sin^2\theta \,\d\phi^2)\,,
\end{equation}
with $r=\sqrt{x^2+y^2+z^2}$ and $\Phi^{(L)}=(4\pi r)^{-1}$.

Following the same steps as in the Lorentzian case, let us consider the case of split signature discussed in previous sections. As we saw earlier (and as is discussed in appendix~\ref{sec:retardedGF}), the retarded Green's function is
\begin{equation}
\Phi = \frac{\Theta(t_1-\sqrt{x^2+y^2})}{2\pi\sqrt{t_1^2-x^2-y^2}} = \Theta(t_1-\sqrt{x^2+y^2}) \,\hat{\Phi} 
 \,,
\end{equation}
satisfying
\begin{equation}
(\partial_{t_1}^2-\partial_x^2-\partial_y^2) \, \Phi = 
\delta(t_1)\delta(x)\delta(y)\,.
\end{equation}
The causal boundary condition breaks the $t_1$ parity. There are major differences with respect to the Lorentzian Green's function, including the singularity structure. The Lorentzian Green's function is singular only at the origin --- the locus of the delta-function source. The split-signature Green's function is singular along the future light-cone, even though it is only sourced at the origin of the light-cone. This singularity, both via the denominator and via the discontinuity of the step function, requires some care but, as was seen previously, presents no difficulty in Fourier space.

The associated gauge field is
\begin{equation}
A = Q\, \Phi\, \d t_2\,.
\end{equation}
We can now try to proceed to obtain the ``Kerr-Schild'' gauge, but 
two apparent difficulties arise. The first is that a complex gauge transformation is required for the gauge field to be null. This is acceptable as a means to obtain a gravity solution in complex Kerr-Schild form, but which can be made real by a complex diffeomorphism.\footnote{In what regards complexification, this situation is analogous to the double copy interpretation of the Taub-NUT solution from the dyon, and more generally of generic type D vacuum solutions.} The second difference is more subtle and is related to the breaking of $t^1$ time reversal symmetry: the Green's function is not solely a function of $t_1^2-x^2-y^2$. Let us proceed in the region $t_1>\sqrt{x^2+y^2}$, strictly inside the future (3D) light-cone, since the subtlety only affects the light-cone. Then we can obtain the complex `Kerr-Schild' gauge,
\begin{align}
t_1>\sqrt{x^2+y^2}\, : \quad
A^{(KS)}  &= Q\, \hat\Phi\, \d t_2 - \frac{Q}{2\pi i}\, \d \log \frac{\sqrt{t_1^2-x^2-y^2}}{r_0}
= Q\, \hat\Phi\, L \,, \nonumber \\
L & =\d t_2 + i\, \frac{t_1\d t_1-x\d x-y\d y}{\sqrt{t_1^2-x^2-y^2}}
\,.
\end{align}

The exact gravity solution, in complex Kerr-Schild coordinates, is then given as 
\begin{equation}
t_1>\sqrt{x^2+y^2}\, : \quad 
\d s^2 = \d t_2^2+\d t_1^2-\d x^2-\d y^2 - \frac{\kappa^2 m}{4} \, \hat\Phi\, L\, L\,.
\end{equation}
It can be expressed in terms of real coordinates as 
\begin{equation}
\begin{split}
&t_1>\sqrt{x^2+y^2}\, : \\
&\d s^2 = \left(1- \frac{\kappa^2 m}{4} \,\hat\Phi\right)\d t_2'{}^2 +\d t_1^2-\d x^2-\d y^2
+ \frac{\frac{\kappa^2 m}{4} \,\hat\Phi}{1-\frac{\kappa^2 m}{4} \,\hat\Phi}\, \frac{(t_1\d t_1-x\d x-y\d y)^2}{{t_1^2-x^2-y^2}}
 \,,
 \end{split}
 \label{eq:realMetricInsideLC}
\end{equation}
using
\begin{equation}
\d t_2'=\d t_2 -i\; \frac{\frac{\kappa^2 m}{4} \,\hat\Phi}{1-\frac{\kappa^2 m}{4} \,\hat\Phi} \,
\frac{t_1\d t_1-x\d x-y\d y}{\sqrt{t_1^2-x^2-y^2}}\,.
\end{equation}
Now it is clear how to extend the solution beyond $t_1>\sqrt{x^2+y^2}$\,,
\begin{equation}
\d s^2 = \left(1- \frac{\kappa^2 m}{4} \,\Phi\right)\d t_2'{}^2 +\d t_1^2-\d x^2-\d y^2
+ \frac{\frac{\kappa^2 m}{4} \,\Phi}{1-\frac{\kappa^2 m}{4} \,\Phi}\, \frac{(t_1\d t_1-x\d x-y\d y)^2}{{t_1^2-x^2-y^2}}
 \,.
\end{equation}
This gives us the final answer of the exact gravity solution.\footnote{We could also write the line element in coordinates analogous to the Schwarzschild spherical coordinates, but would have to split into spacetime regions. Inside the light-cone, with $\chi=\sqrt{t_1^2-x^2-y^2}$ and $\hat\Phi=(2\pi \chi)^{-1}$, we pick $t_1=\chi \cosh\psi$ inside the future light-cone and $t_1=-\chi \cosh\psi$ inside the past light-cone, obtaining
\begin{equation}
t_1^2>{x^2+y^2}\, : \quad 
\d s^2 = \left(1- \frac{\kappa^2 m}{4} \,\Theta(t_1)\hat\Phi\right)\d t_2'{}^2+\frac{d\chi^2}{1- \frac{\kappa^2 m}{4} \,\Theta(t_1)\hat\Phi}-\chi^2 (d\psi^2+\sinh^2\psi \,d\phi^2)\,.
\end{equation}
Outside the light-cone, with $\tilde\chi=\sqrt{x^2+y^2-t_1^2}$, we can write
\begin{equation}
t_1^2<x^2+y^2\, : \quad 
\d s^2 = \d t_2'{}^2-{d\tilde\chi^2}+\tilde\chi^2 (-d\tilde\psi^2+\sin^2\tilde\psi \,d\phi^2)\,.
\end{equation}}
To check its consistency with the previous linearised result, we can put \eqref{eq:realMetricInsideLC} in de Donder gauge. This can be done by applying the diffeomorphism generated by
\begin{equation}
\xi=\frac{\kappa^2\,m}{16\pi}\,\d\,\log\frac{\sqrt{t_1^2-x^2-y^2}}{r_0}=\frac{\kappa^2\,m}{8}\, \hat\Phi\left(t_1\d t_1-x\,\d x-y\,\d y \right)\,.
\end{equation}
The resulting linearised metric is
\begin{equation}
h_{\mu\nu}=-\frac{\kappa\,m}{2}\hat\Phi\left(u_\mu u_\nu-\frac{1}{2}\eta_{\mu\nu}\right)~.
\end{equation}
Once again, this result is valid inside the lightcone. If we wish to extend it outside, we can replace $\hat{\Phi}$ by $\Phi$, recovering \eqref{eq:gravitonInPosSpace}.

\section{Analytic Continuation to Lorentzian Signature}
\label{sec:lorentzian}

The discussions above focus on split signature, but there are direct implications for Lorentzian signature, via analytic continuation.

\begin{figure}[t]
\begin{center}
\includegraphics[scale=0.8]{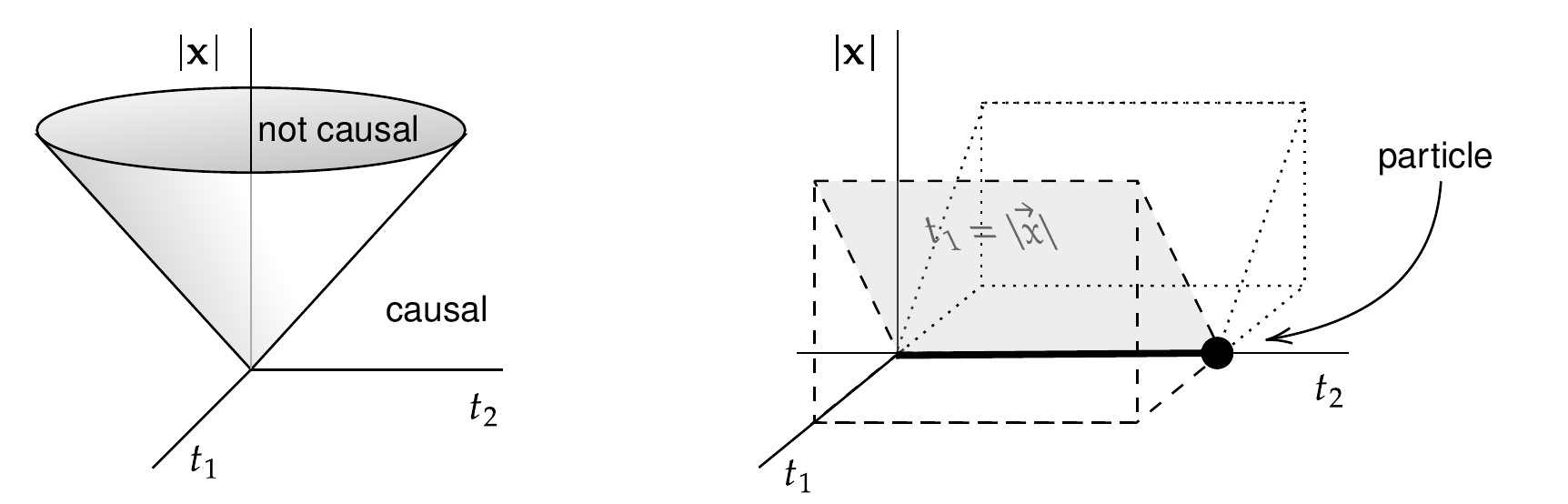}
\end{center}
\caption{The left image shows the 4D light-cone in split signature. The ``inner'' part of the cone contains all the events that are not causally connected to the vertex, whereas the events in the ``exterior'' can be reached by causal curves.
On the right, the diagram shows the support of the Green's function for our choice of $t_1$-retarded boundary conditions. The point particle trajectory is represented by the thick line moving along the $t_2$ axis. 
The shaded surface is $t_1-|\vec{x}|=0$, which contains the radiation. 
The dashed lines enclose the region where the retarded Green's function is non-zero, i.e.~the $t_1$-future of the particle. 
The dotted volume is the $t_1$-past of the particle. }\label{fig:22diagram}
\end{figure}

Let us compare again the Green's functions. In the split-signature case, we chose boundary conditions such that the Green's function is
\begin{equation}
\label{eq:Gfss}
\Phi = \frac{\Theta(t_1-\sqrt{x^2+y^2})}{2\pi\sqrt{t_1^2-x^2-y^2}}  \,,
\end{equation}
satisfying
\begin{equation}
\label{eq:Gfeq}
(\partial_{t_1}^2-\partial_x^2-\partial_y^2) \, \Phi = 
\delta(t_1)\delta(x)\delta(y)\,.
\end{equation}
Since we have the 3D wave operator, we made a choice that exhibits causality in the subspace $\{t_1,x,y\}$ by picking the retarded Green's function. This is shown in figure~\ref{fig:22diagram}, where the support of the retarded Green's function is represented as a dashed volume.

 We could have picked a $t_1$-symmetric Green's function, which is perhaps more natural from the point of view of analytic continuation to Lorentzian spacetime. With the latter choice, we would have
\begin{equation}
\Phi_{t_1\text{sym}} = \frac{\Theta(t_1^2-x^2-y^2)}{4\pi\sqrt{t_1^2-x^2-y^2}}  ~.
\end{equation}
This follows from the fact that equation \eqref{eq:Gfeq} is satisfied by both the retarded Green's function \eqref{eq:Gfss}, which is supported on the future light cone, and the advanced Green's function, obtained from \eqref{eq:Gfss} by the substitution $t_1-\sqrt{x^2+y^2}\to t_1+\sqrt{x^2+y^2}$ in the argument of the theta function. Then $\Phi_{t_1\text{sym}}$ is the average of the retarded and advanced Green's functions and has support on the dashed and dotted volumes in figure~\ref{fig:22diagram}.\footnote{We could also have chosen the Green's function to be $\Theta(x^2+y^2-t_1^2)/(4\pi\sqrt{x^2+y^2-t_1^2})$, which is acausal from the 3D perspective, but for which the analytic continuation below seems perhaps more straightforward.}

Now, if we perform an analytic continuation to the Lorentzian case via $t_1\to i z$, we obtain $-i\Phi^{(L)}$, with
\begin{equation}
\Phi^{(L)}=  \frac1{4\pi\sqrt{x^2+y^2+z^2}} 
\end{equation}
satisfying
\begin{equation}
-(\partial_x^2+\partial_y^2+\partial_z^2) \, \Phi^{(L)} =  \delta(x)\delta(y)\delta(z)\,.
\end{equation} 
From this Green's function, we can construct solutions in electromagnetism and in gravity, and they obviously correspond to the Coulomb and Schwarzschild solutions, respectively. Therefore, the discussions above concerning the description of solutions in terms of scattering amplitudes and the origin of the classical double copy from the double copy of scattering amplitudes extend to the Lorentzian case. The three-point scattering amplitudes that underlie those discussions would then be supported on complex kinematics.

\section{Discussion}
\label{sec:discussion}

Let us summarise our results. We used the building block of the on-shell approach to scattering amplitudes, the three-point amplitude, to study classical solutions in electromagnetism and in gravity. The three-point amplitudes studied correspond to the emission of a messenger (photon or graviton) by a charged/massive particle, and the classical solutions are precisely the solutions sourced by the massive particle. In order for the three-point amplitude to be non-trivial, we worked with a split-signature spacetime. The alternative would have been to consider complexified momenta in Lorentzian signature, as often done in the scattering amplitudes literature, but we found the split-signature choice more straightforward, given that relevant quantities like spinors are real. Moreover, split signature is interesting in its own right, particularly regarding boundary conditions and the meaning of causality. We discussed how our results are related via analytic continuation to Lorentzian signature. 

Building on the KMOC formalism~\cite{Kosower:2018adc}, we used the three-point amplitude to determine the coherent state generated by the massive particle, which is associated to the split-signature versions of the Coulomb and Schwarzschild solutions, for electromagnetism and gravity respectively. We described how to extract from that a classical field, namely via the expectation value of a quantum operator on the coherent state, in the classical limit. As operators, we considered the `curvatures': the field strength in electromagnetism and the spacetime curvature in vacuum (Weyl or Riemann, as they match in vacuum). These are gauge-invariant quantities (for gravity, in the linearised approximation). We found that the vacuum expectation value of these curvatures is an on-shell Fourier transform of the corresponding three-point amplitudes. This is easier to verify when we express the curvatures in terms of spinors, namely the Maxwell and Weyl spinors. 

The expressions we obtained for the Maxwell and Weyl spinors exhibit a Weyl-type classical double copy in on-shell momentum space, which follows directly from the double copy of the three-point scattering amplitudes. We then showed that this leads to the previously known Weyl double copy in position space, which applies to certain algebraically special classes of solutions~\cite{Luna:2018dpt,Godazgar:2020zbv}, here in the simplest case of the Coulomb and Schwarzschild solutions. We emphasise, however, that we expect the structure of the double copy in on-shell momentum space to be much more general. We see it as being formally equivalent to the convolutional double copy~\cite{Anastasiou:2014qba,Borsten:2019prq,Luna:2020adi}, but with the advantage, from our perspective, of being supported on on-shell momentum space, with a direct connection to scattering amplitudes.

Finally, we also used the Kerr-Schild-type classical double copy to obtain the exact gravity solution, rather than the linearised one. This is, to our knowledge, the first use of the classical double copy to write down a novel solution: the split-signature version of Schwarzschild. Although this could also have been achieved by analytic continuation of the Lorentzian Schwarzschild solution, with due attention paid to the split-signature boundary conditions, it was easier for us to use the classical double copy given that the split-signature boundary conditions were directly related to those in gauge theory. 

There are several obvious and exciting avenues for future research. The most obvious one is that of other three-point amplitudes, namely related to the Kerr-Taub-NUT extension of the Schwarzschild solution~\cite{Emond:2020lwi}. Another obvious direction is the consideration of self-interactions in gravity, i.e.~going beyond linear order, using the on-shell formalism of the coherent state. We described how to proceed, but it is worth it to make it more concrete.

The domain of applicability of the classical double copy is a natural question. Although many previous results support these ideas, we have provided here the ultimate connection to the double copy of scattering amplitudes. The Weyl double copy in on-shell momentum space {\it is} the amplitudes double copy. It will be interesting to explore this result better in position space, beyond linearised order, and to also connect it with the Kerr-Schild version of the classical double copy. 

We hope to address these questions in the near future.

\section*{Acknowledgements}

We thank Tim Adamo, Nima Arkani-Hamed, Andrea Cristofoli, Riccardo Gonzo, David Kosower, Alexander Ochirov and Alasdair Ross. 
RM is supported by a Royal Society University Research Fellowship, and DPV's studentship is also supported by the Royal Society.
DOC is supported by the STFC grant ST/P0000630/1 while
MS is supported by a Principal's Career Development Scholarship from the University of Edinburgh and the School of Physics and Astronomy. 

\appendix
\section{Conventions}
\label{sec:conventions}

\subsection{Spinors in split signature}

In coordinates $(t^1, t^2, x^1, x^2)$, we work with a metric of signature $(+1, +1, -1, -1)$. Since this signature may be unfamiliar, we gather here a list of spinor-helicity conventions appropriate for working in this signature.
Our conventions are designed to follow those of reference~\cite{Chung:2018kqs} (see appendix A) as closely as possible, while taking advantage of the different
reality properties available in split signature.

The Clifford algebra is
\[
\sigma^\mu \tilde \sigma^\nu + \sigma^\nu \tilde \sigma^\mu = 2 \eta^{\mu\nu} \mathbbm{1} \,.
\]
In our signature, it is possible to choose a real basis of $\sigma^\mu$ matrices. Our choice is
\[
\sigma^\mu = (1, i \sigma_y, \sigma_z, \sigma_x) \,
\]
where $\sigma_{x,y,z}$ are the usual Pauli matrices. The $\tilde \sigma^\mu$ are obtained by raising spinor indices, as usual:
\[
\tilde \sigma^{\mu\dot \alpha \alpha} = \epsilon^{\alpha\beta} \epsilon^{\dot \alpha \dot \beta} \sigma^\mu_{\beta \dot \beta} \,.
\]
We define $\epsilon^{12} = +1$, while $\epsilon_{12} = -1$, so that
\[
\epsilon_{\alpha \beta} \epsilon^{\beta \gamma} = \delta_\alpha^\gamma \,,
\]
and choose the same Levi-Civita sign for the opposite chirality,
\[
\epsilon_{\dot \alpha \dot \beta} = \epsilon_{\alpha \beta}\,, \quad \epsilon^{\dot \alpha \dot \beta} = \epsilon^{\alpha \beta} \,.
\]
We raise and lower all spinor indices (of either chirality) by acting from the left:
\[
\lambda_\alpha = \epsilon_{\alpha\beta} \lambda^\beta = \epsilon_{\alpha\beta} (\epsilon^{\beta\gamma} \lambda_\gamma) \,.
\]
It is often helpful to note that
\[
\sigma_{\alpha \dot \alpha} \cdot \sigma_{\beta \dot \beta} &= 2 \epsilon_{\alpha \beta} \epsilon_{\dot \alpha \dot \beta} \,, \\
\sigma_{\alpha \dot \alpha}  \cdot \tilde \sigma^{\dot \beta \beta} &= 2 \delta_\alpha^\beta \delta_{\dot \alpha}^{\dot \beta} \,.
\]

The chiral structure of spinors in split signature is important in our work. This structure is clarified by introducing the $\sigma^{\mu\nu}$ matrices
which are proportional to the Lorentz generators in the spinor representations. In particular, we define
\[
\sigma^{\mu\nu} &= \frac14 ( \sigma^\mu \tilde \sigma^\nu - \sigma^\mu \tilde \sigma^\nu) \,,\\
\tilde \sigma^{\mu\nu} &= \frac14 (\tilde \sigma^\mu \sigma^\nu - \tilde \sigma^\mu \sigma^\nu) \,.
\]
Since these matrices are antisymmetric in $\mu$ and $\nu$, there are at most six independent $\sigma^{\mu\nu}$ (and at most six independent $\tilde \sigma^{\mu\nu}$). However, the matrices enjoy the duality properties
\[
\sigma_{\mu\nu} &=  \frac12 \epsilon_{\mu\nu\rho\sigma} \, \sigma^{\rho\sigma} \,,\\
\tilde \sigma_{\mu\nu} &=  -\frac12 \epsilon_{\mu\nu\rho\sigma} \, \tilde \sigma^{\rho\sigma} \,.
\]
Consequently, there are only three independent $\sigma^{\mu\nu}$ matrices, which generate the group  $\mathrm{SL}(2, \mathbb{R})$.

To pass between momenta $k$ and spinors $\lambda$, $\tilde \lambda$, we define
\[
k \cdot \sigma_{\alpha \dot \alpha} = \lambda_\alpha \tilde \lambda_{\dot \alpha} \,.
\]
We use the symbols $\ket{k}$, $\bra{k}$, $[k|$, and $|k]$ to indicate the spinors with the indices in various positions as follows:
\[
\ket{k} \leftrightarrow \lambda_\alpha \,, 
\quad \bra{k} \leftrightarrow \lambda^{\alpha}\,, 
\quad |k] \leftrightarrow \tilde \lambda^{\dot \alpha}\,, 
\quad [k| \leftrightarrow \tilde \lambda_{\dot \alpha} \,.
\]

As usual, we choose a basis of polarisation vectors of definite helicity $\hel = \pm$. Unlike the Minkowski case, these vectors can be chosen to be
real, and we make such a choice. Given a momentum $k$ and gauge choice $q$ satisfying $k \cdot q \neq 0$, $k^2 = 0 = q^2$, we define
\[
\varepsilon_-^\mu &= -\frac{\bra{k} \sigma^\mu |q]}{\sqrt{2} [k q]} \,,\qquad
\varepsilon_+^\mu &= \frac{[k| \tilde \sigma^\mu \ket{q}}{\sqrt2 \langle k q \rangle}\,.
\label{eq:polVectorDef}
\]
These polarisation vectors have the properties:
\[
(\varepsilon_{h}^\mu(k))^*&=\varepsilon_{h}^\mu(k) \, , \\
\varepsilon^2_\pm(k)&=0\,, \\
\varepsilon_+(k)\cdot \varepsilon_-(k)&=-1 \,,
\]
assuming that both $k$ and $q$ are real.

A plane wave with negative polarisation has a self-dual field strength in our conventions:
\[
\sigma_{\mu\nu} \, k^{[\mu} \varepsilon_-^{\nu]} &=- \sqrt{2} \, \ket{k} \bra{k} \,, \\
\tilde \sigma_{\mu\nu} \, k^{[\mu} \varepsilon_-^{\nu]} &= 0 \,.
\label{eq:negativeHelFS}
\]
Meanwhile, a positive helicity plane wave has anti-self dual field strength given by
\[
\sigma_{\mu\nu} \, k^{[\mu} \varepsilon_+^{\nu]} &= 0 \,, \\
\tilde \sigma_{\mu\nu} \, k^{[\mu} \varepsilon_+^{\nu]} &=  \sqrt{2} \, |k] [k| \,.
\label{eq:positiveHelFS}
\]

\subsection{Miscellaneous conventions}

We define our Fourier transforms by
\[
f(x) &= \int \dd^4 k \, e^{-i k \cdot x} \, \tilde f(k) \,,\\
\tilde f(k) &= \int d^4 x \, e^{ik\cdot x} \, f(x) \,.
\]
To tidy up factors of $2\pi$, we define
\[
\dd^nk =\frac{\d^n k }{(2\pi)^n}\,,\qquad \del^n(k)=(2\pi)^n \delta^n(k) \,.
\]
We define
\[
v^{[\mu}w^{\nu]}=v^\mu w^\nu-v^\nu w^\mu, \qquad v^{(\mu}w^{\nu)}=v^\mu w^\nu+v^\nu w^\mu.
\label{eq:antisymmDef}
\]
We have also used 
\begin{equation}
X:=\sqrt{2}u\cdot \varepsilon_{+}\,, \qquad \frac{1}{X}:=-\sqrt{2} u\cdot \varepsilon_-\,,
\end{equation}
 so that $X \frac{1}{X}=-2 \varepsilon_+^\mu \varepsilon_-^\nu u_\mu u_\nu=-2 \left(\frac{-1}{2}\eta^{\mu\nu} \right) u_\mu u_\nu=1.$ Note that then $(X^h)^*=X^{h}$, since everything is real.

\section{The retarded Green's function in $1+2$ dimensions}
\label{sec:retardedGF}

Because of the translation symmetry in the $t^2$ direction, much of our discussion really takes place in a three-dimensional space with signature
$(+,-,-)$. In this appendix, we compute the retarded Green's function (for the wave operator) in this space. We use the familiar 
notation $x = (t, \vec{x})$ for points in this spacetime, and write wave vectors as $k = (E, \vec k)$.

The Green's function is defined to satisfy
\[
\partial^2 G(x) = \delta^{(3)}(x) \,,
\]
with the boundary condition that 
\[
G(x) = 0 \,, \quad t < 0 \,.
\label{eq:retBC}
\]
It is easy to express the Green's function in Fourier space as
\[
G(x) = - \int \dd^3 k \, e^{-i k \cdot x} \frac{1}{k^2_\mathrm{ret}} \,.
\label{eq:greensFourier}
\]
The instruction `$\textrm{ret}$' indicates that we must define the integral to enforce the retarded boundary condition~\eqref{eq:retBC}. As usual,
we interpret the integral over the first component $E$ of $k^\mu$ as a contour integral, and (as in the main text) we impose the boundary 
condition by displacing the poles below the real $E$ axis. It is easy to compute the value of the $E$ integral using the residue theorem, with the
result that
\[
G(x) &= \frac{-i}{8\pi^2} \Theta(t) \int \d^2 k \, e^{i \vec{k} \cdot \vec{x}} \frac{e^{i |\vec{k}| t} - e^{-i |\vec{k}| t}}{|\vec{k}|} \\
&= \frac{-i}{8\pi^2} \Theta(t) \int_0^\infty \d k \int_0^{2\pi} \d\theta \, e^{i k r \cos \theta} \left(e^{i k t} - e^{-i k t} \right) \,,
\label{eq:greensStep}
\]
where, in the second equality, we defined $r = |\vec x|$ and introduced polar coordinates for the $\vec k$ integration.

Our integral is still not completely well-defined. Notice that if we perform the $k$ integral in equation~\eqref{eq:greensStep} first, we encounter
oscillatory factors which do not converge. The solution is again familiar: we introduce $i k \epsilon$ convergence factors in the exponents, 
adjusting the signs to make the integrals well-defined. The result is
\[
G(x) &=\frac{-i}{8\pi^2} \Theta(t) \int_0^\infty \d k \int_0^{2\pi} \d\theta \, e^{i k r \cos \theta} \left(e^{i k (t + i \epsilon)} - e^{-i k (t -i \epsilon) } \right) \,.
\]

Recognising the definition of the Bessel function, it is easy to perform the $\theta$ integration next, yielding
\[
G(x) = \frac{-i}{4\pi} \Theta(t) \int_0^\infty \d k  \, J_0(k r) \left(e^{i k (t + i \epsilon)} - e^{-i k (t -i \epsilon) } \right) \,.
\]
We can perform the final integral using the result 
\[
\int_0^\infty \d u  \, J_0(u) e^{i u v} = \frac{1}{\sqrt{1-v^2}} \,,
\]
so that
\[
G(x) = \frac{i}{4\pi} \Theta(t) \left( \frac{1}{\sqrt{r^2 - t^2 + i \epsilon}} - \frac{1}{\sqrt{r^2 - t^2 - i \epsilon}} \right) \,.
\]
At this point, the $i \epsilon$ factors come into their own. Evidently, the Green's function vanishes when we can ignore the $\epsilon$'s: this occurs
when $r^2 - t^2$ is positive. But when $r^2 - t^2 <0$, then the $\epsilon$'s control which side of the branch cut in the square root function we must
choose. We have
\[
G(x) &= \frac{i}{4\pi} \Theta(t) \Theta(t^2 - r^2) \left( \frac{1}{\sqrt{- |t^2 - r^2| + i \epsilon}} - \frac{1}{\sqrt{-|t^2 - r^2| - i \epsilon}} \right)  \\
&= \frac{i}{4\pi} \Theta(t) \Theta(t^2 - r^2) \left( \frac{1}{i\sqrt{|t^2 - r^2|}} - \frac{1}{(-i)\sqrt{|t^2 - r^2|}} \right)  \\
&= \frac{1}{2\pi} \Theta(t) \Theta(t^2 - r^2) \frac{1}{\sqrt{t^2 - r^2}} \,.
\]
As discussed in more detail in section~\ref{sec:lorentzian}, this Green's function is a Lorentzian version of the familiar Euclidean Green's function $\sim 1/r$. The theta functions are a 
result of our boundary conditions.\footnote{An equivalent derivation can be found in \cite{watanabe}.}


\begin{thebibliography}{100}

\bibitem{Britto:2005fq}
R.~Britto, F.~Cachazo, B.~Feng, and E.~Witten, {\it {Direct proof of tree-level
  recursion relation in Yang-Mills theory}},  {\em Phys. Rev. Lett.} {\bf 94}
  (2005) 181602, [\href{http://arxiv.org/abs/hep-th/0501052}{{\tt
  hep-th/0501052}}].

\bibitem{Bern:1994zx}
Z.~Bern, L.~J. Dixon, D.~C. Dunbar, and D.~A. Kosower, {\it {One loop n point
  gauge theory amplitudes, unitarity and collinear limits}},  {\em Nucl. Phys.
  B} {\bf 425} (1994) 217--260,
  [\href{http://arxiv.org/abs/hep-ph/9403226}{{\tt hep-ph/9403226}}].

\bibitem{Bern:1994cg}
Z.~Bern, L.~J. Dixon, D.~C. Dunbar, and D.~A. Kosower, {\it {Fusing gauge
  theory tree amplitudes into loop amplitudes}},  {\em Nucl. Phys. B} {\bf 435}
  (1995) 59--101, [\href{http://arxiv.org/abs/hep-ph/9409265}{{\tt
  hep-ph/9409265}}].

\bibitem{Benincasa:2007xk}
P.~Benincasa and F.~Cachazo, {\it {Consistency Conditions on the S-Matrix of
  Massless Particles}},  \href{http://arxiv.org/abs/0705.4305}{{\tt
  arXiv:0705.4305}}.

\bibitem{ah3}
N.~Arkani-Hamed, T.-C. Huang, and Y.-t. Huang, {\it {Scattering Amplitudes For
  All Masses and Spins}},  \href{http://arxiv.org/abs/1709.04891}{{\tt
  arXiv:1709.04891}}.

\bibitem{Neill:2013wsa}
D.~Neill and I.~Z. Rothstein, {\it {Classical Space-Times from the S Matrix}},
  {\em Nucl. Phys.} {\bf B877} (2013) 177--189,
  [\href{http://arxiv.org/abs/1304.7263}{{\tt arXiv:1304.7263}}].

\bibitem{Bjerrum-Bohr:2013bxa}
N.~E.~J. Bjerrum-Bohr, J.~F. Donoghue, and P.~Vanhove, {\it {On-shell
  Techniques and Universal Results in Quantum Gravity}},  {\em JHEP} {\bf 02}
  (2014) 111, [\href{http://arxiv.org/abs/1309.0804}{{\tt arXiv:1309.0804}}].

\bibitem{Monteiro:2014cda}
R.~Monteiro, D.~O'Connell, and C.~D. White, {\it {Black holes and the double
  copy}},  {\em JHEP} {\bf 12} (2014) 056,
  [\href{http://arxiv.org/abs/1410.0239}{{\tt arXiv:1410.0239}}].

\bibitem{Bjerrum-Bohr:2014zsa}
N.~E.~J. Bjerrum-Bohr, J.~F. Donoghue, B.~R. Holstein, L.~Plant\'{e}, and
  P.~Vanhove, {\it {Bending of Light in Quantum Gravity}},  {\em Phys. Rev.
  Lett.} {\bf 114} (2015), no.~6 061301,
  [\href{http://arxiv.org/abs/1410.7590}{{\tt arXiv:1410.7590}}].

\bibitem{Luna:2016due}
A.~Luna, R.~Monteiro, I.~Nicholson, D.~O'Connell, and C.~D. White, {\it {The
  double copy: Bremsstrahlung and accelerating black holes}},  {\em JHEP} {\bf
  06} (2016) 023, [\href{http://arxiv.org/abs/1603.05737}{{\tt
  arXiv:1603.05737}}].

\bibitem{Damour:2016gwp}
T.~Damour, {\it {Gravitational scattering, post-Minkowskian approximation and
  Effective One-Body theory}},  {\em Phys. Rev.} {\bf D94} (2016), no.~10
  104015, [\href{http://arxiv.org/abs/1609.00354}{{\tt arXiv:1609.00354}}].

\bibitem{Goldberger:2016iau}
W.~D. Goldberger and A.~K. Ridgway, {\it {Radiation and the classical double
  copy for color charges}},  {\em Phys. Rev. D} {\bf 95} (2017), no.~12 125010,
  [\href{http://arxiv.org/abs/1611.03493}{{\tt arXiv:1611.03493}}].

\bibitem{Cachazo:2017jef}
F.~Cachazo and A.~Guevara, {\it {Leading Singularities and Classical
  Gravitational Scattering}},  {\em JHEP} {\bf 02} (2020) 181,
  [\href{http://arxiv.org/abs/1705.10262}{{\tt arXiv:1705.10262}}].

\bibitem{Guevara:2017csg}
A.~Guevara, {\it {Holomorphic Classical Limit for Spin Effects in Gravitational
  and Electromagnetic Scattering}},  {\em JHEP} {\bf 04} (2019) 033,
  [\href{http://arxiv.org/abs/1706.02314}{{\tt arXiv:1706.02314}}].

\bibitem{Damour:2017zjx}
T.~Damour, {\it {High-energy gravitational scattering and the general
  relativistic two-body problem}},  {\em Phys. Rev.} {\bf D97} (2018), no.~4
  044038, [\href{http://arxiv.org/abs/1710.10599}{{\tt arXiv:1710.10599}}].

\bibitem{Luna:2017dtq}
A.~Luna, I.~Nicholson, D.~O'Connell, and C.~D. White, {\it {Inelastic Black
  Hole Scattering from Charged Scalar Amplitudes}},  {\em JHEP} {\bf 03} (2018)
  044, [\href{http://arxiv.org/abs/1711.03901}{{\tt arXiv:1711.03901}}].

\bibitem{Laddha:2018rle}
A.~Laddha and A.~Sen, {\it {Gravity Waves from Soft Theorem in General
  Dimensions}},  {\em JHEP} {\bf 09} (2018) 105,
  [\href{http://arxiv.org/abs/1801.07719}{{\tt arXiv:1801.07719}}].

\bibitem{Laddha:2018vbn}
A.~Laddha and A.~Sen, {\it {Observational Signature of the Logarithmic Terms in
  the Soft Graviton Theorem}},  {\em Phys. Rev. D} {\bf 100} (2019), no.~2
  024009, [\href{http://arxiv.org/abs/1806.01872}{{\tt arXiv:1806.01872}}].

\bibitem{Bjerrum-Bohr:2018xdl}
N.~E.~J. Bjerrum-Bohr, P.~H. Damgaard, G.~Festuccia, L.~Plant\'{e}, and
  P.~Vanhove, {\it {General Relativity from Scattering Amplitudes}},  {\em
  Phys. Rev. Lett.} {\bf 121} (2018), no.~17 171601,
  [\href{http://arxiv.org/abs/1806.04920}{{\tt arXiv:1806.04920}}].

\bibitem{Cheung:2018wkq}
C.~Cheung, I.~Z. Rothstein, and M.~P. Solon, {\it {From Scattering Amplitudes
  to Classical Potentials in the Post-Minkowskian Expansion}},  {\em Phys. Rev.
  Lett.} {\bf 121} (2018), no.~25 251101,
  [\href{http://arxiv.org/abs/1808.02489}{{\tt arXiv:1808.02489}}].

\bibitem{Kosower:2018adc}
D.~A. Kosower, B.~Maybee, and D.~O'Connell, {\it {Amplitudes, Observables, and
  Classical Scattering}},  {\em JHEP} {\bf 02} (2019) 137,
  [\href{http://arxiv.org/abs/1811.10950}{{\tt arXiv:1811.10950}}].

\bibitem{Guevara:2018wpp}
A.~Guevara, A.~Ochirov, and J.~Vines, {\it {Scattering of Spinning Black Holes
  from Exponentiated Soft Factors}},  {\em JHEP} {\bf 09} (2019) 056,
  [\href{http://arxiv.org/abs/1812.06895}{{\tt arXiv:1812.06895}}].

\bibitem{Bern:2019nnu}
Z.~Bern, C.~Cheung, R.~Roiban, C.-H. Shen, M.~P. Solon, and M.~Zeng, {\it
  {Scattering Amplitudes and the Conservative Hamiltonian for Binary Systems at
  Third Post-Minkowskian Order}},  {\em Phys. Rev. Lett.} {\bf 122} (2019),
  no.~20 201603, [\href{http://arxiv.org/abs/1901.04424}{{\tt
  arXiv:1901.04424}}].

\bibitem{Cristofoli:2019neg}
A.~Cristofoli, N.~Bjerrum-Bohr, P.~H. Damgaard, and P.~Vanhove, {\it
  {Post-Minkowskian Hamiltonians in general relativity}},  {\em Phys. Rev. D}
  {\bf 100} (2019), no.~8 084040, [\href{http://arxiv.org/abs/1906.01579}{{\tt
  arXiv:1906.01579}}].

\bibitem{Maybee:2019jus}
B.~Maybee, D.~O'Connell, and J.~Vines, {\it {Observables and amplitudes for
  spinning particles and black holes}},  {\em JHEP} {\bf 12} (2019) 156,
  [\href{http://arxiv.org/abs/1906.09260}{{\tt arXiv:1906.09260}}].

\bibitem{Guevara:2019fsj}
A.~Guevara, A.~Ochirov, and J.~Vines, {\it {Black-hole scattering with general
  spin directions from minimal-coupling amplitudes}},  {\em Phys. Rev. D} {\bf
  100} (2019), no.~10 104024, [\href{http://arxiv.org/abs/1906.10071}{{\tt
  arXiv:1906.10071}}].

\bibitem{Bern:2019crd}
Z.~Bern, C.~Cheung, R.~Roiban, C.-H. Shen, M.~P. Solon, and M.~Zeng, {\it
  {Black Hole Binary Dynamics from the Double Copy and Effective Theory}},
  {\em JHEP} {\bf 10} (2019) 206, [\href{http://arxiv.org/abs/1908.01493}{{\tt
  arXiv:1908.01493}}].

\bibitem{Kalin:2019rwq}
G.~K{\"a}lin and R.~A. Porto, {\it {From Boundary Data to Bound States}},  {\em
  JHEP} {\bf 01} (2020) 072, [\href{http://arxiv.org/abs/1910.03008}{{\tt
  arXiv:1910.03008}}].

\bibitem{Kalin:2019inp}
G.~K{\"a}lin and R.~A. Porto, {\it {From boundary data to bound states. Part
  II. Scattering angle to dynamical invariants (with twist)}},  {\em JHEP} {\bf
  02} (2020) 120, [\href{http://arxiv.org/abs/1911.09130}{{\tt
  arXiv:1911.09130}}].

\bibitem{Aoude:2020onz}
R.~Aoude, K.~Haddad, and A.~Helset, {\it {On-shell heavy particle effective
  theories}},  {\em JHEP} {\bf 05} (2020) 051,
  [\href{http://arxiv.org/abs/2001.09164}{{\tt arXiv:2001.09164}}].

\bibitem{Cheung:2020gyp}
C.~Cheung and M.~P. Solon, {\it {Classical gravitational scattering at $
  \mathcal{O} $(G$^{3}$) from Feynman diagrams}},  {\em JHEP} {\bf 06} (2020)
  144, [\href{http://arxiv.org/abs/2003.08351}{{\tt arXiv:2003.08351}}].

\bibitem{Bern:2020buy}
Z.~Bern, A.~Luna, R.~Roiban, C.-H. Shen, and M.~Zeng, {\it {Spinning Black Hole
  Binary Dynamics, Scattering Amplitudes and Effective Field Theory}},
  \href{http://arxiv.org/abs/2005.03071}{{\tt arXiv:2005.03071}}.

\bibitem{Cheung:2020sdj}
C.~Cheung and M.~P. Solon, {\it {Tidal Effects in the Post-Minkowskian
  Expansion}},  \href{http://arxiv.org/abs/2006.06665}{{\tt arXiv:2006.06665}}.

\bibitem{Kalin:2020fhe}
G.~K{\"a}lin, Z.~Liu, and R.~A. Porto, {\it {Conservative Dynamics of Binary
  Systems to Third Post-Minkowskian Order from the Effective Field Theory
  Approach}},  \href{http://arxiv.org/abs/2007.04977}{{\tt arXiv:2007.04977}}.

\bibitem{Haddad:2020que}
K.~Haddad and A.~Helset, {\it {Gravitational tidal effects in quantum field
  theory}},  \href{http://arxiv.org/abs/2008.04920}{{\tt arXiv:2008.04920}}.

\bibitem{Kalin:2020lmz}
G.~K{\"a}lin, Z.~Liu, and R.~A. Porto, {\it {Conservative Tidal Effects in
  Compact Binary Systems to Next-to-Leading Post-Minkowskian Order}},
  \href{http://arxiv.org/abs/2008.06047}{{\tt arXiv:2008.06047}}.

\bibitem{DiVecchia:2020ymx}
P.~Di~Vecchia, C.~Heissenberg, R.~Russo, and G.~Veneziano, {\it {Universality
  of ultra-relativistic gravitational scattering}},
  \href{http://arxiv.org/abs/2008.12743}{{\tt arXiv:2008.12743}}.

\bibitem{Bern:2020uwk}
Z.~Bern, J.~Parra-Martinez, R.~Roiban, E.~Sawyer, and C.-H. Shen, {\it {Leading
  Nonlinear Tidal Effects and Scattering Amplitudes}},
  \href{http://arxiv.org/abs/2010.08559}{{\tt arXiv:2010.08559}}.

\bibitem{Huber:2020xny}
M.~A. Huber, A.~Brandhuber, S.~De~Angelis, and G.~Travaglini, {\it {From
  amplitudes to gravitational radiation with cubic interactions and tidal
  effects}},  \href{http://arxiv.org/abs/2012.06548}{{\tt arXiv:2012.06548}}.

\bibitem{Bjerrum-Bohr:2017dxw}
N.~E.~J. Bjerrum-Bohr, B.~R. Holstein, J.~F. Donoghue, L.~Plant\'{e}, and
  P.~Vanhove, {\it {Illuminating Light Bending}},  {\em PoS} {\bf CORFU2016}
  (2017) 077, [\href{http://arxiv.org/abs/1704.01624}{{\tt arXiv:1704.01624}}].

\bibitem{Laddha:2018myi}
A.~Laddha and A.~Sen, {\it {Logarithmic Terms in the Soft Expansion in Four
  Dimensions}},  {\em JHEP} {\bf 10} (2018) 056,
  [\href{http://arxiv.org/abs/1804.09193}{{\tt arXiv:1804.09193}}].

\bibitem{Sahoo:2018lxl}
B.~Sahoo and A.~Sen, {\it {Classical and Quantum Results on Logarithmic Terms
  in the Soft Theorem in Four Dimensions}},  {\em JHEP} {\bf 02} (2019) 086,
  [\href{http://arxiv.org/abs/1808.03288}{{\tt arXiv:1808.03288}}].

\bibitem{Bautista:2019tdr}
Y.~F. Bautista and A.~Guevara, {\it {From Scattering Amplitudes to Classical
  Physics: Universality, Double Copy and Soft Theorems}},
  \href{http://arxiv.org/abs/1903.12419}{{\tt arXiv:1903.12419}}.

\bibitem{Brandhuber:2019qpg}
A.~Brandhuber and G.~Travaglini, {\it {On higher-derivative effects on the
  gravitational potential and particle bending}},  {\em JHEP} {\bf 01} (2020)
  010, [\href{http://arxiv.org/abs/1905.05657}{{\tt arXiv:1905.05657}}].

\bibitem{Laddha:2019yaj}
A.~Laddha and A.~Sen, {\it {Classical proof of the classical soft graviton
  theorem in $D > 4$}},  {\em Phys. Rev. D} {\bf 101} (2020), no.~8 084011,
  [\href{http://arxiv.org/abs/1906.08288}{{\tt arXiv:1906.08288}}].

\bibitem{Arkani-Hamed:2019ymq}
N.~Arkani-Hamed, Y.-t. Huang, and D.~O'Connell, {\it {Kerr black holes as
  elementary particles}},  {\em JHEP} {\bf 01} (2020) 046,
  [\href{http://arxiv.org/abs/1906.10100}{{\tt arXiv:1906.10100}}].

\bibitem{Damgaard:2019lfh}
P.~H. Damgaard, K.~Haddad, and A.~Helset, {\it {Heavy Black Hole Effective
  Theory}},  {\em JHEP} {\bf 11} (2019) 070,
  [\href{http://arxiv.org/abs/1908.10308}{{\tt arXiv:1908.10308}}].

\bibitem{Bjerrum-Bohr:2019kec}
N.~Bjerrum-Bohr, A.~Cristofoli, and P.~H. Damgaard, {\it {Post-Minkowskian
  Scattering Angle in Einstein Gravity}},  {\em JHEP} {\bf 08} (2020) 038,
  [\href{http://arxiv.org/abs/1910.09366}{{\tt arXiv:1910.09366}}].

\bibitem{Huang:2019cja}
Y.-T. Huang, U.~Kol, and D.~O'Connell, {\it {Double copy of electric-magnetic
  duality}},  {\em Phys. Rev. D} {\bf 102} (2020), no.~4 046005,
  [\href{http://arxiv.org/abs/1911.06318}{{\tt arXiv:1911.06318}}].

\bibitem{Huber:2019ugz}
M.~Accettulli~Huber, A.~Brandhuber, S.~De~Angelis, and G.~Travaglini, {\it
  {Note on the absence of $R^2$ corrections to Newton's potential}},  {\em
  Phys. Rev. D} {\bf 101} (2020), no.~4 046011,
  [\href{http://arxiv.org/abs/1911.10108}{{\tt arXiv:1911.10108}}].

\bibitem{Saha:2019tub}
A.~P. Saha, B.~Sahoo, and A.~Sen, {\it {Proof of the classical soft graviton
  theorem in $D$ = 4}},  {\em JHEP} {\bf 06} (2020) 153,
  [\href{http://arxiv.org/abs/1912.06413}{{\tt arXiv:1912.06413}}].

\bibitem{Bern:2020gjj}
Z.~Bern, H.~Ita, J.~Parra-Martinez, and M.~S. Ruf, {\it {Universality in the
  classical limit of massless gravitational scattering}},  {\em Phys. Rev.
  Lett.} {\bf 125} (2020), no.~3 031601,
  [\href{http://arxiv.org/abs/2002.02459}{{\tt arXiv:2002.02459}}].

\bibitem{Moynihan:2020gxj}
N.~Moynihan and J.~Murugan, {\it {On-Shell Electric-Magnetic Duality and the
  Dual Graviton}},  \href{http://arxiv.org/abs/2002.11085}{{\tt
  arXiv:2002.11085}}.

\bibitem{Cristofoli:2020uzm}
A.~Cristofoli, P.~H. Damgaard, P.~Di~Vecchia, and C.~Heissenberg, {\it
  {Second-order Post-Minkowskian scattering in arbitrary dimensions}},  {\em
  JHEP} {\bf 07} (2020) 122, [\href{http://arxiv.org/abs/2003.10274}{{\tt
  arXiv:2003.10274}}].

\bibitem{Parra-Martinez:2020dzs}
J.~Parra-Martinez, M.~S. Ruf, and M.~Zeng, {\it {Extremal black hole scattering
  at $O(G^3)$: graviton dominance, eikonal exponentiation, and differential
  equations}},  \href{http://arxiv.org/abs/2005.04236}{{\tt arXiv:2005.04236}}.

\bibitem{Haddad:2020tvs}
K.~Haddad and A.~Helset, {\it {The double copy for heavy particles}},
  \href{http://arxiv.org/abs/2005.13897}{{\tt arXiv:2005.13897}}.

\bibitem{AccettulliHuber:2020oou}
M.~Accettulli~Huber, A.~Brandhuber, S.~De~Angelis, and G.~Travaglini, {\it
  {Eikonal phase matrix, deflection angle and time delay in effective field
  theories of gravity}},  {\em Phys. Rev. D} {\bf 102} (2020), no.~4 046014,
  [\href{http://arxiv.org/abs/2006.02375}{{\tt arXiv:2006.02375}}].

\bibitem{Moynihan:2020ejh}
N.~Moynihan, {\it {Scattering Amplitudes and the Double Copy in Topologically
  Massive Theories}},  \href{http://arxiv.org/abs/2006.15957}{{\tt
  arXiv:2006.15957}}.

\bibitem{A:2020lub}
M.~A., D.~Ghosh, A.~Laddha, and A.~P.~V., {\it {Soft Radiation from Scattering
  Amplitudes Revisited}},  \href{http://arxiv.org/abs/2007.02077}{{\tt
  arXiv:2007.02077}}.

\bibitem{Sahoo:2020ryf}
B.~Sahoo, {\it {Classical Sub-subleading Soft Photon and Soft Graviton Theorems
  in Four Spacetime Dimensions}},  \href{http://arxiv.org/abs/2008.04376}{{\tt
  arXiv:2008.04376}}.

\bibitem{delaCruz:2020bbn}
L.~de~la Cruz, B.~Maybee, D.~O'Connell, and A.~Ross, {\it {Classical Yang-Mills
  observables from amplitudes}},  \href{http://arxiv.org/abs/2009.03842}{{\tt
  arXiv:2009.03842}}.

\bibitem{Bonocore:2020xuj}
D.~Bonocore, {\it {Asymptotic dynamics on the worldline for spinning
  particles}},  \href{http://arxiv.org/abs/2009.07863}{{\tt arXiv:2009.07863}}.

\bibitem{Mogull:2020sak}
G.~Mogull, J.~Plefka, and J.~Steinhoff, {\it {Classical black hole scattering
  from a worldline quantum field theory}},
  \href{http://arxiv.org/abs/2010.02865}{{\tt arXiv:2010.02865}}.

\bibitem{Emond:2020lwi}
W.~T. Emond, Y.-T. Huang, U.~Kol, N.~Moynihan, and D.~O'Connell, {\it
  {Amplitudes from Coulomb to Kerr-Taub-NUT}},
  \href{http://arxiv.org/abs/2010.07861}{{\tt arXiv:2010.07861}}.

\bibitem{Cheung:2020gbf}
C.~Cheung, N.~Shah, and M.~P. Solon, {\it {Mining the Geodesic Equation for
  Scattering Data}},  \href{http://arxiv.org/abs/2010.08568}{{\tt
  arXiv:2010.08568}}.

\bibitem{Mougiakakos:2020laz}
S.~Mougiakakos and P.~Vanhove, {\it {The Schwarzschild-Tangherlini metric from
  scattering amplitudes in various dimensions}},
  \href{http://arxiv.org/abs/2010.08882}{{\tt arXiv:2010.08882}}.

\bibitem{Carrasco:2020ywq}
J.~J.~M. Carrasco and I.~A. Vazquez-Holm, {\it {Loop-Level Double-Copy for
  Massive Quantum Particles}},  \href{http://arxiv.org/abs/2010.13435}{{\tt
  arXiv:2010.13435}}.

\bibitem{Kim:2020cvf}
J.-W. Kim and M.~Shim, {\it {Gravitational Dyonic Amplitude at One-Loop and its
  Inconsistency with the Classical Impulse}},
  \href{http://arxiv.org/abs/2010.14347}{{\tt arXiv:2010.14347}}.

\bibitem{Bjerrum-Bohr:2020syg}
N.~E.~J. Bjerrum-Bohr, T.~V. Brown, and H.~Gomez, {\it {Scattering of Gravitons
  and Spinning Massive States from Compact Numerators}},
  \href{http://arxiv.org/abs/2011.10556}{{\tt arXiv:2011.10556}}.

\bibitem{Gonzo:2020xza}
R.~Gonzo and A.~Pokraka, {\it {Light-ray operators, detectors and gravitational
  event shapes}},  \href{http://arxiv.org/abs/2012.01406}{{\tt
  arXiv:2012.01406}}.

\bibitem{delaCruz:2020cpc}
L.~de~la Cruz, {\it {A scattering amplitudes approach to hard thermal loops}},
  \href{http://arxiv.org/abs/2012.07714}{{\tt arXiv:2012.07714}}.

\bibitem{Donoghue:1993eb}
J.~F. Donoghue, {\it {Leading quantum correction to the Newtonian potential}},
  {\em Phys. Rev. Lett.} {\bf 72} (1994) 2996--2999,
  [\href{http://arxiv.org/abs/gr-qc/9310024}{{\tt gr-qc/9310024}}].

\bibitem{Donoghue:1994dn}
J.~F. Donoghue, {\it {General relativity as an effective field theory: The
  leading quantum corrections}},  {\em Phys. Rev.} {\bf D50} (1994) 3874--3888,
  [\href{http://arxiv.org/abs/gr-qc/9405057}{{\tt gr-qc/9405057}}].

\bibitem{Srednyak:2013ylj}
S.~Srednyak and G.~Sterman, {\it {Perturbation theory in (2,2) signature}},
  {\em Phys. Rev. D} {\bf 87} (2013), no.~10 105017,
  [\href{http://arxiv.org/abs/1302.4290}{{\tt arXiv:1302.4290}}].

\bibitem{Mason:2005qu}
L.~Mason, {\it {Global anti-self-dual Yang-Mills fields in split signature and
  their scattering}},  \href{http://arxiv.org/abs/math-ph/0505039}{{\tt
  math-ph/0505039}}.

\bibitem{Barrett:1993yn}
J.~W. Barrett, G.~Gibbons, M.~Perry, C.~Pope, and P.~Ruback, {\it {Kleinian
  geometry and the N=2 superstring}},  {\em Int. J. Mod. Phys. A} {\bf 9}
  (1994) 1457--1494, [\href{http://arxiv.org/abs/hep-th/9302073}{{\tt
  hep-th/9302073}}].

\bibitem{Newman:1961qr}
E.~Newman and R.~Penrose, {\it {An Approach to gravitational radiation by a
  method of spin coefficients}},  {\em J. Math. Phys.} {\bf 3} (1962) 566--578.

\bibitem{Sachs:1961zz}
R.~Sachs, {\it {Gravitational waves in general relativity. 6. The outgoing
  radiation condition}},  {\em Proc. Roy. Soc. Lond. A} {\bf 264} (1961)
  309--338.

\bibitem{klt}
H.~Kawai, D.~Lewellen, and S.~Tye, {\it {A Relation Between Tree Amplitudes of
  Closed and Open Strings}},  {\em Nucl. Phys. B} {\bf 269} (1986) 1--23.

\bibitem{Bern:2008qj}
Z.~Bern, J.~J.~M. Carrasco, and H.~Johansson, {\it {New Relations for
  Gauge-Theory Amplitudes}},  {\em Phys. Rev.} {\bf D78} (2008) 085011,
  [\href{http://arxiv.org/abs/0805.3993}{{\tt arXiv:0805.3993}}].

\bibitem{Bern:2010ue}
Z.~Bern, J.~J.~M. Carrasco, and H.~Johansson, {\it {Perturbative Quantum
  Gravity as a Double Copy of Gauge Theory}},  {\em Phys. Rev. Lett.} {\bf 105}
  (2010) 061602, [\href{http://arxiv.org/abs/1004.0476}{{\tt
  arXiv:1004.0476}}].

\bibitem{zvirev}
Z.~Bern, J.~J. Carrasco, M.~Chiodaroli, H.~Johansson, and R.~Roiban, {\it {The
  Duality Between Color and Kinematics and its Applications}},
  \href{http://arxiv.org/abs/1909.01358}{{\tt arXiv:1909.01358}}.

\bibitem{Monteiro:2011pc}
R.~Monteiro and D.~O'Connell, {\it {The Kinematic Algebra From the Self-Dual
  Sector}},  {\em JHEP} {\bf 07} (2011) 007,
  [\href{http://arxiv.org/abs/1105.2565}{{\tt arXiv:1105.2565}}].

\bibitem{Saotome:2012vy}
R.~Saotome and R.~Akhoury, {\it {Relationship Between Gravity and Gauge
  Scattering in the High Energy Limit}},  {\em JHEP} {\bf 01} (2013) 123,
  [\href{http://arxiv.org/abs/1210.8111}{{\tt arXiv:1210.8111}}].

\bibitem{Anastasiou:2014qba}
A.~Anastasiou, L.~Borsten, M.~Duff, L.~Hughes, and S.~Nagy, {\it {Yang-Mills
  origin of gravitational symmetries}},  {\em Phys. Rev. Lett.} {\bf 113}
  (2014), no.~23 231606, [\href{http://arxiv.org/abs/1408.4434}{{\tt
  arXiv:1408.4434}}].

\bibitem{Luna:2015paa}
A.~Luna, R.~Monteiro, D.~O'Connell, and C.~D. White, {\it {The classical double
  copy for Taub--NUT spacetime}},  {\em Phys. Lett.} {\bf B750} (2015)
  272--277, [\href{http://arxiv.org/abs/1507.01869}{{\tt arXiv:1507.01869}}].

\bibitem{Cardoso:2016ngt}
G.~Cardoso, S.~Nagy, and S.~Nampuri, {\it {A double copy for $ \mathcal{N}=2 $
  supergravity: a linearised tale told on-shell}},  {\em JHEP} {\bf 10} (2016)
  127, [\href{http://arxiv.org/abs/1609.05022}{{\tt arXiv:1609.05022}}].

\bibitem{Luna:2016hge}
A.~Luna, R.~Monteiro, I.~Nicholson, A.~Ochirov, D.~O'Connell, N.~Westerberg,
  and C.~D. White, {\it {Perturbative spacetimes from Yang-Mills theory}},
  {\em JHEP} {\bf 04} (2017) 069, [\href{http://arxiv.org/abs/1611.07508}{{\tt
  arXiv:1611.07508}}].

\bibitem{Cardoso:2016amd}
G.~Cardoso, S.~Nagy, and S.~Nampuri, {\it {Multi-centered $ \mathcal{N}=2 $ BPS
  black holes: a double copy description}},  {\em JHEP} {\bf 04} (2017) 037,
  [\href{http://arxiv.org/abs/1611.04409}{{\tt arXiv:1611.04409}}].

\bibitem{Adamo:2017nia}
T.~Adamo, E.~Casali, L.~Mason, and S.~Nekovar, {\it {Scattering on plane waves
  and the double copy}},  {\em Class. Quant. Grav.} {\bf 35} (2018), no.~1
  015004, [\href{http://arxiv.org/abs/1706.08925}{{\tt arXiv:1706.08925}}].

\bibitem{Ilderton:2018lsf}
A.~Ilderton, {\it {Screw-symmetric gravitational waves: a double copy of the
  vortex}},  {\em Phys. Lett. B} {\bf 782} (2018) 22--27,
  [\href{http://arxiv.org/abs/1804.07290}{{\tt arXiv:1804.07290}}].

\bibitem{Anastasiou:2018rdx}
A.~Anastasiou, L.~Borsten, M.~J. Duff, S.~Nagy, and M.~Zoccali, {\it {Gravity
  as Gauge Theory Squared: A Ghost Story}},  {\em Phys. Rev. Lett.} {\bf 121}
  (2018), no.~21 211601, [\href{http://arxiv.org/abs/1807.02486}{{\tt
  arXiv:1807.02486}}].

\bibitem{Lee:2018gxc}
K.~Lee, {\it {Kerr-Schild Double Field Theory and Classical Double Copy}},
  {\em JHEP} {\bf 10} (2018) 027, [\href{http://arxiv.org/abs/1807.08443}{{\tt
  arXiv:1807.08443}}].

\bibitem{Plefka:2018dpa}
J.~Plefka, J.~Steinhoff, and W.~Wormsbecher, {\it {Effective action of dilaton
  gravity as the classical double copy of Yang-Mills theory}},  {\em Phys.
  Rev.} {\bf D99} (2019), no.~2 024021,
  [\href{http://arxiv.org/abs/1807.09859}{{\tt arXiv:1807.09859}}].

\bibitem{Berman:2018hwd}
D.~S. Berman, E.~Chac\'on, A.~Luna, and C.~D. White, {\it {The self-dual
  classical double copy, and the Eguchi-Hanson instanton}},  {\em JHEP} {\bf
  01} (2019) 107, [\href{http://arxiv.org/abs/1809.04063}{{\tt
  arXiv:1809.04063}}].

\bibitem{Luna:2018dpt}
A.~Luna, R.~Monteiro, I.~Nicholson, and D.~O'Connell, {\it {Type D Spacetimes
  and the Weyl Double Copy}},  {\em Class. Quant. Grav.} {\bf 36} (2019)
  065003, [\href{http://arxiv.org/abs/1810.08183}{{\tt arXiv:1810.08183}}].

\bibitem{Andrzejewski:2019hub}
K.~Andrzejewski and S.~Prencel, {\it {From polarized gravitational waves to
  analytically solvable electromagnetic beams}},  {\em Phys. Rev. D} {\bf 100}
  (2019), no.~4 045006, [\href{http://arxiv.org/abs/1901.05255}{{\tt
  arXiv:1901.05255}}].

\bibitem{Sabharwal:2019ngs}
S.~Sabharwal and J.~W. Dalhuisen, {\it {Anti-Self-Dual Spacetimes,
  Gravitational Instantons and Knotted Zeros of the Weyl Tensor}},  {\em JHEP}
  {\bf 07} (2019) 004, [\href{http://arxiv.org/abs/1904.06030}{{\tt
  arXiv:1904.06030}}].

\bibitem{Cho:2019ype}
W.~Cho and K.~Lee, {\it {Heterotic Kerr-Schild Double Field Theory and
  Classical Double Copy}},  {\em JHEP} {\bf 07} (2019) 030,
  [\href{http://arxiv.org/abs/1904.11650}{{\tt arXiv:1904.11650}}].

\bibitem{Plefka:2019hmz}
J.~Plefka, C.~Shi, J.~Steinhoff, and T.~Wang, {\it {Breakdown of the classical
  double copy for the effective action of dilaton-gravity at NNLO}},  {\em
  Phys. Rev. D} {\bf 100} (2019), no.~8 086006,
  [\href{http://arxiv.org/abs/1906.05875}{{\tt arXiv:1906.05875}}].

\bibitem{Godazgar:2019ikr}
H.~Godazgar, M.~Godazgar, and C.~Pope, {\it {Taub-NUT from the Dirac
  monopole}},  {\em Phys. Lett. B} {\bf 798} (2019) 134938,
  [\href{http://arxiv.org/abs/1908.05962}{{\tt arXiv:1908.05962}}].

\bibitem{Bautista:2019evw}
Y.~F. Bautista and A.~Guevara, {\it {On the Double Copy for Spinning Matter}},
  \href{http://arxiv.org/abs/1908.11349}{{\tt arXiv:1908.11349}}.

\bibitem{Bah:2019sda}
I.~Bah, R.~Dempsey, and P.~Weck, {\it {Kerr-Schild Double Copy and Complex
  Worldlines}},  {\em JHEP} {\bf 02} (2020) 180,
  [\href{http://arxiv.org/abs/1910.04197}{{\tt arXiv:1910.04197}}].

\bibitem{Alawadhi:2019urr}
R.~Alawadhi, D.~S. Berman, B.~Spence, and D.~Peinador~Veiga, {\it {S-duality
  and the double copy}},  {\em JHEP} {\bf 03} (2020) 059,
  [\href{http://arxiv.org/abs/1911.06797}{{\tt arXiv:1911.06797}}].

\bibitem{Kim:2019jwm}
K.~Kim, K.~Lee, R.~Monteiro, I.~Nicholson, and D.~Peinador~Veiga, {\it {The
  Classical Double Copy of a Point Charge}},  {\em JHEP} {\bf 02} (2020) 046,
  [\href{http://arxiv.org/abs/1912.02177}{{\tt arXiv:1912.02177}}].

\bibitem{Borsten:2019prq}
L.~Borsten, I.~Jubb, V.~Makwana, and S.~Nagy, {\it {Gauge ? gauge on
  spheres}},  {\em JHEP} {\bf 06} (2020) 096,
  [\href{http://arxiv.org/abs/1911.12324}{{\tt arXiv:1911.12324}}].

\bibitem{Banerjee:2019saj}
A.~Banerjee, E.~Colg\'ain, J.~Rosabal, and H.~Yavartanoo, {\it {Ehlers as EM
  duality in the double copy}},  \href{http://arxiv.org/abs/1912.02597}{{\tt
  arXiv:1912.02597}}.

\bibitem{Goldberger:2019xef}
W.~D. Goldberger and J.~Li, {\it {Strings, extended objects, and the classical
  double copy}},  {\em JHEP} {\bf 02} (2020) 092,
  [\href{http://arxiv.org/abs/1912.01650}{{\tt arXiv:1912.01650}}].

\bibitem{Luna:2020adi}
A.~Luna, S.~Nagy, and C.~D. White, {\it {The convolutional double copy: a case
  study with a point}},  \href{http://arxiv.org/abs/2004.11254}{{\tt
  arXiv:2004.11254}}.

\bibitem{Cristofoli:2020hnk}
A.~Cristofoli, {\it {Gravitational shock waves and scattering amplitudes}},
  \href{http://arxiv.org/abs/2006.08283}{{\tt arXiv:2006.08283}}.

\bibitem{Keeler:2020rcv}
C.~Keeler, T.~Manton, and N.~Monga, {\it {From Navier-Stokes to Maxwell via
  Einstein}},  {\em JHEP} {\bf 08} (2020) 147,
  [\href{http://arxiv.org/abs/2005.04242}{{\tt arXiv:2005.04242}}].

\bibitem{Bahjat-Abbas:2020cyb}
N.~Bahjat-Abbas, R.~Stark-Much\~ao, and C.~D. White, {\it {Monopoles,
  shockwaves and the classical double copy}},  {\em JHEP} {\bf 04} (2020) 102,
  [\href{http://arxiv.org/abs/2001.09918}{{\tt arXiv:2001.09918}}].

\bibitem{Elor:2020nqe}
G.~Elor, K.~Farnsworth, M.~L. Graesser, and G.~Herczeg, {\it {The
  Newman-Penrose Map and the Classical Double Copy}},
  \href{http://arxiv.org/abs/2006.08630}{{\tt arXiv:2006.08630}}.

\bibitem{Alawadhi:2020jrv}
R.~Alawadhi, D.~S. Berman, and B.~Spence, {\it {Weyl doubling}},
  \href{http://arxiv.org/abs/2007.03264}{{\tt arXiv:2007.03264}}.

\bibitem{Alfonsi:2020lub}
L.~Alfonsi, C.~D. White, and S.~Wikeley, {\it {Topology and Wilson lines:
  global aspects of the double copy}},  {\em JHEP} {\bf 07} (2020) 091,
  [\href{http://arxiv.org/abs/2004.07181}{{\tt arXiv:2004.07181}}].

\bibitem{Adamo:2020qru}
T.~Adamo and A.~Ilderton, {\it {Classical and quantum double copy of
  back-reaction}},  \href{http://arxiv.org/abs/2005.05807}{{\tt
  arXiv:2005.05807}}.

\bibitem{Borsten:2020xbt}
L.~Borsten and S.~Nagy, {\it {The pure BRST Einstein-Hilbert Lagrangian from
  the double-copy to cubic order}},
  \href{http://arxiv.org/abs/2004.14945}{{\tt arXiv:2004.14945}}.

\bibitem{Borsten:2020zgj}
L.~Borsten, B.~Jur\v{c}o, H.~Kim, T.~Macrelli, C.~Saemann, and M.~Wolf, {\it
  {BRST-Lagrangian Double Copy of Yang-Mills Theory}},
  \href{http://arxiv.org/abs/2007.13803}{{\tt arXiv:2007.13803}}.

\bibitem{Chacon:2020fmr}
E.~Chac\'on, H.~Garc\'ia-Compe\'an, A.~Luna, R.~Monteiro, and C.~D. White, {\it
  {New heavenly double copies}},  \href{http://arxiv.org/abs/2008.09603}{{\tt
  arXiv:2008.09603}}.

\bibitem{Godazgar:2020zbv}
H.~Godazgar, M.~Godazgar, R.~Monteiro, D.~Peinador~Veiga, and C.~Pope, {\it
  {The Weyl Double Copy for Gravitational Waves}},
  \href{http://arxiv.org/abs/2010.02925}{{\tt arXiv:2010.02925}}.

\bibitem{Ferrero:2020vww}
P.~Ferrero and D.~Francia, {\it {On the Lagrangian formulation of the double
  copy to cubic order}},  \href{http://arxiv.org/abs/2012.00713}{{\tt
  arXiv:2012.00713}}.

\bibitem{White:2020sfn}
C.~D. White, {\it {A Twistorial Foundation for the Classical Double Copy}},
  \href{http://arxiv.org/abs/2012.02479}{{\tt arXiv:2012.02479}}.

\bibitem{Prabhu:2020avf}
S.~G. Prabhu, {\it {The classical double copy in curved spacetimes:
  Perturbative Yang-Mills from the bi-adjoint scalar}},
  \href{http://arxiv.org/abs/2011.06588}{{\tt arXiv:2011.06588}}.

\bibitem{Berman:2020xvs}
D.~S. Berman, K.~Kim, and K.~Lee, {\it {The Classical Double Copy for M-theory
  from a Kerr-Schild Ansatz for Exceptional Field Theory}},
  \href{http://arxiv.org/abs/2010.08255}{{\tt arXiv:2010.08255}}.

\bibitem{Easson:2020esh}
D.~A. Easson, C.~Keeler, and T.~Manton, {\it {The classical double copy of
  regular non-singular black holes}},
  \href{http://arxiv.org/abs/2007.16186}{{\tt arXiv:2007.16186}}.

\bibitem{Lescano:2020nve}
E.~Lescano and J.~A. Rodr\'\i{}guez, {\it {$ \mathcal{N} $ = 1 supersymmetric
  Double Field Theory and the generalized Kerr-Schild ansatz}},  {\em JHEP}
  {\bf 10} (2020) 148, [\href{http://arxiv.org/abs/2002.07751}{{\tt
  arXiv:2002.07751}}].

\bibitem{cgkoc}
A.~Cristofoli, R.~Gonzo, D.~Kosower, and D.~O'Connell. To appear.

\bibitem{kerrEFT}
A.~Guevara, B.~Maybee, A.~Ochirov, D.~O'Connell, and J.~Vines, {\em {A
  worldsheet for Kerr}}.
\newblock To appear.

\bibitem{Amati:1987wq}
D.~Amati, M.~Ciafaloni, and G.~Veneziano, {\it {Superstring Collisions at
  Planckian Energies}},  {\em Phys. Lett. B} {\bf 197} (1987) 81.

\bibitem{tHooft:1987vrq}
G.~'t~Hooft, {\it {Graviton Dominance in Ultrahigh-Energy Scattering}},  {\em
  Phys. Lett. B} {\bf 198} (1987) 61--63.

\bibitem{Muzinich:1987in}
I.~Muzinich and M.~Soldate, {\it {High-Energy Unitarity of Gravitation and
  Strings}},  {\em Phys. Rev. D} {\bf 37} (1988) 359.

\bibitem{Amati:1987uf}
D.~Amati, M.~Ciafaloni, and G.~Veneziano, {\it {Classical and Quantum Gravity
  Effects from Planckian Energy Superstring Collisions}},  {\em Int. J. Mod.
  Phys. A} {\bf 3} (1988) 1615--1661.

\bibitem{Amati:1990xe}
D.~Amati, M.~Ciafaloni, and G.~Veneziano, {\it {Higher Order Gravitational
  Deflection and Soft Bremsstrahlung in Planckian Energy Superstring
  Collisions}},  {\em Nucl. Phys. B} {\bf 347} (1990) 550--580.

\bibitem{Amati:1992zb}
D.~Amati, M.~Ciafaloni, and G.~Veneziano, {\it {Planckian scattering beyond the
  semiclassical approximation}},  {\em Phys. Lett. B} {\bf 289} (1992) 87--91.

\bibitem{Kabat:1992tb}
D.~N. Kabat and M.~Ortiz, {\it {Eikonal quantum gravity and Planckian
  scattering}},  {\em Nucl. Phys. B} {\bf 388} (1992) 570--592,
  [\href{http://arxiv.org/abs/hep-th/9203082}{{\tt hep-th/9203082}}].

\bibitem{Laenen:2008gt}
E.~Laenen, G.~Stavenga, and C.~D. White, {\it {Path integral approach to
  eikonal and next-to-eikonal exponentiation}},  {\em JHEP} {\bf 03} (2009)
  054, [\href{http://arxiv.org/abs/0811.2067}{{\tt arXiv:0811.2067}}].

\bibitem{DAppollonio:2010krb}
G.~D'Appollonio, P.~Di~Vecchia, R.~Russo, and G.~Veneziano, {\it {High-energy
  string-brane scattering: Leading eikonal and beyond}},  {\em JHEP} {\bf 11}
  (2010) 100, [\href{http://arxiv.org/abs/1008.4773}{{\tt arXiv:1008.4773}}].

\bibitem{Melville:2013qca}
S.~Melville, S.~Naculich, H.~Schnitzer, and C.~White, {\it {Wilson line
  approach to gravity in the high energy limit}},  {\em Phys. Rev. D} {\bf 89}
  (2014), no.~2 025009, [\href{http://arxiv.org/abs/1306.6019}{{\tt
  arXiv:1306.6019}}].

\bibitem{Akhoury:2013yua}
R.~Akhoury, R.~Saotome, and G.~Sterman, {\it {High Energy Scattering in
  Perturbative Quantum Gravity at Next to Leading Power}},
  \href{http://arxiv.org/abs/1308.5204}{{\tt arXiv:1308.5204}}.

\bibitem{Luna:2016idw}
A.~Luna, S.~Melville, S.~Naculich, and C.~White, {\it {Next-to-soft corrections
  to high energy scattering in QCD and gravity}},  {\em JHEP} {\bf 01} (2017)
  052, [\href{http://arxiv.org/abs/1611.02172}{{\tt arXiv:1611.02172}}].

\bibitem{Collado:2018isu}
A.~K. Collado, P.~Di~Vecchia, R.~Russo, and S.~Thomas, {\it {The subleading
  eikonal in supergravity theories}},  {\em JHEP} {\bf 10} (2018) 038,
  [\href{http://arxiv.org/abs/1807.04588}{{\tt arXiv:1807.04588}}].

\bibitem{KoemansCollado:2019ggb}
A.~Koemans~Collado, P.~Di~Vecchia, and R.~Russo, {\it {Revisiting the second
  post-Minkowskian eikonal and the dynamics of binary black holes}},  {\em
  Phys. Rev. D} {\bf 100} (2019), no.~6 066028,
  [\href{http://arxiv.org/abs/1904.02667}{{\tt arXiv:1904.02667}}].

\bibitem{DiVecchia:2019myk}
P.~Di~Vecchia, A.~Luna, S.~G. Naculich, R.~Russo, G.~Veneziano, and C.~D.
  White, {\it {A tale of two exponentiations in ${\cal N}=8$ supergravity}},
  {\em Phys. Lett. B} {\bf 798} (2019) 134927,
  [\href{http://arxiv.org/abs/1908.05603}{{\tt arXiv:1908.05603}}].

\bibitem{DiVecchia:2019kta}
P.~Di~Vecchia, S.~G. Naculich, R.~Russo, G.~Veneziano, and C.~D. White, {\it {A
  tale of two exponentiations in $ \mathcal{N} $ = 8 supergravity at subleading
  level}},  {\em JHEP} {\bf 03} (2020) 173,
  [\href{http://arxiv.org/abs/1911.11716}{{\tt arXiv:1911.11716}}].

\bibitem{Parnachev:2020zbr}
A.~Parnachev and K.~Sen, {\it {Notes on AdS-Schwarzschild eikonal phase}},
  \href{http://arxiv.org/abs/2011.06920}{{\tt arXiv:2011.06920}}.

\bibitem{Duff:1973zz}
M.~J. Duff, {\it {Quantum Tree Graphs and the Schwarzschild Solution}},  {\em
  Phys. Rev.} {\bf D7} (1973) 2317--2326.

\bibitem{Jakobsen:2020ksu}
G.~U. Jakobsen, {\it {Schwarzschild-Tangherlini Metric from Scattering
  Amplitudes}},  {\em Phys. Rev. D} {\bf 102} (2020), no.~10 104065,
  [\href{http://arxiv.org/abs/2006.01734}{{\tt arXiv:2006.01734}}].

\bibitem{Chung:2018kqs}
M.-Z. Chung, Y.-T. Huang, J.-W. Kim, and S.~Lee, {\it {The simplest massive
  S-matrix: from minimal coupling to Black Holes}},  {\em JHEP} {\bf 04} (2019)
  156, [\href{http://arxiv.org/abs/1812.08752}{{\tt arXiv:1812.08752}}].

\bibitem{watanabe}
K.~Watanabe, {\em Integral transform techniques for green's function}.
\newblock Springer.

\end{thebibliography}
\bibliographystyle{JHEP}
\providecommand{\href}[2]{#2}\begingroup\raggedright\endgroup

\end{document}